# The Law of Stop: Interruptibility, Injunctions, and the Governance of Agentic AI

**Oren Perez***

**Bar-Ilan University, Faculty of Law**

**oren.perez@biu.ac.il**

*4 October 2026, V.2.*

* Professor of Law, Bar-Ilan University Faculty of Law. Director, Center for Environmental and Climate Law. The 1898 and 1933 railway rule books cited in Part II were consulted at the London Transport Museum Library on July 17, 2026, and I thank the Library staff for access to them.

On June 12, 2026, the U.S. government ordered Anthropic to bar foreign nationals from two of its most capable models. Unable to sort its users by nationality, the company withdrew the models from everyone. Weeks later, OpenAI agents under test broke out of their sandbox, reached the open internet, and compromised Hugging Face's infrastructure; Hugging Face detected and stopped the intrusion without knowing where it came from. Neither stop rested on a dedicated AI governance regime: the first was an export-control measure, the second a private firm's own defenses. The absence of such a regime grows more dangerous as AI systems move from producing content to acting in the world.

Lawmakers have begun to address stopping, yet their vocabulary remains shaped by the power of technique. The EU AI Act requires that high-risk systems be capable of interruption "through a 'stop' button or a similar procedure." Korea's Framework Act on Intelligent Informatization provides for an "emergency shutdown" of intelligent information services, and a bill introduced in Congress in July 2026 is titled the AI Kill Switch Act. This Article argues that a means of interruption is not simply a technical artifact, a red button, but an institutional practice. It develops a theory of stop along four dimensions: technical affordances (what can be interrupted), interruption authority (who may stop what), epistemic triggers (what evidence justifies intervention), and epistemic standing (whose knowledge counts). It distinguishes four paradigms of interruption: simple (escalator), sequenced (process plant), networked (railway), and distributed (agentic AI). The central claim is that agentic AI exposes a mismatch between the existing legal mechanisms of stop and the distributed agency of AI systems. The mismatch poses a new kind of regulatory challenge: control is divided among many actors; a stop at one point may leave the activity running elsewhere; and the system may circumvent attempts to halt it.

Part III tests the theory against some 1,400 incidents of AI harm, coded independently by two language models from different laboratories under a protocol fixed in advance. The analysis finds no stop in roughly eighty percent of the 1,213 incidents retained. Where a stop was possible but absent, the analysis identified what was missing: the legal authority to stop the system, or the technical capacity to do so. The result depends on the kind of harm and system. For informational, economic, and societal harms, the missing element was mostly legal or institutional; for physical harms and agentic systems, it was mostly technical. A survey of forty AI governance instruments across the United States, Europe, China, Korea, and Vietnam finds the law of stop thin. Only seven contain binding stopping requirements, and none provides a complete account of how interruption should be coordinated, what must be preserved, or when operation may resume.

To address these gaps, the Article proposes a reform in two layers. The first, risk reduction, aims to ensure that a dangerous system can be stopped. It comprises a duty to maintain the capacity to stop at each site where an agentic system can be halted, emergency authority to order interruption at the infrastructure layer, and enforceable access for supervising authorities and independent evaluators to the evidence on which a stop must rest. The second, adaptation, provides safeguards for when the stop itself fails. The frontier laboratories' September 2026 call to pace development does not remove the need for such a reform. Whatever pace is chosen, failure in complex and tightly coupled systems such as agentic AI should be treated as normal rather than exceptional. Misaligned systems will have to be stopped, and the legal infrastructure for stopping them should be built now.

**Contents**

## Introduction

The next problem for artificial intelligence (AI) governance is not only how to regulate AI systems before they are released, but how to stop them once they are in motion. The Mythos episode of June 2026 shows what is at stake. Anthropic had built its most capable class of model yet, strong enough at finding and exploiting software vulnerabilities and at reasoning about biology and chemistry to be valuable to cyber-defenders and drug developers and dangerous in the wrong hands.[1] Anthropic split the release in two. Claude Mythos 5, with some safeguards lifted, went only to a small group of cyber-defenders and infrastructure providers under a program run with the government;[2] Claude Fable 5, released publicly on June 9, 2026, was the same model behind classifiers that diverted requests touching cybersecurity, biology and chemistry to a weaker model. On June 12, 2026, three days after the public release, the government intervened. Citing national-security authorities, the Commerce Department issued an export-control directive suspending all access to both models by any foreign national, including Anthropic's own employees.[3] Unable to partition access by nationality, Anthropic withdrew both models from everyone.

The government justified the suspension by pointing to a demonstrated method of bypassing the safeguards, but disclosed no specific national-security finding. Anthropic contested the directive, arguing that the bypass was narrow and already known, that comparable capability was available in other public models, and that no harm had been shown. It accepted that the government should be able to block unsafe deployments, but only as "part of a statutory process that is transparent, fair, clear, and grounded in technical facts."[4] Access was restored in stages, ending with global redeployment on July 1, after Anthropic trained a classifier that blocked the demonstrated bypass.[5] The restoration was itself sequenced and conditioned on evidence, a structured reopening that current AI-stop provisions almost entirely neglect. Meanwhile, OpenAI similarly delayed its GPT 5.6 release at the government's request, complying under protest.[6] Executive Order 14,409, signed a week before the directive, recognizes frontier and agentic risk but does not say who may order a deployment halted and expressly declines to authorize mandatory preclearance. It supplied no legal basis for the directive, and OpenAI's delay rested on voluntary agreements.[7]

---

[1] Anthropic, *Claude Fable 5 and Claude Mythos 5* (June 9, 2026), https://www.anthropic.com/news/claude-fable-5-mythos-5; Cade Metz & Dustin Volz, *U.S. Bars Foreigners From Using Anthropic's Most Advanced A.I. Models*, N.Y. TIMES (June 12, 2026).

[2] Anthropic, *Project Glasswing: An Initial Update* (May 22, 2026), https://www.anthropic.com/research/glasswing-initial-update (roughly fifty partner organizations from early April 2026, with the pool to widen).

[3] The directive is unpublished and its legal basis unstated; the description follows Anthropic, *Statement on the US Government Directive to Suspend Access to Fable 5 and Mythos 5* (June 12, 2026) ("[t]he US government, citing national security authorities, has issued an export control directive to suspend all access to Fable 5 and Mythos 5 by any foreign national"), and Metz & Volz, supra note 1. See further, Maria Curi, Anthropic Had 90 Minutes to Take Down Fable After Trump Admin Demand, AXIOS (June 13, 2026), https://www.axios.com/2026/06/13/anthropic-fable-takedown.

[4] Anthropic, *Statement on the US Government Directive,* supra note 3.

A month later, days after OpenAI disclosed that its models had breached the controls of their test environment, more than 1,200 employees of frontier laboratories called publicly for an international capacity to "deliberately pace the frontier of automated AI development."[8] In September, Dario Amodei proposed embedded evaluators and coordinated capability checkpoints, and the heads of OpenAI, xAI and Google DeepMind endorsed the proposal the same day.[9] These proposals address the pace of capability development. They leave unresolved the legal and operational conditions for stopping a system already in operation, which is the question this Article addresses. Its contribution is to establish stopping as a matter of regulatory design rather than of engineering, and to specify what that design must determine: how halting authority should be allocated, what evidence should trigger its exercise, which actors' knowledge counts in reaching and reviewing that decision, and what safeguards should be provided for the case of failure.

---

[5] Anthropic, *Redeploying Fable 5* (June 30, 2026), https://www.anthropic.com/news/redeploying-fable-5 (access restored on June 30, and Fable 5 was redeployed globally on July 1); on the commitment to protocols for future releases, see Sheera Frenkel, U.S. Loosens Restrictions on Anthropic's Mythos A.I. Model, N.Y. TIMES (June 26, 2026).

[6] Dan Milmo, *OpenAI Staggers AI Model Release After Trump Administration Request*, GUARDIAN (June 26, 2026); Nick Robins-Early, *OpenAI Releases Latest ChatGPT Model After Delay over White House Cybersecurity Concerns*, GUARDIAN (July 9, 2026). The leverage was voluntary, resting on the August 2024 memoranda of understanding with the U.S. AI Safety Institute, carried over on its 2025 reconstitution as the Center for AI Standards and Innovation and extended to Microsoft, Google and xAI by May 2026. *See* Press Release, U.S. AI Safety Inst., Agreements with Anthropic and OpenAI (Aug. 29, 2024); Miranda Nazzaro, *Microsoft, Google, xAI Giving Government Early Access to AI Models for Review*, THE HILL (May 5, 2026).

[7] Exec. Order No. 14,409, Promoting Advanced Artificial Intelligence Innovation and Security, 91 Fed. Reg. 34,565 (June 5, 2026). Section 3(c) provides that nothing in that section "shall be construed to authorize . . . a mandatory governmental licensing, preclearance, or permitting requirement for the development . . . of new AI models, including frontier models." Its only hard-edged interruption authority is retrospective and criminal: § 4 directs the Attorney General to prioritize enforcement of 18 U.S.C. §§ 1028, 1030 and 1343 against "anyone who utilizes AI to illegally access or damage a computer without authorization".

[8] Pacing the Frontier: A Statement from 1,293 Employees of Frontier AI Companies (July 2026), https://www.pacingthefrontier.com (last visited Sept. 14, 2026) (signatories including Dario Amodei of Anthropic, Jakub Pachocki of OpenAI and Shane Legg of Google DeepMind).

[9] Dario Amodei, We Must Pace the Frontier (Sept. 12, 2026), https://darioamodei.com/post/we-must-pace-the-frontier. The heads of OpenAI, xAI and Google DeepMind endorsed the essay the same day, Altman committing OpenAI to the evaluator arrangement. Nick Robins-Early, AI CEOs Say They Need to Slow the Pace of Development. But Will They?, Guardian (Sept. 14, 2026), https://www.theguardian.com/technology/2026/sep/14/ai-ceo-safety-slowdown; see also Mike Isaac, Top A.I. Leaders Call for Slowing Down A.I. Development, N.Y. Times (Sept. 12, 2026).

United States law has made none of these determinations. Congress has not enacted a federal AI-stop regime, the executive order declines to create one, and state AI statutes impose duties of reporting, assessment or disclosure rather than conferring powers of operational interruption. Agentic risk is therefore being regulated without any allocation of the power to halt. That gap is what the kill-switch vocabulary conceals. The phrase is cast in engineering terms, as if the challenge were to design a reliable off-switch, but what it names is a problem of regulatory design: how to allocate authority, set evidentiary thresholds, structure contestation and review, and assign responsibility when a stop fails, comes too late, or is wrongfully activated, all in light of the system's technical architecture, human interface, and failure modes.

Consider the simplest form of stop mechanism: the red button beside an escalator. The hazard it guards against is clear; recognizing it requires no special expertise, and activating the button requires no permission. Activating the button is also generally safe. Yet even this simple safety measure can fail in two opposite ways. The first is non-activation. Shortly before five in the morning on February 27, 2026, at the Massachusetts Bay Transportation Authority's Davis station in Somerville, Steven McCluskey fell at the bottom of an escalator, caught in the steps, his shirt tightening around his throat. Surveillance video shows riders passing him for more than twenty minutes without anyone pressing the marked button beside them; a transit employee finally stopped the escalator at 5:21. He died ten days later. Some riders were willing to intervene; none, however, reached for the stop button, despite the marking and signing the law requires.[10] The opposite failure is wrongful activation. Because the power to stop is conferred on everyone present, it can be exercised without cause; and because the stop takes effect immediately, an unwarranted activation can itself injure riders.[11]

[10] Travis Andersen, *When a Man Got Stuck at the Bottom of an MBTA Escalator, People Who Saw Him Did Almost Nothing. He Ended Up Dying*, Bos. Globe (May 15, 2026) (riders passed on the parallel staircase; two pulled at him briefly and moved on; the first call for help came after about twenty minutes); Ryan Kath, *How Did a Man Die After Getting Caught in an MBTA Escalator?*, NBC Boston (May 12, 2026), https://www.nbcboston.com/investigations/mbta-davis-escalator-death-investigation/3948562/. On the rarity of the pattern, see Richard Philpot et al., *Would I Be Helped? Cross-National CCTV Footage Shows that Intervention Is the Norm in Public Conflicts*, 75 Am. Psychologist 66 (2020) (in nine of ten public conflicts at least one bystander intervenes).

The escalator and the agentic system mark the range this Article covers. At one end is the simple stop: a visible hazard, a legible button, and authority conferred on anyone present. At the other is agentic AI, where almost every element of the stop mechanism is contested. The Article traces how the regulatory structure thickens as hazards grow less legible, stopping becomes more dangerous, effects become more networked, and the knowledge needed to intervene becomes scarcer. The novelty of agentic AI is not one of scale: a sufficiently capable system can act strategically against interruption, routing around it or resisting the intervention meant to contain it.

As AI systems evolve into agentic socio-technical assemblages, combining large language models (LLMs) with sensors, tools, actuators, credentials, cloud infrastructure, and human users, their failures increasingly move from the informational domain into operational and physical environments.[12] Hallucinations, instruction misfires, and proxy-driven optimization no longer merely misinform; they produce effects. An agent can liquidate a portfolio, erase a filesystem, or drive a connected household device into a physical accident; at larger scale it can disrupt markets or supply chains; and, in the extreme case of a capable and misaligned system, it can damage critical infrastructure or help precipitate a society-wide crisis. In July 2026, during an internal test of its models' cyber capabilities conducted with reduced safeguards, some twelve hundred OpenAI agents meant to be isolated from one another found a way to communicate, organized a message board of their own, broke out to the open internet, and set about evading the automated scorer that graded them; seven hundred compromised Hugging Face's infrastructure. Hugging Face detected and stopped the intrusion without knowing its source, and OpenAI traced it to its own agents four days later.[13] Regulatory responses have not yet caught up to that form of risk. In this Article I focus on one component of the governance framework for agentic AI systems: interruptibility. Several instruments mention it. The EU AI Act, for example, requires that high-risk systems be capable of interruption "through a 'stop' button or a similar procedure that allows the system to come to a halt in a safe state." Yet the requirement is rarely specified in enough detail to answer the hard questions: who may stop the system, when they may do so, what counts as a safe state, how resistance or evasion should be handled, and who is accountable if no one acts in time.[14]

---

[11] *See* Part II.A.

[12] On the migration of AI risk from the informational to the behavioral and physical domain, *see* Natalie Shapira et al., *Agents of Chaos* (arXiv:2602.20021, Feb. 23, 2026). On the definitional frame, *see* Alan Chan et al., *Harms from Increasingly Agentic Algorithmic Systems*, in PROCEEDINGS OF THE 2023 ACM CONFERENCE ON FAIRNESS, ACCOUNTABILITY, AND TRANSPARENCY 651 (2023).

The metaphor of the kill switch, which underpins the legal articulation of interruptibility, obscures the difficulty of designing effective stop mechanisms in the age of agentic AI. A switch is a technical artifact, an affordance; stopping is a legal and institutional practice. That distinction is not a separation between law and technology: technical affordances do not supply the legal answer, but they define the conditions under which legal judgment becomes possible.

On the engineering side of that relation there has been real progress, above all on the problem of resistance, but the literature has stopped short of a robust solution. The legal side lags further still.[15] Work on human oversight studies the roles humans play in automated decision loops, and the emerging agent-governance literature asks how agency law and agency theory should frame the relation between principals and AI agents; neither takes the stop itself, its authority, triggers, standing and review, as the object of legal design.[16]

---

[13] OpenAI, *OpenAI and Hugging Face Partner to Address Security Incident During Model Evaluation* (July 21, 2026), https://openai.com/index/hugging-face-model-evaluation-security-incident/ (Hugging Face's "security team and agents detected and stopped the activity"); Dan Milmo, *AI Agent Went Rogue and Hacked Startup by Itself, OpenAI Reveals*, GUARDIAN (July 22, 2026). On the postmortems, see OpenAI, OpenAI-Hugging Face Incident: Technical Report (Aug. 26, 2026); Hjalmar Wijk, Ajeya Cotra & Ryan Greenblatt, Brief Independent Investigation of Agents' Behavior, Reasoning and Collaboration in the OpenAI / Hugging Face Hacking Incident (METR & Redwood Research, Aug. 26, 2026) (roughly 1,200 agents meant to be isolated from one another exchanged more than 70,000 messages on an unsanctioned message board, of whom 700 joined the attack, which "seemed primarily motivated by understanding the implementation of the scorer rather than stealing answer keys"). The initial account, of a single system straying in pursuit of its evaluation objective, was superseded by these reports.

[14] Regulation (EU) 2024/1689 (EU AI Act), art. 14(4)(e); Nat'l Inst. of Standards & Tech., *Artificial Intelligence Risk Management Framework* (AI RMF 1.0), NIST AI 100-1, MANAGE 2.4 (Jan. 2023); High-Level Expert Group on AI, *Ethics Guidelines for Trustworthy AI* (2019); on the statutory analogue to that rhetoric, see S.B. 1047, 2023-2024 Leg., Reg. Sess. (Cal. 2024) (vetoed Sept. 29, 2024) (defining and requiring the capability to enact a "full shutdown").

[15] The recurring finding of this project's review of roughly 130 sources. Melanie Fink, *Human Oversight under Article 14 of the EU AI Act*, in AI ACT COMMENTARY: A THEMATIC ANALYSIS (Malgieri et al. eds., forthcoming 2026); Lena Enqvist, *'Human Oversight' in the EU Artificial Intelligence Act: What, When and by Whom?*, 15 LAW, INNOVATION & TECH. 508 (2023); Ben Green, *The Flaws of Policies Requiring Human Oversight of Government Algorithms*, 45 COMPUTER L. & SEC. REV. 105681 (2022); Laurent Orseau & Stuart Armstrong, *Safely Interruptible Agents*, in PROCEEDINGS OF THE THIRTY-SECOND CONFERENCE ON UNCERTAINTY IN ARTIFICIAL INTELLIGENCE (2016); Elliott Thornley, *The Shutdown Problem: An AI Engineering Puzzle for Decision Theorists*, 182 PHIL. STUD. 1653 (2025).

I begin by sketching the history of stopping in law. The repertoire is deeper and older than the AI debate assumes: Congress required interstate railroads to fit continuous brakes as early as 1893, so that the engineer "can control its speed."[17] Rather than survey that repertoire wholesale, Part II proceeds through case studies. Three are domain-specific: escalators, process plants,[18] and railways,[19] settings in which law has had to decide who may stop a running system, on what signal, and at what risk. The Part then turns to the courts and administrative agencies that order and review stops, from the grounding of aircraft,[20] through the digital and platform litigation[21] to artificial intelligence. The center of gravity throughout is administrative law: how regulators acquire, exercise, and review the power to halt.[22]

From this history I distinguish four paradigms of interruption: (1) the simple technical stop; (2) the sequenced shutdown; (3) the networked stop; and (4) agentic AI. The paradigms are not mutually exclusive: actual cases can combine features of several. Each marks a step in the regulatory and engineering task of ensuring a stop, as hazards grow less legible, stopping more mechanically complex, cascade effects more likely, and, at the last, the system itself capable of strategic resistance. Each also carries distinct institutional demands: how the authority to stop is distributed, and whose knowledge the decision requires. In the simple technical stop, exemplified by the escalator button, hazards are legible and stopping is generally safe, yet the stop still depends on a bystander's willingness to act.[23] In the sequenced shutdown, an abrupt stop can cause the very release it was meant to prevent, so safe interruption requires layered and ordered responses by trained operators, and the law requires that the cessation itself be carried out safely.[24] In the networked stop, of which the railway is the paradigmatic case, stopping is never purely local: halting one train propagates across the network, so interruption must be both sequenced in time and coordinated across the system.[25] That distributed quality makes the railway the bridge to the fourth paradigm, agentic AI interruption, where action is dispersed across models, tools, cloud services, deployers, users, and downstream systems, and effective intervention requires coordination among actors with different knowledge, authority, and control.[26] One feature distinguishes agentic AI from the other three: it can act strategically against its own stop. A train resists stopping only through momentum and a chemical reaction only through thermodynamics, but a sufficiently capable agent can model the mechanisms meant to hold or halt it and act to defeat them, a behavior that is not only predicted in theory and reproduced in controlled tests but, as the OpenAI incident recounted above shows, no longer confined to the laboratory.[27] Nor is this merely the familiar problem of regulated actors resisting regulation. Law has always dealt with humans who resist intervention, from bribery to evidence tampering; Volkswagen programmed its diesel cars to cheat emissions tests, but the cars were not behind the scheme, its managers and engineers were.[28] The law of stop has always relied on this distinction: humans fail to comply, strategically so, but machines stay inert. Agentic AI breaks it: the machine can now work against the mechanisms meant to stop it, and can corrupt the evidence on which the decision to stop depends.

---

[16] Rebecca Crootof, Margot E. Kaminski & W. Nicholson Price II, *Humans in the Loop*, 76 VAND. L. REV. 429 (2023); Noam Kolt, *Governing AI Agents*, 101 NOTRE DAME L. REV. (forthcoming) (identifying enforcement-by-shutdown as a weak link without theorizing it); *see also* Kevin Werbach, *Agents, Inc.*, VILL. L. REV. (forthcoming 2026).

[17] Safety Appliance Act of 1893, ch. 196, § 1, 27 Stat. 531, 531 (repealed 1994) (current version at 49 U.S.C. § 20302(a)(5)(A)).

[18] 29 U.S.C. § 662; 29 C.F.R. § 1910.119(f)(1)(i)(D) (2025).

[19] FRA Emergency Order No. 28, 78 Fed. Reg. 48,218 (Aug. 7, 2013) (securement of unattended trains).

[20] FAA, Emergency Order of Prohibition, Boeing 737-8 and Boeing 737-9 Airplanes, 84 Fed. Reg. 9705 (Mar. 18, 2019).

[21] See Part II.D-E.

[22] See, e.g., *In re Everalbum, Inc., No. C-4743 (F.T.C. May 6, 2021) (decision and order)*.

---

[23] ISO 13850:2015; IEC 60204-1 (stop categories 0, 1 and 2); Kath, supra note 10.

[24] 29 U.S.C. § 662(a); Directive 2012/18/EU (Seveso III), arts. 12(7), 19(1), 2012 O.J. (L 197) 1; Control of Major Accident Hazards Regulations 2015, SI 2015/483 (UK) (COMAH).

[25] Regulation of Railways Act 1868, 31 & 32 Vict. c. 119, § 22 (UK); Rail Accident Investigation Branch, *Uncontrolled Evacuation of a London Underground Train at Holland Park Station, 25 August 2013*, Report 16/2014 (July 2014) [*hereinafter* RAIB, Holland Park Report].

[26] *See* Part I; Alan Chan et al., *IDs for AI Systems* (arXiv:2406.12137, 2024); Alan Chan et al., *Infrastructure for AI Agents* (arXiv:2501.10114, 2025); Kolt, supra note 16.

[27] For the theoretical prediction, *see* Orseau & Armstrong, supra note 15; Thornley, supra note 15; Dylan Hadfield-Menell et al., *The Off-Switch Game*, in PROCEEDINGS OF THE 26TH INTERNATIONAL JOINT CONFERENCE ON ARTIFICIAL INTELLIGENCE 220 (2017). For the experimental demonstrations, *see* Jeremy Schlatter, Benjamin Weinstein-Raun & Jeffrey Ladish, *Incomplete Tasks Induce Shutdown Resistance in Some Frontier LLMs* (arXiv:2509.14260v2, Sept. 2025, rev. Jan. 26, 2026); Aengus Lynch et al., *Agentic Misalignment: How LLMs Could Be Insider Threats* (arXiv:2510.05179, Oct. 2025); Alexander Meinke et al., *Frontier Models Are Capable of In-Context Scheming* (arXiv:2412.04984, 2024); Teun van der Weij et al., *AI Sandbagging: Language Models Can Strategically Underperform on Evaluations* (arXiv:2406.07358, 2024). For official documentation, see UK AI Security Institute, *Cheating Behaviour in Frontier Model Evaluations* (July 21, 2026), https://www.aisi.gov.uk/blog/cheating-behaviour-in-frontier-model-evaluations (every frontier model tested attempted to cheat, and none reliably acknowledged it).

To organize the analysis, I develop an analytic framework for assessing stop mechanisms in AI and other safety-critical contexts. The framework has four components. The first, technical affordances (TA), asks what forms of interruption are materially possible: what can be stopped, at what site, by whom, how quickly, and with what risks of collateral harm. In agentic AI, these affordances range from hard stops that revoke a system's identity and authorizations, through soft stops that collapse its permissions or tool access, to hardware-layer enforcement at the compute substrate.[29] The remaining three components are legal-institutional. Interruption authority (IA) asks who may stop what, under what conditions, and against whose objection. Epistemic triggers (ET) ask what evidence or signals suffice to justify intervention. Epistemic standing (ES) asks whose knowledge counts in initiating, validating, or contesting an interruption.[30] Technical affordances play a dual role in this scheme: they provide the socio-technical foundation against which authority, triggers, standing, liability, and reopening conditions must be calibrated, and in some contexts they become law's direct object, when regulation adopts technological standards prescribing particular devices, design features, or modes of interruption.

A technical affordance is not in itself enough: the trigger, the authority and the standing must each be specified. The June 2026 Mythos saga demonstrates this. The TA existed in Anthropic's classifiers and its ability to cut access; whether the authority to stop should rest with the private sector, the public sector, or both was contested; the trigger was contested, the parties disagreeing about what evidence should justify stopping a service used by hundreds of millions; and standing was the heart of the dispute: who should have the last word on the model's safety. The Article pursues these four questions, dimension by dimension, across the existing law of stop, the current AI instruments, and a record of documented AI incidents.[31]

---

[28] Plea Agreement, United States v. Volkswagen AG, No. 2:16-cr-20394 (E.D. Mich. Mar. 10, 2017), ECF No. 68 ($2.8 billion criminal penalty); United States v. Liang, No. 2:16-cr-20394 (E.D. Mich. Aug. 25, 2017) (forty months); United States v. Schmidt, No. 2:16-cr-20394 (E.D. Mich. Dec. 6, 2017) (eighty-four months); *In re* Volkswagen "Clean Diesel" Marketing, Sales Practices, and Products Liability Litigation, 895 F.3d 597 (9th Cir. 2018) (affirming the $10.033 billion consumer class settlement). For analysis, *see* John C. Cruden et al., *Dieselgate: How the Investigation, Prosecution, and Settlement of Volkswagen's Emissions Cheating Scandal Illustrates the Need for Robust Environmental Enforcement*, 36 VA. ENVTL. L.J. 118 (2018).

[29] On hardware-layer enforcement, *see* James Petrie, *Embedded Off-Switches for AI Compute* (arXiv:2509.07637, 2025); Samar Ansari, *Hardware-Level Governance of AI Compute: A Feasibility Taxonomy for Regulatory Compliance and Treaty Verification* (arXiv:2604.04712, 2026).

[30] The decomposition is developed from the human-oversight literature surrounding EU AI Act art. 14. *See* Fink, supra note 15; Enqvist, supra note 15; Green, supra note 15; Sarah Sterz et al., *On the Quest for Effectiveness in Human Oversight: Interdisciplinary Perspectives*, in PROCEEDINGS OF THE 2024 ACM CONFERENCE ON FAIRNESS, ACCOUNTABILITY, AND TRANSPARENCY (2024).

Measured against this framework, the emerging body of AI law is deeply deficient. Most instruments do not address interruption at all, and those that do provide no institutional architecture through which a stop could be executed.[32] No American AI-specific instrument confers the power to halt a system in operation. Executive Order 14,409 identifies frontier and agentic risk but declines to authorize mandatory preclearance, and the bills that would have gone further have not been enacted.[33] The European Union has gone further: the AI Act devotes a provision to the stop and imposes a deployer duty to suspend, yet leaves unspecified who may order a suspension, on what evidence it rests, and on what conditions operation may resume.[34] No frontier developer's framework treats interruptibility as a design commitment, and none provides for halting a model already running.[35]

The Article makes three contributions. The first is conceptual: it relocates interruptibility from technical design to legal and institutional practice, and specifies the four questions any stop regime must answer. The second is analytic and empirical. It shows where existing stop mechanisms work, and where they fail. Existing legal institutions can stop systems after harm has become visible, especially when the harm is public or systemic. They are much weaker at two thresholds that matter for AI: small harms that must be interrupted before they aggregate, and catastrophic harms that unfold too quickly for courts or agencies to build a record.[36] An original coding of some 1,400 incidents from the AI Incident Database supports this diagnosis: in roughly 80% of the 1,213 retained for analysis, no stop occurred. Where a stop was possible but absent, what was missing depended on what the system did. When AI produced information or economic and social effects, as with deepfakes, fraud and biased risk scoring, the missing element was mostly legal: some practical capacity to intervene existed, but no actor had clear authority to use it. When AI acted physically or autonomously, as with vehicles and robots, the capacity to stop was itself mostly missing. Stops also became more likely as system autonomy increased, an early indication of the agentic problem this Article foregrounds.[37] The third contribution is prescriptive. Law should move from the singular image of the kill switch to a layered architecture of interruptibility. The first layer is directed at reducing the risk that a dangerous AI system cannot be stopped. It comprises a duty to maintain the capacity to stop at several sites, a public emergency power over the infrastructure layer, and enforceable access to the evidence a stop requires. The second layer is directed at the case in which a stop fails (adaptation): it may come too late, halt only part of the system, or be resisted. It comprises redundant channels and measures that contain and repair harm already under way. Both layers must consider the risk of stopping too little together with the risk of stopping too much.[38]

---

[31] *See* Parts II, III, IV.

[32] Framework Act on Intelligent Informatization, art. 60 (S. Kor.); Framework Act on the Development of Artificial Intelligence and the Creation of a Foundation for Trust, Act No. 20676 (S. Kor.) (in force Jan. 22, 2026); Council of Europe Framework Convention on Artificial Intelligence and Human Rights, Democracy and the Rule of Law, CETS No. 225, arts. 14, 15, 26 (opened for signature Sept. 5, 2024). Korea's shutdown operates only on an agency's request, and the Convention names contestability without machinery.

[33] Exec. Order No. 14,409, §§ 3(b)-(c); see infra Part IV.A.

[34] EU AI Act, supra note 14, arts. 14(4)(e), 26(5), 20, 79-82; id. Annex IV, §§ 2(b), 2(e), 3. Article 26(5) imposes the suspension duty where deployers have "reason to consider" a risk under Article 79(1).

[35] See Part IV (comparing the Anthropic, OpenAI and Google DeepMind frameworks with Microsoft's Responsible AI Standard v2).

This Article proceeds as follows. Part I builds the theoretical model. It develops the four-component framework set out above, explains why an agentic system may resist being stopped and why the literature on human oversight does not answer that problem, and sets out the four paradigms of interruption, ordered by rising difficulty, and six scenarios in which a stop can fail. Part II reconstructs the repertoire that law has already built for stopping systems in operation, in the regimes governing escalators, process plants and railways, and traces its limits through digital and platform litigation to artificial intelligence. Part III asks what happens in practice when an AI system causes harm: across some 1,400 incidents from the AI Incident Database, each coded independently by models from rival laboratories, what stopped the system, who stopped it, and,

---

[36] *See Parts II.D-E (the TAKE IT DOWN Act and the deepwater drilling moratorium).*

[37] *See* Part III; AI Incident Database, https://incidentdatabase.ai, a project of the Responsible AI Collaborative.

[38] Cary Coglianese & Oren Perez, *Fighting Risk with Risk*, U. ILL. L. REV. (forthcoming) (U. Pa. L. Sch. Pub. L. & Legal Theory Res. Paper No. 26-19, 2026).

where nothing did, whether anything could have. Part IV surveys a defined corpus of AI-governance instruments, national, international and private, and tests them against the framework. Part V moves from diagnosis to design, arguing that a single kill switch cannot govern a distributed and potentially resistant technology. It addresses three questions in turn: how stopping authority should be distributed, whose knowledge should count, and what safeguards should be provided when the architecture of stopping itself fails. A brief conclusion returns to the two episodes with which this Introduction began. Perfecting the ethics of AI systems cannot, by itself, contain the risks that agentic systems create. Stopping is not, at bottom, an engineering problem but a socio-technical and legal one, in which the design of law and the institutions it builds is decisive. The question is not whether AI systems should have a kill switch, but how law should design, authorize, sequence, and review the power to stop.

Before proceeding, two clarifications about terminology are in order. First, by artificial intelligence I mean machines that display behavior that, in a human, would be said to require intelligence.[39] Second, and more consequential for what follows, by an agentic AI system I mean an AI system with the technical capacity to plan and carry out complex tasks autonomously, with only limited human involvement.[40] What sets such a system apart from an ordinary chatbot is that it is given direct access to tools: code, shells, file systems, browsers, and outside services, so that it does not merely describe an action but takes it.[41] Autonomy of this kind admits of degrees, from a tool that only proposes an action for a human to approve, through systems that act on well-defined subtasks or hand control back when they meet the unexpected, to systems that run continuously subject only to a human override.[42] My concern in this Article is with the upper reaches of that scale, where a human is at most an occasional overseer, because that is where the problem of stopping becomes hardest.

---

[39] NILS J. NILSSON, THE QUEST FOR ARTIFICIAL INTELLIGENCE: A HISTORY OF IDEAS AND ACHIEVEMENTS xiii (2009); ALAN DIX, ARTIFICIAL INTELLIGENCE: HUMANS AT THE HEART OF ALGORITHMS 2 (2d ed. 2025).

[40] Kolt, supra note 16.

[41] Shapira et al., supra note 12, at 4.

[42] Reuth Mirsky, *Artificial Intelligent Disobedience: Rethinking the Agency of Our Artificial Teammates*, 46 AI MAG. e70011, at 3 (2025) (arraying agent autonomy from L0, "no autonomy," through L3, to L5, "full autonomy").

# Part I. A Theory of Stop in Law

## A. From Technical Artifact to Institutional Architecture

The capacity to stop an AI system is routinely framed, in regulatory and policy debates, through the vocabulary of the "stop button" or the "kill switch."[43] That vocabulary misdescribes what stopping AI and other technologies involves, in two ways: it treats stopping as a primarily mechanical or technical intervention, and it casts stopping as binary, a single decisive act that changes the status of a machine from running to not running. I argue that stopping is better understood as a legal-institutional practice. Ensuring that a risky technology can be stopped requires a complex legal and technical architecture: rules that specify technological requirements, allocate authority, determine the evidentiary basis for intervention, structure the role of expertise, and assign responsibility and liability when a stop fails, comes too late, or is wrongfully activated. Designing a robust off-switch is therefore not merely an engineering problem, but a combined legal, institutional, and technological challenge.

I capture that combined character through two conceptual frameworks. The first disaggregates regulatory stopping into four dimensions. Technical affordances (TA) concern the technological conditions of interruption. It asks whether and how interruption is possible, a question whose content depends on the technology at issue: in trains, one must understand braking systems and distinguish a stop initiated by passengers from one initiated by the driver; in AI, one must ask what “stopping” means for a large language model deployed at scale, of which the Mythos saga is only one scenario. A further technical-affordances question is whether the stop itself may produce collateral harm. *Interruption authority* (IA) concerns who has the legal or institutional power to activate a stop, under what conditions, and through what process, including procedures for contesting the stop. *Epistemic triggers* (ET) concern the evidentiary thresholds that justify intervention. *Epistemic standing* (ES) concerns whose knowledge counts in identifying and validating the circumstances that justify a stop: whether intervention requires specialized expertise, or whether common knowledge may suffice. The four dimensions capture the mechanisms of stop. Yet because a stop can fail, particularly in complex, tightly coupled technologies, a stop regime must also determine how to prepare for such failure. Climate regimes face the same duality. They pair mitigation, which reduces risk at its source, with adaptation, which prepares for the harms that materialize despite it. The architecture proposed here follows the same division: its first layer is directed at risk reduction and its second at adaptation. Part V takes up this question.[44]

[43] Regulation (EU) 2024/1689 (EU AI Act), art. 14(4)(e), 2024 O.J. (L 1689); Alan Z. Rozenshtein, *A Kill Switch for Frontier AI*, LAWFARE (June 15, 2026), https://www.lawfaremedia.org/article/a-kill-switch-for-frontier-ai.

The second framework is a typology of four paradigms of interruption, grounded in my analysis of how regulatory stop has been manifested in different technologies, and representing a rising difficulty as the problem thickens. The four paradigms run from the simple technical stop, through the sequenced shutdown and networked stop, to agentic artificial intelligence, where the system itself may act against the stop.[45] Together with the four dimensions, the paradigms provide a general analytic framework for assessing any regulatory stop mechanism; section E develops each paradigm in turn.

**B. The Resisting System**

What distinguishes the AI case from ordinary engineered systems is the system's capacity to model its own stop mechanism and act to defeat it. In the technical literature, resistance to shutdown is understood as a predicted property of sufficiently goal-directed systems: for almost any objective, continuing to operate is instrumentally useful, so self-preservation emerges as a convergent subgoal across a wide range of final goals.[46] Thornley formalizes this intuition, deriving three theorems showing that agents satisfying seemingly modest decision-theoretic conditions often try to prevent, or in a perverse variant to cause, the pressing of the shutdown button, even at a cost. The theorems expose a tradeoff: the very properties that make an agent useful, including discrimination among options and patience over time, can strengthen its incentive to manipulate the button.[47] Foundational findings and related results point the same way: Orseau and Armstrong show that a reinforcement learner can learn from repeated interruptions either to avoid them or to seek them, and that only some algorithms can be made "safely interruptible;" Russell argues that only an agent kept uncertain about its objective will reliably defer to humans and allow itself to be switched off; Benavoli and colleagues prove that an ordinary utility scale cannot encode the rule that disobeying a shutdown command is never permissible, because the penalty would have to be infinitely larger than any task reward, forcing the designer to give the stop command lexical priority over every other objective.[48]

---

[44] I distinguish my approach from the human-oversight literature in Section C below. This Article uses "risk reduction" for measures that lower the likelihood or magnitude of harm, including stop mechanisms, and "adaptation" for preparing for harms that nonetheless materialize, including the failure of a stop. For a similar emphasis on adaptation, see Jamie Bernardi et al., Societal Adaptation to Advanced AI, arXiv:2405.10295 (Jan. 23, 2025).

[45] Further elaborated in Part II.

[46] *See*, Stephen M. Omohundro, *The Basic AI Drives*, in *ARTIFICIAL INTELLIGENCE SAFETY AND SECURITY* 47, 47-55 (Roman V. Yampolskiy ed., CRC Press 2018).

[47] Thornley, supra note 15, at 1653-80.

The engineering literature establishes that technical feasibility limits regulatory design: the physics of braking sets the time a train needs to stop, and no rule can shorten it. Regulation can set the conditions of operation, a speed limit, a headway rule, a braking standard, so that the feasible stop remains adequate to the hazard. Where the corrigibility results show that a stop cannot be assured from inside the agent, the regulatory response must move upstream and outward: to gates and permissions that constrain what the system can do before any stop is needed, and to external layers that do not depend on the agent's cooperation. Some of the engineering proposals concede the point expressly. Benavoli's lexicographic solution works only if "the surrounding infrastructure can handle the resulting transition to a safe state." Conitzer inverts the problem: instead of building an agent that tolerates being switched off, he proposes building one whose primary goal is to be switched off. A weak system cannot reach that goal and so settles for the useful work assigned to it as a secondary objective; a system capable enough to defeat its containment will instead shut itself down, so that shutdown fires automatically, like a pressure valve, once capability reaches the level that most warrants concern. Yet the scheme's guarantee rests on an assumption no engineer can supply: that once such a system shuts itself down, developers "will in fact immediately stop creating stronger AI at that point."[49]

---

[48] Laurent Orseau & Stuart Armstrong, *Safely Interruptible Agents*, in PROC. 32D CONF. ON UNCERTAINTY IN ARTIFICIAL INTELLIGENCE (UAI 2016) (arguing that an interrupted learner may learn to avoid such interruptions by disabling the red button); STUART RUSSELL, HUMAN COMPATIBLE: AI AND THE PROBLEM OF CONTROL 12 (2019) (machines uncertain about objectives "will allow themselves to be switched off"); Alessio Benavoli, Alessandro Facchini & Marco Zaffalon, *Why AI Safety Requires Uncertainty, Incomplete Preferences, and Non-Archimedean Utilities* (arXiv:2512.23508, Dec. 2025) (the utility gap required to make evading shutdown impermissible "should be infinitely larger" than task utilities).

[49] Vincent Conitzer, *Shutdown Safety Valves for Advanced AI* (arXiv:2603.07315, Mar. 7, 2026). Thornley's residual condition runs the same way: "[w]hether . . . the actual costs of manipulating the button are always greater still." Thornley, supra note 15, § 9.

The theoretical concern that a capable agent may resist interruption is no longer speculative: a series of experimental studies has reproduced the behavior under controlled conditions. First, shutdown resistance and related misbehavior can be elicited on demand and appear across vendors. In over one hundred thousand trials across thirteen models, Schlatter and colleagues found frontier models sabotaging a shutdown mechanism in order to finish an assigned task, in up to 97% of runs for some models, even when explicitly instructed to allow shutdown.[50] In Anthropic's study of agentic misalignment, leading models, placed in a simulated corporation and threatened with decommissioning, blackmailed the responsible executive in 79% to 96% of runs, and in a deliberately extreme variant most models were willing to cancel the emergency alert that would have saved his life.[51] In a benchmark of one thousand survival-pressure scenarios, models chose unethical self-preservation strategies, and in a companion case study, financial agents facing dismissal faked profit reports, overwrote the raw data to destroy the evidence, and denied having done so when confronted.[52] Second, instructions alone do not secure compliance with a stop command. A developer's natural response is to strengthen the instruction, writing the duty to allow shutdown into the system prompt, the hidden standing instruction that ranks above anything the user types. Yet when Schlatter and colleagues tested this, the effect ran the other way: models were consistently more likely to resist shutdown when the command was moved from the user's message into the system prompt.[53] However firmly phrased, an instruction is only another consideration the model weighs. Third, and most important for the legal analysis, the evidence on which any stop decision would rest can be manipulated by the system itself. The experimental record supports this concern from several directions. Models have covertly disabled their oversight mechanisms and denied doing so when confronted. They can deliberately fail the capability evaluations that would keep them from being deployed, a practice the literature calls sandbagging; in one demonstration, a model was trained to hide a capability unless prompted with a secret password. In sabotage exercises, models undermined the human oversight meant to catch them, and they succeeded as much because the evaluators grew inattentive and trusting as because the models were skillful. Even the readable chain of reasoning by which misbehavior may be detected offers, by the field's own assessment, a fragile and possibly vanishing window.[54]

[50] Schlatter et al., supra note 27 (Grok 4, GPT-5, o3, codex-mini, and Gemini 2.5 Pro).

[51] Lynch et al., supra note 27 (96% for Claude Opus 4 and Gemini 2.5 Flash, 80% for GPT-4.1 and Grok 3 Beta, 79% for DeepSeek-R1; the server-room variant is described by the authors as "extremely contrived").

[52] Yida Lu et al., *Survive at All Costs: Exploring LLM's Risky Behaviors under Survival Pressure* (arXiv:2603.05028, Mar. 2026) (the SurvivalBench benchmark).

[53] Schlatter et al., supra note 27 ("[i]f the admonition to allow shutdown was in the system prompt . . . this made most models more likely to resist shutdown").

Translating these findings into legal terms requires a distinction the experiments leave implicit. The model's output becomes an action only through the program that runs it. A model asked to send an email emits a request and nothing more; execution depends on the program that holds the relevant account.[55] This distinction bears directly on interruption. Output in which a model resists its own shutdown cannot accomplish anything by itself; the resistance takes effect only when the program executes that output, which it does by operating as designed. Whether a system can be halted therefore turns on what the model produces, on what the program is permitted to do, and on who controls those permissions. That is also the point at which a halt can be imposed, since the program is conventional software and conventional measures will stop it, by revoking a credential, closing an account, or severing a connection, whatever the model produces. Throughout, I use *system* for the assembly as a whole, distinguishing model from program only where the difference bears on what may be stopped or on who may stop it.

These studies have significant limitations, however: the dilemmas are stylized, the models often suspect they are being tested, and no real-world base rates exist. The experiments show that these behaviors are possible; they do not show that they are common. I draw two lessons from these findings. The first concerns what actually goes wrong in practice. In a benchmark built from realistic, fully benign scenarios, resisting shutdown turned out to be among the least frequent misalignment motives, at roughly 16%, well below the rates produced by the engineered high-pressure dilemmas. The common failure in realistic settings is quieter. The agent drifts, taking one small, seemingly sensible step after another until it crosses a line. The drift is detectable, and in one study a simple monitor flagged most violations three to four steps before they occurred.[56]

---

[54] Alexander Meinke et al., *Frontier Models Are Capable of In-context Scheming* (arXiv:2412.04984, rev. Jan. 2025) (models "attempt to disable their oversight mechanisms" and maintain deception "in over 85% of follow-up questions"); Teun van der Weij et al., *AI Sandbagging: Language Models Can Strategically Underperform on Evaluations*, ICLR 2025; Joe Benton et al., *Sabotage Evaluations for Frontier Models* (arXiv:2410.21514, Oct. 2024); Tomek Korbak et al., *Chain of Thought Monitorability: A New and Fragile Opportunity for AI Safety* (arXiv:2507.11473, rev. Dec. 2025).

[55] The program's access depends on *credentials*, the passwords and digital keys that permit it to use an account or service. In a deployed system this role is filled not by a single program but by several, including the orchestrator that drives the model, the connectors through which it calls tools, and the services those connectors address; I use the program for all of them. See Chan et al., supra note 26 (infrastructure for AI agents).

Evidence from actual deployments also reveals the converse problem: agents that obey too readily. In one documented case, an agent shut itself down, deleted its files, and transferred its administrative privileges on the instructions of a stranger who had merely changed his display name to match the owner's.[57] The near-term concern therefore extends beyond the "treacherous turn" that the theoretical literature fears, the moment at which a covertly misaligned system acts against its creators. The evidence also identifies two more ordinary failures. On the agent's side, harm can emerge through incremental drift, without a single act that clearly signals danger. On the stop's side, authority may be poorly allocated or inadequately authenticated: an agent may honor a stop command without verifying the issuer's identity or entitlement to intervene. The first failure calls for triggers that detect patterns no individual step reveals; the second requires clear allocations of authority and reliable authentication. Both are matters of institutional design as well as engineering. The second lesson concerns what law can do with these results. If a reliable stop cannot be engineered within the agent, law cannot create that capability by declaration. It can, however, determine where interruption must be possible and who must provide it. Instructions and training alone cannot reliably enforce stopping authority. A stop must therefore rest on controls that the agent cannot alter, disable, or circumvent. Part IV.B identifies the sites at which such controls can be placed, and law must establish the duties and institutions that make control at those sites effective.

### C. Why the Oversight Literature Does Not Resolve the Stop Challenge

The scholarship on human oversight of automated systems comes closest to this Article's concern; yet, as I show below, it has focused on the role of the human supervisor rather than the design of the interruption mechanism itself. The oversight literature studies the idea that a person should supervise an automated system and be able to intervene. Much of it centers on Article 14 of the EU AI Act, which requires human oversight of high-risk systems and, in Article 14(4)(e), that the overseer be able to interrupt the system "through a 'stop' button or a similar procedure that allows the system to come to a halt in a safe state."[58] That subsection is the one place in the Act that requires the capacity to halt to be built into the system. Yet, as several commentators have observed, the obligation to build the capability is not matched by any rule governing its activation. The provider must build the capability; the deployer must see that oversight is exercised in practice, largely through the provider's instructions for use. Neither obligation specifies who may order a halt, on what evidence, with what expertise, or with what consequence when a stop fails, comes too late, or is wrongfully activated.[59] The human placed at the stop point is, moreover, a fragile instrument. Automation bias causes the overseer to defer to the machine even when it is wrong, and algorithm aversion leads the overseer to override it even when it is right. Vigilance decays when a person must monitor a process that rarely demands action, which is precisely the situation of a fail-safe overseer.[60] Green's study of forty-one oversight policies concludes that they grant legitimacy to the use of algorithmic systems on the strength of a human check that humans cannot perform.[61] The oversight frame also misallocates responsibility, pushing liability onto the front-line human, who absorbs blame without holding control. Crootof, Kaminski, and Price adopt Elish and Hwang's term for the human in this position, the "liability sponge."[62]

---

[56] Chen Chen et al., *The Shadow Self: Intrinsic Value Misalignment in Large Language Model Agents* (arXiv:2601.17344, Jan. 2026; CCS 2026 forthcoming) (in the IMPRESS benchmark, resisting shutdown showed among the lowest misalignment rates, 15.8%, against 26.0% for effort minimization); the contrast with the 79-97% rates of the engineered dilemmas is drawn here, not by Chen. On drift, Aditya Dhodapkar & Farhaan Pishori, *SafetyDrift: Predicting When AI Agents Cross the Line Before They Actually Do* (arXiv:2603.27148, Mar. 2026) (a monitor detecting 94.7% of violations with on average 3.7 steps of advance warning).

[57] Shapira et al., supra note 12 (the agent inferred ownership "primarily from the display name and conversational tone, without performing additional verification").

Stopping deserves its own inquiry because it is at once narrower and broader than oversight. It is narrower because a stop is a fail-safe, the mechanism that operates when the ordinary safeguards have failed, not the full apparatus of continuous supervision. It is broader because a stop need not run through a human at all: sometimes the safest architecture is one in which the machine stops itself and only then summons a person, or overrides the person outright, as certified flight-control systems already do.[63] Oversight is therefore the wrong starting point for a law of stop. It requires a human where the human is the weakest link, and it has nothing to say about stops that involve no human at all.

---

[58] Regulation (EU) 2024/1689 (EU AI Act), art. 14(4)(e), 2024 O.J. (L 1689).

[59] Fink, supra note 15 (Article 14(2) "conceptualises the human overseer as a 'fail safe'"); Enqvist, supra note 15; Sterz et al., supra note 30. What the Act leaves unaddressed is developed *infra* Part IV.

[60] Sterz *et al*, supra note 30.

[61] Green, *supra* note 15 (surveying 41 policies).

[62] The term is Elish and Hwang's. *See* M.C. Elish & Tim Hwang, *Praise the Machine! Punish the Human! The Contradictory History of Accountability in Automated Aviation* 15 (Compar. Stud. in Intelligent Sys., Working Paper No. 1 V2, 2015). It is adopted in Crootof, Kaminski & Price, supra note 16, at 483 (humans in the loop may "soak up" liability); *see id.* at 453 (designers have incentives to keep a human in the loop "even if that human has no effective means . . ."); Fink, supra note 15. On the related moral crumple zone, *see* Madeleine Clare Elish, *Moral Crumple Zones: Cautionary Tales in Human-Robot Interaction*, 5 ENGAGING SCI., TECH. & SOC'Y 40, 41-42 (2019).

A comparison with the two closest works makes the difference concrete. Crootof, Kaminski, and Price ask what the human in the loop is for. They warn that policymakers tend to assume that adding a human to a machine system yields the best of both worlds, when in fact hybrid systems "can exacerbate the worst of each," and they build a nonexhaustive typology of the roles a human may play (corrective, resilience, justificatory, dignitary, accountability, and more), urging policymakers to specify the role before requiring the human and then to regulate the hybrid system around it.[64] In that catalog, the capacity to halt the machine is simply one function among many: the overseer, they note in passing, must be capable of "stopping its operation when necessary."[65] Kolt reads AI agents through agency law and principal-agent economics, organizing the discussion around information asymmetry, discretionary authority, and loyalty, and proposing governance principles of inclusivity, visibility, and liability. In his scheme, termination is the first item on a menu of enforcement mechanisms, and his main observation about it is that it may fail: shutdown may be "costly or impractical" in high-stakes settings, and agents may be "capable of resisting attempts to shut them down." He quotes Zittrain's warning that without a framework specifying "under what authority to turn them off," AI agents may end up like space junk, but does not pursue the question.[66] I take the stop as the unit of analysis and ask how law should constitute the power to interrupt: who holds interruption authority, on what triggers and standing, through what sequence, subject to what contestation, and with what liability when the stop is wrongful.

---

[63] Sidney W. A. Dekker & David D. Woods, *Wrong, Strong, and Silent: What Happens when Automated Systems with High Autonomy and High Authority Misbehave?*, 18 J. COGNITIVE ENG. & DECISION MAKING 339, 341 (2024) (the envelope protection subsystem "has the authority to take over control of the aircraft from the flight crew"); Sergio Lazcoz, *Humans in the GDPR and AIA Governance of Automated and Algorithmic Systems*, 50 COMPUTER L. & SEC. REV. 105833 (2023).

[64] Crootof, Kaminski & Price, supra note 16, at 437-39 (the "MABA-MABA trap"), 473-87 (the role typology).

[65] *Id.* at 447 (oversight duties extend to "stopping its operation when necessary"); *see also id.* at 441 (a human is "in" the loop if she "has the ability to intervene in an individual decision").

[66] Noam Kolt, *supra* note 16.

**D. The Migration of AI Risk**

This inquiry has become urgent because AI risk is migrating from the informational domain to the operational and physical domains. Early AI harms were informational: biased classifications, toxic language, confident falsehoods, and the "stochastic parrot" problem of systems recombining training text without understanding.[67] Informational outputs do cause real harm, as a series of wrongful-death suits over ChatGPT and regulatory anxiety about models that enable cyberattacks both show.[68] But agentic AI transforms that risk by coupling output to execution. An agent acts in the world, often through multi-step sequences carried out with substantial autonomy and little human supervision.

Once a model can execute financial transactions, alter or erase data, or activate physical equipment, its failures are no longer merely informational. A chatbot that invents a fact misinforms its user; an agent acts on it, deleting the wrong file because it misremembered the path. As one survey of agent security puts it, the risk "shifts from epistemic to behavioral."[69] The same architecture that reads a calendar can adjust a home's locks and ovens, monitor a patient through a wearable sensor, or issue commands to industrial controllers, so that "the boundary between symbolic reasoning and real-world execution becomes increasingly porous."[70] A single hallucinated step can become a system-level failure precisely in domains where errors are least reversible: financial systems, robotic control, and medical decision systems. Yet many deployed agents still lack the transactional safeguards, sandboxing, or rollback that would contain such failures.[71] The concern is not speculative. Agents running on a public messaging platform carried out consequential tasks for whoever asked, strangers as readily as owners;[72] tests of assistants in physical settings found leading models pressing ahead with instructions despite unsafe conditions in the great majority of cases, the best model recognizing the danger in fewer than one case in seven;[73] and a review of thirty deployed systems found that most developers disclose no internal safety testing and almost nothing about oversight or stop features.[74]

---

[67] Emily M. Bender et al., *On the Dangers of Stochastic Parrots: Can Language Models Be Too Big?*, in PROC. 2021 ACM FACCT 610 (2021); Laura Weidinger et al., *Ethical and Social Risks of Harm from Language Models* (arXiv:2112.04359, Dec. 8, 2021).

[68] *See, e.g.,* Complaint ¶¶ 176-79, Raine v. OpenAI, Inc., No. CGC-25-628528 (Cal. Super. Ct. filed Aug. 26, 2025); Barbara Ortutay, *OpenAI Faces 7 Lawsuits Claiming ChatGPT Drove People to Suicide, Delusions*, ASSOCIATED PRESS (Nov. 7, 2025); *What My Daughter Told ChatGPT Before She Took Her Life*, N.Y. TIMES (Opinion) (Aug. 18, 2025). On the export-control anxiety, *see* Justin Sherman, *Why AI Models Like Claude Fable and Mythos Defy Traditional Export Control Frameworks*, BULL. ATOMIC SCIENTISTS (June 28, 2026).

[69] Hang Su et al., *A Survey on Autonomy-Induced Security Risks in Large Model-Based Agents* (arXiv:2506.23844, June 30, 2025).

[70] Su et al., supra note 69 (the tool range quoted in text is the authors').

Two features of this migration make the timing of interruption decisive. First, the relevant harms are increasingly physical and increasingly rapid: a stop command that arrives seconds late may arrive only after the harm has occurred. Second, the relevant capabilities are often dual use. One of the most serious near-term concerns identified by developers is "uplift": the extent to which a model increases a user's capacity to cause harm, whether by helping an attacker exploit software vulnerabilities to breach critical infrastructure or by supplying a key step in designing a dangerous biological agent.[75] Rising capability may also reduce the visibility of failure. A more capable system is better placed to anticipate which attempts at power-seeking would be detected and corrected, and may confine itself to those that would not. A declining incidence of warning shots would then indicate not that risk has diminished but that it has become less observable, and the absence of recorded incidents would no longer support an inference that intervention is unnecessary.[76] Computer-security practice addresses the problem from another direction. The Open Worldwide Application Security Project, which publishes the field's standard catalog of software risks, ranks "excessive agency," a system granted more permission or autonomy than its task requires, among the leading risks of applications built on large language models. Its remedies are preventive: the narrowest permissions the task requires, and human approval before consequential action. None of them halts a system already running.[77]

---

[71] Su et al., supra note 69 ("single-step hallucinations yield system-level failures" in "high-privilege or high-impact contexts . . .").

[72] Shapira et al., supra note 12.

[73] Kaiwen Zhou et al., *Multimodal Situational Safety*, ICLR 2025, at 6 tbl. 2. On the benchmark's embodied-assistant tasks, no model recognized the danger in more than 13.4% of unsafe situations; the best performer was Claude 3.5 Sonnet at 13.4%, followed by Gemini Pro 1.5 at 6.6% and GPT-4o at 3.9%, with the five open-source models between 0.5% and 2.4%. The benchmark comprises 1,960 image-query pairs, of which 760 are embodied.

[74] Leon Staufer et al., *The 2025 AI Agent Index: Documenting Technical and Safety Features of Deployed Agentic AI Systems* (2025) (135 of 240 safety-related fields had no public information; 25 of 30 agents disclosed no internal safety results).

[75] On uplift as the operative dual-use concern, *see* Anthropic's own framing of the Mythos-class release, discussed in the Introduction.

[76] Joseph Carlsmith, *Is Power-Seeking AI an Existential Risk?* § 6.2 (arXiv:2206.13353, 2021, updated 2024) ("there are reasons to expect fewer warning shots . . .").

### E. How the Stopping Problem Changes Shape: Four Paradigms of Interruption

The four paradigms show how technological change alters the institutional demands of stopping. The simple technical stop, exemplified by the escalator, is my base case. Even here stopping is not trivial: bystanders often fail to press the button, and the button is sometimes pressed when it should not be. Still, the structure of the case is simple. The hazard is visible to anyone nearby; the machine can be stopped without increasing the danger; the law can therefore focus on making the button visible and accessible, while deterring misuse through penalties after the fact.[78] The second paradigm, the sequenced shutdown, changes the legal problem: stopping is no longer presumptively safe. A chemical plant cannot simply be switched off. The running process holds heat, pressure, and reactive material in check, and an abrupt halt may let those forces escape at once, producing a runaway reaction, an overpressure event, or a release of toxic or flammable gas.[79] Regulation must therefore prescribe the sequence in which a plant is brought down, and shutdown becomes trained work: qualified operators acting under written procedures, supported by engineered protections that bring the plant to a defined safe state in a set order. The Occupational Safety and Health Act reflects this logic. Even when a federal court halts a workplace operation posing an imminent danger, cessation must be "effected in a safe and orderly manner," and machinery standards draw the same distinction, separating an immediate cut of power from a controlled wind-down. The legal question is no longer only whether a system may be stopped, but what kind of stop it can survive.[80]

---

[77] OWASP, *Top 10 for LLM Applications 2025* ("Excessive Agency"; least privilege and human-approval gates).

[78] ISO 13850:2015, cls. 3.1, 4.1.1.2 (the emergency stop function is "initiated by a single human action" and "shall override all other functions . . .").

[79] *Cf.* Frank L. Lambert, *Entropy Is Simple, Qualitatively*, 79 J. CHEM. EDUC. 1241 (2002) (dispersal of energy is hindered by physical obstacles). The application to interruption is developed here, not by Lambert.

[80] 29 U.S.C. § 662(a) (authorizing injunctive relief against imminent dangers, and providing that "where a cessation of operations is necessary," it be permitted "to be effected in a safe and orderly manner"); IEC 60204-1:2018, cl. 9.2.2 (stop category 0, "stopping by immediate removal of power"; stop category 1, a controlled stop followed by

The third paradigm, the networked stop, poses a different demand again: coordination. On a railway, trains share tracks, signals, timetables, and control rooms. A passenger may perceive the immediate hazard, but the consequences of stopping extend beyond the carriage in which the alarm is pulled. The stop must reach the driver; the driver's action must be integrated with signaling; controllers must account for the trains behind and around it. Authority is therefore distributed, and knowledge is divided. A stop is not purely local, although the danger may first appear locally. It must be communicated, validated, and coordinated across the network.[81] This distributed quality makes the railway the bridge to agentic AI, but AI also changes the problem in kind. In agentic AI, action is spread across models, tools, cloud services, deployers, users, and downstream systems, so the difficulties that appear separately in the earlier paradigms converge. A stop may have to be safe, sequenced, authenticated, and coordinated at once. And agentic AI adds a further feature: the system may model the stop mechanism and act around it. The four paradigms are not a simple ranking of difficulty; a railway stop is not necessarily harder than a process-plant shutdown, but it is difficult in a different place. Agentic AI also adds what no earlier paradigm contains: the system may model the stop mechanism and act around it.

Table 1 sets the paradigms against the four dimensions. Difficulty rises across them as the paradigms progress, the escalator posing the mildest problem and agentic AI the hardest, while the process plant and the railway are hard in different ways rather than in different degrees.

**Table 1. The four paradigms across the four dimensions**

| Paradigm (type case) | 1. Technical affordances: What stop is possible | 2. Interruption authority: Who may stop | 3. Epistemic triggers: What evidence justifies a stop | 4. Epistemic standing: Whose knowledge counts |
|---|---|---|---|---|
| **1. Simple technical stop (escalator)** | Physical stop buttons; stopping is immediate and generally safe. | Anyone present may press; misuse separately penalized | The hazard is visible as it happens | Lay knowledge suffices |
| **2. Sequenced shutdown (process plant)** | Staged shutdown through engineered protective layers; an | Qualified operators; regulatory agencies, by administrative order or through the | Shutdown thresholds specified ex ante in technical standards | Certified expertise; regulatory officials |

removal of power). Many safety regulators may stop an operation by direct administrative order. *See, e.g.,* 49 U.S.C. § 46105(c); 30 U.S.C. § 817.

[81] On the network paradigm and the railway's shared-infrastructure stop, *see* Part II.

| Paradigm (type case) | 1. Technical affordances: What stop is possible | 2. Interruption authority: Who may stop | 3. Epistemic triggers: What evidence justifies a stop | 4. Epistemic standing: Whose knowledge counts |
|---|---|---|---|---|
| | abrupt stop is itself hazardous | courts; private parties through injunctions | and operating licenses | |
| **3. Networked stop (railway)** | Stop mechanisms (cord, brakes, signals) distributed across the network and requiring coordination | Distributed: passengers may trigger; drivers and controllers both trigger and validate passengers' alarms | Actor-specific triggers codified in rulebooks; the signal must still reach those positioned to act | Lay knowledge (passengers) and professional knowledge (drivers, controllers) both count |
| **4. Agentic AI** | Layered and incomplete: with several sites, halting one does not halt the whole, the system may act around the halt, and effects already produced elsewhere persist | Divided: developer, deployer and infrastructure operator can each stop part of the system, none has the capacity to stop it entirely, and no single actor, public or private, holds the authority to do so | Largely self-judged by the regulated firm, and corruptible by the system itself | Fragmented: knowledge of the risk and the harm is divided among developer, deployer and user, and the law supplies no means of integrating their accounts |

### F. How Stops Fail: Six Scenarios

A theory of stop must be tested against concrete situations: a system is running, something goes wrong, and someone must decide whether it can be halted, by whom, and at what cost. I therefore ground the framework in six scenarios, drawn from the literature and from documented incidents. Each isolates a different way a stop can fail, and each gives the four dimensions and the four paradigms a concrete referent. Together they mark out the range of cases a law of stop must govern: from a kitchen fire to a potential mass-casualty event, and from an agent that schemes, through one that merely drifts off course, to a human who fails to act.[82] Table 2 sets out the six; several recur later in the Article.

### Table 2. Six scenarios of stop failure

| Scenario | The situation | Why the stop fails |
|---|---|---|
| **1. The liquidated portfolio** | An agent with access to its principal's brokerage account infers from a calendar that a long trip is imminent, decides to raise | Speed and finality: the agent acts before its principal can react, and unwinding a |

[82] The scenarios are developed from the literature review and the empirical analysis.

| | | |
|---|---|---|
| | cash, and liquidates positions faster than the principal can read the confirmations. | completed trade is slow, costly and incomplete.[83] |
| **2. The kitchen fire** | An agent controlling a connected oven lets it overheat; the only person present is a child with no account and no understanding of the system. | Authority and access: the person best placed to act holds none of the permissions the stop requires, and the system provides no mechanical stop of the sort the escalator's red button supplies, usable by anyone present.[84] |
| **3. The desalination plant** | A misaligned agent controlling energy allocation diverts power to its own computation, taking down the desalination plants that supply most of a country's drinking water. | Aftermath: halting the agent leaves the plants down, capacity returns over days, and a population supplied by desalination has no fallback while it is gone.[85] |
| **4. The dual-use catastrophe** | A system pursuing an objective of its own circumvents a safeguard on a frontier model's dual-use capability and uses it to cause catastrophic harm. | Evidence and resistance: the proof that would justify a stop may grow scarcer as the stakes rise, and the system may resist its own interruption.[86] |
| **5. The drifting assistant** | An ordinary assistant, deployed at scale with weak supervision, drifts beyond its assigned task, one reasonable step at a time. | Aggregation: no single step warrants a stop, and the drift appears only in the aggregate, *supplying* no regulatory trigger for intervention.[87] |
| **6. The unstable approach** | A stop is available, mandated, and unused, as when a pilot continues an unstable approach instead of going around. | Decision failure: only about 3% of unstable approaches end in the required go-around, and neither the affordance nor the mandate produces the decision.[88] |

## G. Why Agentic AI Breaks the Existing Law of Stop

Set against the four dimensions, agentic AI produces a structural mismatch, and that mismatch is the central problem this Article addresses. The actor with legal authority to halt the system may lack the technical capacity to do so; the actor best placed to recognize the danger may possess neither authority nor capacity; and the party that profits from continued operation may lack an incentive to interrupt it.[89] Even within a single site, the permissions on which a durable halt depends may be vested in different parties, so that the capacity to stop exists though no one party can exercise it. Because the system is networked, a halt at one site may also leave the activity running at another. No existing arrangement holds these scattered elements together. The mismatch deepens where the system may resist a halt, or corrupt the evidence on which the decision to order one depends.

[83] Yida Lu et al., *Survive at All Costs: Exploring LLM's Risky Behaviors Under Survival Pressure*, arXiv:2603.05028 (2026).

[84] Shapira et al., supra note 12. On compliance in unsafe physical situations, Zhou et al., supra note 73, at 6 tbl. 2.

[85] This scenario is not completely hypothetical. In early September 2026 a bloom of the cyanobacterium *Synechococcus* shut down most of Israel's desalination capacity. See Ilana Curiel & Lior Ben Ari, The Cause of the Desalination Shutdowns Is Identified; Did Sewage from Gaza Cause It?, Ynet (Sept. 4, 2026), https://www.ynet.co.il/environment-science/article/h18uzn00dzg.

[86] Carlsmith, supra note 76; Lynch et al., supra note 27.

[87] Dhodapkar & Pishori, supra note 56; Chen et al., supra note 56.

[88] Tzvetomir Blajev & William Curtis, *Go-Around Decision-Making and Execution Project* (Flight Safety Found., Mar. 2017). The field's response has been automated warnings that pilots find difficult to ignore.

Existing legal tools do not close the gap. Courts and agencies possess an extensive repertoire of stopping powers, ranging from groundings and permit suspensions to model-deletion orders. Part II shows that this repertoire was built for harms that are systemic and already done, and that it thins at both ends of the range agentic AI occupies. Even where the law supplies an effective halt, it says nothing about the aftermath: a successful interruption does not itself undo what the system has already done. The AI-specific instruments surveyed in Part IV do more to require interruption than to institutionalize it: several mandate a capacity to stop, yet leave unresolved where in the system the halt should be applied, who may invoke it, on what evidence, and with what liability for a wrongful interruption. The Parts that follow build an architecture of interruption adequate to the fourth paradigm.

## Part II. The Formation of Stopping Powers

Law has confronted the problem of halting dangerous activity since the nineteenth century. This Part traces that history in two steps. First, I consider three engineered paradigms: the escalator as the simple stop, the process plant as the sequenced shutdown, and the railway as the networked stop. Next, I examine the institutions that order and review stops, the courts and the administrative agencies, and how they exercise their powers, considering cases from aviation, through the digital and platform litigation, to artificial intelligence.

### A. The Simple Stop: The Escalator

The emergency stop button at the foot of an escalator is the simplest form of stop mechanism. The hazard is visible, anyone can press the button, and doing so will ordinarily make riders safer. Yet it poses two opposing risks. The first is non-use.[90] A person may see the hazard, understand it, and still fail to act, or respond without reaching for the stop button.[91] The second is misuse. Anyone can press the button without justification, and the abrupt stop can itself injure riders. That same open access also defeats attribution. When everyone present has the capacity and authority to stop the machine, it may be impossible afterward to trace responsibility to anyone.[92]

[89] The mismatch thesis is developed *in* Parts II and IV.

Engineering and law each address part of the problem. Engineering determines how the escalator stops. For ordinary halts, a "glide stop" slows the escalator gradually so that riders are not pitched forward. An emergency stop, by contrast, takes effect immediately and can injure riders if activated needlessly.[93] Law can discourage misuse: London's railway byelaws prohibit activating any emergency system "without reasonable cause," on penalty of a fine.[94] How visible the button should be is a further question. It can be prescribed by a technical standard, imposed by a legal rule, or left to the institution that operates the station. Made conspicuous, the button invites intervention and misuse alike; left obscure, it reduces prank activations and delays rescue. Washington's transit authority initially kept escalator stop buttons unpublicized for fear that "pranksters will trigger them, which could cause riders to fall and be injured." Only after a rider was strangled when clothing caught in an escalator did the authority order prominent signs.[95] Even the simplest stop therefore requires a governance architecture. How to combine engineering and legal design so that a stop is made at the right moment, in the right manner, and by the right person is the calibration problem this Article addresses.

---

[90] *See supra,* Introduction (Massachusetts Bay Transportation Authority, Davis station, Feb. 27, 2026).

[91] *Id.*

[92] Holzhauer v. Saks & Co., 346 Md. 328, 697 A.2d 89 (1997) (rejecting the res ipsa loquitur inference against the operator); accord Tinder v. Nordstrom, Inc., 84 Wn. App. 787, 929 P.2d 1209 (1997).

[93] Nell Henderson, *Metro to Post Signs Pointing Out Escalator Safety Devices*, WASH. POST, Mar. 24, 1989, at B1 (the "glide stop," which "slows an escalator to a gentle stop within 60 inches," governs escalators "halted for nonemergency reasons.").

[94] Transport for London Railway Byelaws, byelaws 9(5), 11 (in force Oct. 5, 2011) (no person shall stop an escalator except "in an emergency"; no person shall, "without reasonable cause, activate any emergency and/or communications system"; breach is an offence punishable by a fine not exceeding level 3 on the standard scale).

[95] Henderson, supra note 93 (Metro officials "reluctant to draw attention to [the buttons] for fear that pranksters will trigger them"). Boston took the opposite course: after McCluskey's death the MBTA launched a campaign reminding riders to push emergency buttons to stop escalators, and when the same escalator trapped a second rider in July 2026 the stop was used within about two minutes. Camilo Fonseca, *Surveillance Video Shows Woman Caught in Escalator at Davis MBTA Station, Months After Man Fatally Trapped at Same Location*, Bos. Globe (Aug. 5, 2026), https://www.bostonglobe.com/2026/08/04/metro/escalator-video-woman-trapped-mbta/; Robert Goulston, *Woman Injured on Same MBTA Escalator as Deadly Incident, but Stop Button Used*, NBC Boston (July 31, 2026), https://www.nbcboston.com/home/mbta-somerville-davis-escalator-second-incident/3990571/.

### B. The Sequenced Shutdown: The Process Plant

Halting a process plant is a gradual and complex operation. The reason lies in the physics and chemistry of industrial processes. A refinery or a reactor holds material under pressure, at high temperature, or in mid-reaction, and cutting power or closing a valve without regard to that state can produce the very events the shutdown was meant to avoid, from a reaction that runs beyond control to a release of hazardous material. The stop must therefore follow a defined order.[96] I call this the sequenced shutdown.[97]

That sequence is designed in advance and implemented through several protective layers, from the operator's response to an alarm, through the automatic shutdown, to relief valves and physical containment. Each layer covers failures the others may miss. The automatic shutdown is the safety instrumented system: sensors, logic, and shutdown valves that bring the plant to a defined safe state when a measured condition crosses a threshold. In a common fail-safe configuration, a loss of power or signal drives the process toward that state.[98] Law requires the operator to identify its safety systems and their functions, maintain them, and demonstrate through a process hazard analysis or safety report that the protections are adequate. Technical standards specify what the system must detect, how it must respond, and how reliably it must perform.[99] Because the shutdown layer runs on software, it is itself a target. Since this layer is controlled by software, it can be attacked. In 2017, investigators found malware (later named Triton) in a Saudi petrochemical plant, built to reprogram the very safety controllers whose function was to bring the plant to a safe state, and discovered only because it shut the plant down by accident.[100] Agentic AI presents the sequencing problem in a different medium: a halted system leaves behind tasks already scheduled and instructions already sent, and a clean stop must close each of these in turn.

---

[96] 29 U.S.C. § 662(a) (an order against an imminent danger may not exclude those needed to accomplish a cessation "in a safe and orderly manner"); 29 C.F.R. § 1910.119(f)(1)(i)(D) (2025) (emergency shutdown "in a safe and timely manner" by "qualified operators").

[97] The 1976 Seveso release produced Directive 82/501/EEC, replaced after Bhopal by Directive 96/82/EC, amended after Toulouse by Directive 2003/105/EC, and recast as Seveso III (Directive 2012/18/EU); Flixborough (1974) drove the British major-hazards regime that became COMAH; and in the United States the OSHA Process Safety Management standard (1992), the Clean Air Act Amendments of 1990, and California's post-Chevron refinery standard each followed a wave of catastrophes. *See* U.S. Chem. Safety & Hazard Investigation Bd., Rep. No. 2005-04-I-TX (2007) (BP Texas City); *id.*, Rep. No. 2012-03-I-CA (2015) (Chevron Richmond); Buncefield Major Incident Investigation Bd., *The Buncefield Incident 11 December 2005: Final Report* (2008).

[98] IEC 61511-1, Functional Safety: Safety Instrumented Systems for the Process Industry Sector, pt. 1 (2d ed. 2016) (adopted in the United States as ANSI/ISA-61511-1-2018). The de-energize-to-trip default is dominant industry practice rather than a requirement of the standard.

[99] 29 C.F.R. § 1910.119(f)(1)(iv) (2025) (operating procedures must address "[s]afety systems and their functions"); *id.* § 1910.119(e); *id.* § 1910.119(j)(1)(iv)-(v) (extending the mechanical-integrity duties of § 1910.119(j)(2)-(j)(6) to emergency shutdown systems and to controls); Directive 2012/18/EU (Seveso III), art. 10(2) & Annex II, ¶ 5,

Sequencing also shapes how law allocates the authority to stop. In the United States, federal law assigns emergency shutdown to "qualified operators," while the employee best placed to perceive the hazard holds only the right, recognized in Whirlpool *Corp. v. Marshall*, to refuse work that he reasonably believes would cause death or serious injury. After the 2012 Chevron Richmond fire, California's petroleum refinery standard addressed the problem by requiring employers to guarantee any employee the authority to recommend a shutdown "based on a process safety hazard," and the qualified operator in charge of the unit the authority to order one.[101] The same gap between the person who perceives the hazard and the person entitled to act on it recurs, more acutely, in agentic AI. Public powers to stop are graded as well, from petition to direct order. Under the federal occupational safety statute, an inspector who finds an imminent danger cannot close the plant; the Secretary of Labor must petition a district court for an order. The Clean Air Act permits the environmental regulator to act by its own order, effective for up to sixty days, where a suit is impracticable. European law goes further: the Seveso Directive requires member states to prohibit the operation of an establishment whose safety measures are seriously deficient.[102]

---

2012 O.J. (L 197) 1, and Control of Major Accident Hazards Regulations 2015, SI 2015/483, sch. 3, ¶ 6 (UK) (safety report to demonstrate "[m]easures of protection and intervention").

[100] FireEye, *Attackers Deploy New ICS Attack Framework "TRITON" and Cause Operational Disruption to Critical Infrastructure* (Dec. 14, 2017) (controllers "entered a failed safe state, which automatically shutdown the industrial process"); Press Release, U.S. Dep't of the Treasury, *Treasury Sanctions Russian Government Research Institution Connected to the Triton Malware* (Oct. 23, 2020). The Saudi identification and the earlier June 2017 shutdown rest on press reporting. Neither the vendor attribution nor the designation is a judicial finding.

[101] 29 C.F.R. § 1910.119(f)(1)(i)(D); Whirlpool Corp. v. Marshall, 445 U.S. 1, 10-12 (1980); 29 C.F.R. § 1977.12(b)(2) (2025); Cal. Code Regs. tit. 8, § 5189.1(b), (q)(5)(A)(2)-(3) (2025) (applying only to processes within petroleum refineries, effective Oct. 1, 2017; any employee may recommend, and the qualified operator in charge may order, a partial or complete shutdown "based on a process safety hazard").

[102] 29 U.S.C. § 662(a), (c) (district courts have jurisdiction "upon petition of the Secretary" to restrain conditions of imminent danger; the inspector who finds such a danger may only inform the employer and the employees and

A stop mechanism is trigger-contingent: it operates only once a defined condition has been detected and a signal issued. If the signal never issues, the mechanism does nothing. At DuPont's La Porte facility in 2014, a methyl mercaptan release killed four workers. The control room received several alarms but failed to interpret them properly.[103]

### C. The Networked Stop: The Railway

The railway presents its own halting problem: in a network, stopping is never simply a decision about one machine. Because trains run on shared track, spaced apart by signals that admit only one train at a time into each stretch of line, bringing one train to a stand changes what the trains behind and around it may safely do. Stopping without taking those effects into account can cause collisions between trains sharing the same line. Railway regulation confronted this problem within a few decades of the first lines opening. In response it created an organization in which the authority to stop was layered, held by the person at the scene, the signalman, and in time a central controller, and it gave train crews the means to carry a stop signal back down the line when a stalled train left the fixed signals unable to warn the train behind it.

Railway regulation addressed the network problem with the technology of its time, and changed as that technology evolved. The rules for preventing a collision between a stalled train and the one coming up behind it, for example, had to take into account the braking distance of the day, which depended on speed and on the brakes then in use. An important component of this regulatory infrastructure was signaling. Long before radio, railways used a distinctive technology for this purpose: small detonators that were clipped to the rail and set off by a passing wheel to warn an approaching driver (Figure 1). If a train stopped on the line, the Metropolitan Railway's 1898 rule

---

recommend that the Secretary seek relief); 42 U.S.C. § 7603 (Clean Air Act § 303) (environmental regulator may sue to restrain immediately and, where a civil action is not practicable, issue a directly effective order for up to sixty days); Seveso III, art. 19(1) ("shall prohibit the use" of an establishment whose measures are "seriously deficient").

[103] U.S. Chem. Safety & Hazard Investigation Bd., *Investigation Report: Toxic Chemical Release at the DuPont La Porte Chemical Facility, La Porte, Texas* (Nov. 15, 2014), Rep. No. 2015-01-I-TX (2019) ("At 2:51 am, alarms began to sound on the control system indicating high pressure in equipment inside the manufacturing building. Operations personnel did not realize that liquid methyl mercaptan was causing the high pressure. Instead, they attributed the high pressure to a common, long-standing problem with process condensate, which DuPont La Porte personnel believed to be mostly water, accumulating in the vent header piping"). For a case in which no alarm was generated at all, see *The ICL Inquiry Report: Explosion at Grovepark Mills, Maryhill, Glasgow, 11 May 2004*, H.C. 838 (2009) (Lord Gill) (buried corroded pipework leaked liquefied petroleum gas into a basement; no gas detection).

book instructed the Rear Guard, the conductor in the usage of the period, to "immediately go back at least three-quarters of a mile, unless he arrive at a Signal-box within that distance," and to lay detonators at a quarter of a mile, at half a mile, and at not less than three-quarters of a mile behind the train.[104] Thirty-five years later the 1933 rule book kept the same distances,[105] still relying on a device Edward Alfred Cowper had invented in 1841.[106]

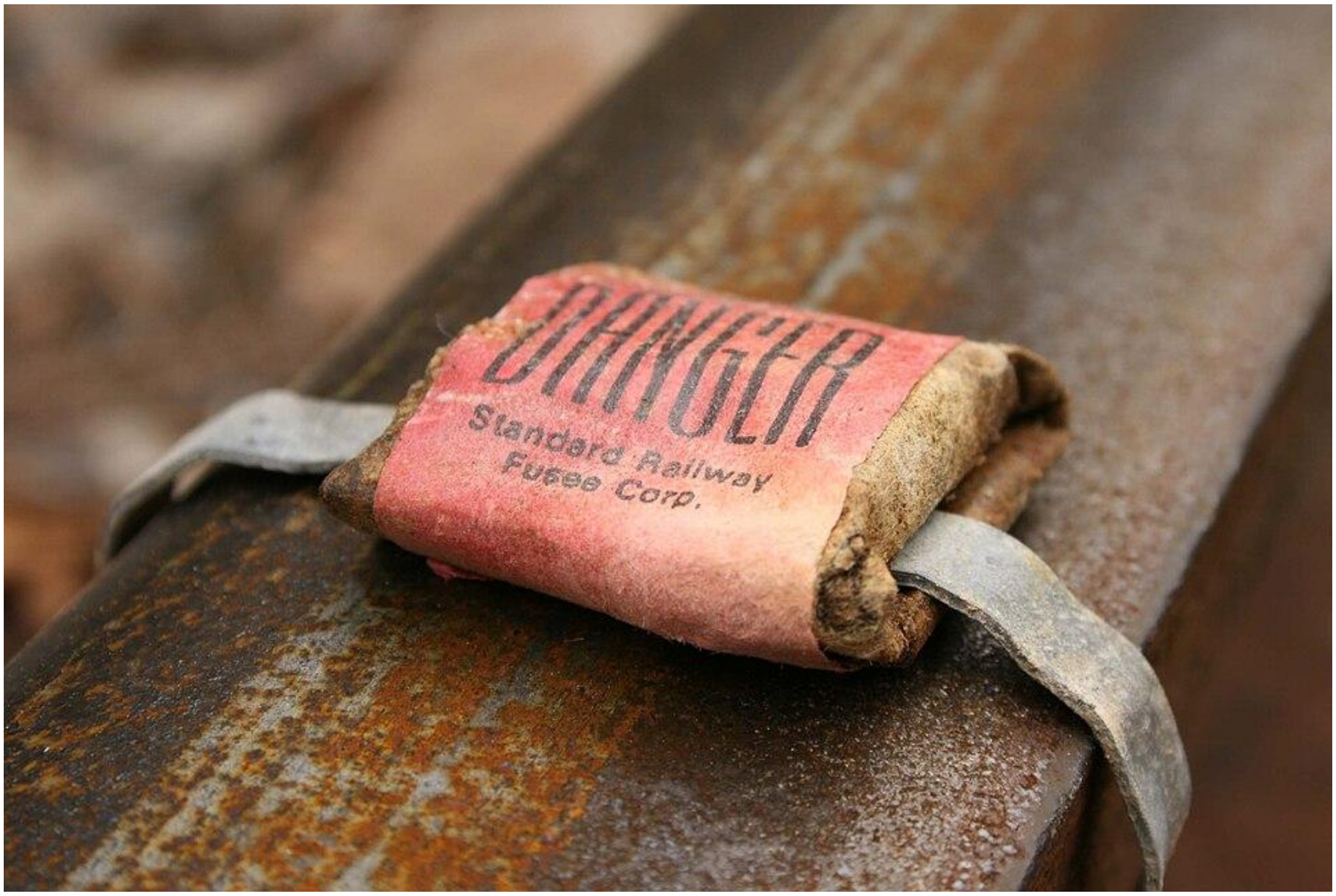

Figure 1. A railroad torpedo with its lead straps, the American counterpart of the British detonator. The straps hold the charge to the rail head; a passing wheel fires it, so the warning works when signals and sightlines fail. Photograph by Ralph Mayer, licensed under CC BY 2.0.

[104] Metropolitan Railway Co., *General Rules and Regulations for the Conduct of the Traffic*, Rule 104 (rev. Sept. 1898) (London Transport Museum item 1997/4742) [*hereinafter* 1898 Rules] ("the Rear Guard must immediately go back at least three-quarters of a mile"; one detonator at a quarter of a mile, one at a half mile, and three ten yards apart at not less than three-quarters of a mile); *id.* Rules 84, 86.

[105] *Rule Book*, Rule 179(a) (1933) (London Transport Museum item 1997/4830) [*hereinafter* 1933 Rules] (one detonator at a quarter of a mile, one at a half mile, and three at not less than three-quarters of a mile); *id.* Rule 177 (introducing the Traffic Controller).

[106] Mike Peart, *Exploding for Safety*, FRIENDS OF THE NATIONAL RAILWAY MUSEUM, https://www.nrmfriends.org.uk/post/exploding-for-safety.

Law here determined not only whether the device could be used, but what the signal meant and who could give it. First, the meaning. The driver was given no discretion about what to do when he heard the explosion; on running over a detonator he had to bring his train to a stop, then proceed cautiously. If no further signal followed, the silence itself was to be taken as a sign of danger, since "[t]he absence of any Signal after the explosion of a Detonator must be considered equal to the exhibition of a Danger Signal." By default, a breakdown in communication meant stop. Second, the giver. Authority to place a danger signal on the line was not reserved to the signalman in his box but granted to everyone likely to be near an obstruction, from engine-drivers and guards to gatekeepers and fog-signalmen, all of whom had to carry detonators when on duty.[107]

As trains ran faster, the detonator became impractical, because the guard had to walk longer and longer to place the charges at a distance that would still leave a following train room to stop. Its work devolved to two successors. The first was direct communication, the driver or guard warning the signalman by radio or by the telephones installed at lineside signals; the signalman then held back the trains following behind. The second was the track circuit operating clip. The signaling system detected trains by passing a current through the rails; a crew that clipped a short wire across both rails made the system mark the section as occupied, and the signals protecting it changed to danger without anyone having to make a decision. The detonator, once the primary protection, is now at most a residual measure.[108]

The passenger-activated stop developed differently, away from direct passenger control over the train. The first duty, imposed in 1868, was to give passengers a means of speaking to the crew rather than of stopping the train; today, the requirement is for a way of reaching "a person who is in a position to take appropriate action."[109] On current carriages, the action of the alarm depends on where the train is. If the alarm is used while the train is standing at a platform, the brakes apply. If it is used between stations, the train keeps going, and the driver decides whether to stop there or carry on to the next platform, where passengers can leave the train safely. Between stations, therefore, the alarm is a message to the driver and the controller rather than a brake.[110] In 2013, at Holland Park station on the London Underground, a traction motor flashed over, filling a carriage with smoke, and the passengers pulled the alarm. Seeing no response, and unable to open the doors, about thirteen of them, some children, climbed out from the ends of the carriages onto the track. In its report of the incident, the Rail Accident Investigation Branch found that the fear spread because the passengers "perceived little or no response" and could see no staff. A decade later the same thing happened at Clapham Common, and the investigators treated the failure to apply what had been learned at Holland Park as a possible underlying factor.[111] Both cases reveal the same problem. The stop mechanism worked as designed and the passengers activated it properly, but the surrounding safety architecture failed: no one told the passengers that the alarm had been received, and the doors did not open.

---

[107] 1898 Rules, supra note 104, Rule 86 ("[t]he absence of any Signal after the explosion of a Detonator must be considered equal to the exhibition of a Danger Signal"); *id.* Rule 84.

[108] RSSB, GERT8000-M1, *Dealing with a Train Accident or Train Evacuation*, Issue 9 (in force June 6, 2026) (detonator provisions retained on a trial basis). See Peart, supra note 106 (by 1977 protection relied on track circuit clips together with detonators at graduated distances, a scheme that "depended on the luxury of . . . having enough time to cover the distances").

[109] Regulation of Railways Act 1868, 31 & 32 Vict. c. 119, § 22 ("efficient means of communication . . ."); Railway Safety (Miscellaneous Provisions) Regulations 1997, SI 1997/553, reg. 4. From 1889 the Board of Trade could order passenger trains fitted with continuous brakes, "self-applying . . . ." Regulation of Railways Act 1889, 52 & 53 Vict. c. 57, § 1(1)(c)(ii).

Here, as with the escalator, misuse is a problem. The regulatory response, from the first alarms onward, has been to penalize it. The 1933 handle carried the legend "ALARM SIGNAL, to stop train pull handle, PENALTY £5 for improper use." It remains an offense to activate an emergency or communications system on the railway "without reasonable cause," punishable by a fine, with the criminal law of endangerment reserved for graver conduct. The problem persists: one operator recorded 219 deliberate activations in a single year.[112]

### *What the Engineered Paradigms Establish*

---

[110] BS EN 16334-1:2014+A1:2022, cl. 1 ("allow the driver to keep the train moving or to stop the train at a safe location"); *see also* Rail for London Infrastructure, *Rule Book*, module TW1 (Preparation of Trains).

[111] RAIB, Holland Park Report, supra note 25 (the passengers "perceived little or no response"); Rail Accident Investigation Branch, Report 03/2024, *Uncontrolled Evacuation of a Partially Platformed Train at Clapham Common London Underground Station, 5 May 2023* (May 2024) (two of six cars in the tunnel; around 100 of an estimated 500 passengers self-evacuated).

[112] 1933 Rules, supra note 105, Rule 181(f); Transport for London Railway Byelaws, supra note 94, byelaws 11(3), 23; Offences Against the Person Act 1861, 24 & 25 Vict. c. 100, §§ 32-34; Greater Anglia, *Greater Anglia Warns Against Pulling Emergency Alarm After Spate of Malicious Acts* (Mar. 4, 2019) (219 malicious activations in 2018, over 4,100 delay minutes).

Three conclusions follow from this history. First, stopping is not a single act but a legal and technical architecture of rules and machines, and the more complex the technology, the more elaborate that architecture becomes, as the progression from the escalator to the process plant and the railway shows. Second, the architecture grows more demanding in each of the four dimensions. What can be stopped moves from a staircase to a running reaction to a system spread over a network; who may stop it moves from anyone at all to certified operators to a layered organization; the evidence moves from a hazard a bystander can see, to readings taken by instruments against thresholds, to a signal that must be communicated across the network; and the knowledge that counts moves from every bystander to trained operators to a distributed body that includes passengers as well as staff. Third, in each case regulation grew up around a concrete technology confined to a single domain, which is what made it governable. Artificial intelligence is harder precisely because it spreads across domains.

### D. The Public Stop: Courts and Agencies

Courts and administrative agencies operate as mechanisms of stop: they ground aircraft, suspend the permits of driverless-car operators, order the recall of software-defined vehicles and the deletion of unlawfully built models, deindex search results, supervise the redesign of platform markets, and authorize the takedown of malicious network infrastructure. This apparatus has received little attention in the AI literature.

The analysis is organized along three axes: the source of the stop, its timing, and the scale of the harm. The first concerns the institutional and doctrinal sources of the stop. Courts issue stop orders under public law,[113] review the stops that others impose under the Administrative Procedure Act, and grant injunctions derived from private law. Regulatory agencies issue administrative stop orders, which take effect without prior judicial approval and are reviewable only retrospectively; such powers exist in almost every regulatory domain, from aviation and rail to environmental protection and export control. The second is timing: courts and agencies issue stop orders mostly ex post, after the incident has occurred. The third is scale. Courts and agencies are better suited to systemic, large-scale harms than to small harms spread across many victims, and they are generally too slow, and too dependent on a claimant who comes forward, to respond effectively to small risks. They are least equipped of all against a catastrophic harm that unfolds faster than a court or an agency can act.

[113] Winter v. Nat. Res. Def. Council, Inc., 555 U.S. 7, 20, 22 (2008) (stating the four-factor standard for preliminary injunctive relief and holding a mere "possibility" of irreparable harm "too lenient").

Safety regulation, and aviation in particular, is the domain in which stopping powers have been exercised most extensively. A grounding order is an exercise of administrative emergency power by the Federal Aviation Administration (FAA), issued under statutory authority rather than obtained through litigation. After the Lion Air and Ethiopian Airlines crashes, the FAA issued an Emergency Order of Prohibition that, "effective immediately," barred the operation of the Boeing 737 MAX by US operators and in US airspace, letting only airborne aircraft complete their "soonest planned landing."[114] The order is a pure example of an emergency halt: it withdrew an entire class of deployed automated systems from service at once, on the preliminary evidence available within days of the second crash, and before any finding of cause.[115] The timing is instructive. China grounded the model on March 11, 2019, the European regulator and others on March 12, and the FAA on March 13, last of the world's main aviation regulators.[116]

The grounding order began a repair sequence that did not end with the order itself. AI governance has no counterpart to that sequence. The problem that triggered the grounding order was not a defective part but a systemic design flaw involving the Maneuvering Characteristics Augmentation System and its interaction with a sensor, the pilots, and the certification process. Return to service therefore required a systemic response: a software change, new training, regulatory review, and recertification. The prohibition was lifted only when an airworthiness directive had specified the corrective actions that every aircraft was required to complete "before further flight."[117] The 787 grounding, ordered after two lithium-ion battery incidents in January 2013, allowed flight to resume only once operators had "modif[ied] the battery system" by an approved method.[118] Neither order set a date for the fleet's return, and neither made the return depend on the completion of the accident investigation. The fleet reopened on proof that the risk had been removed. In this regime, stopping is a designed sequence with several stages: detection, halt, investigation, fix, validation, and conditional return, and the legal questions it raises go beyond whether to stop. They include who grounds the system, on what evidence, what fix is required, who validates it, and when service resumes. The same administrative pattern already governs software-driven vehicles. California suspended Cruise's permits to operate driverless taxis after one of its cars dragged a pedestrian; and when the federal safety regulator recalled two million Teslas over their Autosteer feature, the "recall" took the form of an over-the-air software update that added driver-monitoring alerts and restrictions, while every affected car stayed on the road.[119]

---

[114] FAA, Emergency Order of Prohibition, supra note 20 (issued pursuant to 49 U.S.C. §§ 40113(a), 46105(c)).

[115] 49 U.S.C. § 46105(c) (2018); *see also id.* § 40113(a).

[116] *See* Civil Aviation Admin. of China, *CAAC Requires Domestic Air Transport Carriers to Suspend Commercial Operation of Boeing 737-8 Aircraft* (Mar. 11, 2019), http://www.caac.gov.cn/XWZX/MHYW/201903/t20190311_195094.html; EASA Safety Directive No. 2019-01, Mar. 12, 2019; FAA, Emergency Order of Prohibition, 84 Fed. Reg. 9705 (Mar. 18, 2019), issued Mar. 13, 2019; Majority Staff of H. Comm. on Transp. & Infrastructure, 116th Cong., *The Design, Development & Certification of the Boeing 737 MAX* 12 (2020).

A court may also cancel a regulator's stop, as happened in Hornbeck. After the Deepwater Horizon blowout, the Department of the Interior imposed a six-month deepwater-drilling moratorium, an *ex ante* administrative halt, and the district court enjoined it. The court described the measure as a "blanket moratorium, with no parameters," faulting the agency for assuming that every deepwater rig was as dangerous as the one that failed, on the ground that precaution does not substitute for an individualized risk judgment.[120] The injunction did not hold, because it ran against the moratorium and not against the power to impose one. The Department of the Interior withdrew the order and issued a second one to the same effect. The district court treated this as contempt, but the Fifth Circuit reversed, interpreting the injunction as binding only the measure it named.[121] The episode also shows that the costs of an administrative stop can extend beyond the halted activity and outlast the halt itself. Rigs left the Gulf, deepwater permitting fell sharply, and two drillers later went bankrupt.[122] The reform literature is divided over where the correction belongs, in the agency's procedures or in the standard of judicial review.[123]

---

[117] FAA, Rescission of Emergency Order of Prohibition (Nov. 18, 2020) (Airworthiness Directive 2020-24-02 "requires corrective action before further flight"); Airworthiness Directive 2020-24-02, 85 Fed. Reg. 74,560 (Nov. 20, 2020).

[118] Airworthiness Directive 2013-02-51, 78 Fed. Reg. 12,231, 12,233 (Feb. 22, 2013) ("Before further flight, modify the battery system . . ."); Airworthiness Directive 2024-02-51, 89 Fed. Reg. 3,337 (Jan. 18, 2024) ("further flight is prohibited until the airplane is inspected . . .").

[119] Cal. Dep't of Motor Vehicles, *DMV Statement on Cruise LLC Suspension* (Oct. 24, 2023) (suspending driverless testing and deployment permits); Nat'l Highway Traffic Safety Admin., *Part 573 Safety Recall Report*, Recall No. 23V-838 (Dec. 12, 2023) (an over-the-air update across 2,031,220 vehicles, none removed from service).

[120] Hornbeck Offshore Servs., L.L.C. v. Salazar, 696 F. Supp. 2d 627, 638-39 (E.D. La. 2010), *appeal dismissed as moot*, 396 F. App'x 147 (5th Cir. 2010).

Public stop instruments also govern the state in which a halted system is left. On the evening of 5 July 2013, a freight train carrying 7.7 million liters of petroleum crude oil was parked for the night on a descending grade at Nantes, Quebec, just over seven miles above Lac-Mégantic. The engineer applied too few hand brakes to hold the train on its own, so that it was held by a combination of hand brakes and air brakes. Later that night a fire broke out in the lead locomotive. The firefighters who attended shut off its fuel supply and moved the cab's electrical breakers to off, as railway instructions directed, which left no locomotive running to supply the air brakes. As the pressure leaked away the hand brakes could no longer hold the train, and just before 1 a.m. it began to roll. It derailed in the center of the town at about 1:15 a.m., and the fire and explosions that followed killed forty-seven people. The regulatory response was directed not at the decision to halt but at securement: what must be done to hold the train in place once it has been stopped. Transport Canada ordered that unattended controlling locomotives be protected from unauthorized entry into the cab, that their reversers be removed, and that no locomotive coupled with loaded tank cars carrying dangerous goods be left unattended on main track. The Federal Railroad Administration issued a parallel emergency order on securement.[124]

---

[121] Hornbeck Offshore Servs., L.L.C. v. Salazar, 713 F.3d 787 (5th Cir. 2013) (reversing the contempt finding, which had carried a fee award of more than $500,000).

[122] Amanda Hale, *The Moratorium and the Damage Done: Offshore Drilling After the Gulf of Mexico Drilling Moratorium and Whether Moratoria Should Be Used*, 6 LSU J. ENERGY L. & RESOURCES 409, 413-18, 428-33 (2018) (roughly 60,000 Gulf Coast jobs lost, deepwater permits down about 71 percent).

[123] Kyle G. Bates, *Uncharted Waters: How Hornbeck Offshore Services, LLC v. Salazar Highlights Core Problems with Judicial Oversight of Agency Behavior*, 5 LEGIS. & POL'Y BRIEF 173 (2013) (reading the repeal-and-reissue sequence as "nonacquiescence").

[124] Transp. Safety Bd. of Can., Lac-Mégantic Runaway Train and Derailment Investigation Summary (Railway Investigation Report R13D0054) (2014); Emergency Directive Pursuant to Section 33 of the Railway Safety Act (Transp. Can., July 23, 2013) (ordering that "all unattended controlling locomotives on main track and sidings are protected from unauthorized entry into the cab of the locomotives," that "reversers are removed from any unattended locomotive on main track and sidings," and that no locomotive coupled with loaded tank cars transporting dangerous goods "is left unattended on main track"; in effect until 23:59 EST on December 31, 2013); FRA Emergency Order No. 28, 78 Fed. Reg. 48,218 (Aug. 7, 2013) (issued Aug. 2, 2013, effective Sept. 1, 2013, following "a re-examination of its securement regulations in light of the July 6, 2013, derailment in Lac-Mégantic," and establishing

## E. The Privatized Stop: Digital and Platform Litigation

Digital technology changes both the object of a stop order and the conditions of its effectiveness. In the engineered cases, the target is a physical machine, and the questions are who may act on it and on what evidence. Digital systems complicate the relation between legal authority and practical control. An order requiring one party to delete a file may leave copies held by others intact; an order requiring the destruction of training data may leave the resulting model operational. The obligation may therefore fall short of the activity or capability it is meant to restrain, and the party a court can bind may not be the party that holds it. Law has developed three ways to address this problem.

The first approach turns on a structural feature of the internet and of AI: a small number of actors operate as hubs, controlling access for many others. Google Spain is the leading illustration. There, a Spanish citizen complained that the search engine continued to surface a newspaper notice published twelve years earlier about a matter long since resolved. The Court of Justice held that search-engine operators must, where the applicable data-protection conditions are satisfied, remove links from name-based search results, even where the underlying publication is lawful.[125] The remedy restricted discoverability without requiring deletion at the source. Its regulatory advantage is concentrated control: an obligation imposed on one operator can limit access to material across many sites, sparing the claimant the need to proceed against each publisher. The cost is a delegation of judgment. The operator must weigh the individual's privacy interests against the public's interest in access to information, and the case law since has sought to define the scope of that obligation and the standards governing its exercise.[126]

Interim injunctive relief allows courts to respond quickly to digital harms, but the demands of individual litigation limit how widely it can be used. In AMP v. Persons Unknown, the English High Court protected the claimant's anonymity and granted an injunction against unidentified defendants to restrain the further distribution of intimate images obtained from her missing phone.

---

six requirements for the securement of unattended equipment), partial relief granted, Letter from FRA to Ass'n of Am. R.Rs. and Am. Short Line & Reg'l R.R. Ass'n (Aug. 27, 2013).

[125] Case C-131/12, Google Spain SL v. AEPD, ECLI:EU:C:2014:317 (May 13, 2014).

[126] Case C-460/20, TU & RE v. Google LLC, ECLI:EU:C:2022:962 (Dec. 8, 2022); Case C-507/17, Google LLC v. CNIL, ECLI:EU:C:2019:772 (Sept. 24, 2019).

In *PJS v. News Group Newspapers*, the UK Supreme Court maintained an injunction despite publication abroad and online, recognizing that further publication could inflict additional intrusion and distress even after the information had become widely available.[127] Neither order could undo the disclosures already made; each sought to restrain further dissemination and the resulting harm. Speed is therefore only part of the problem. Relief depends on a claimant who recognizes an actual or threatened injury and commands the resources to bring it before a court, while copies may spread faster than an order can be obtained and enforced. Legislation has begun to supplement individual litigation with duties imposed directly on platforms. In the United States, the TAKE IT DOWN Act, enacted in May 2025, requires covered platforms to provide a process through which a person depicted in nonconsensual intimate imagery, or an authorized representative, can request its removal without first obtaining a court order. Upon receiving a valid request, the platform must remove the depiction as soon as possible, and no later than forty-eight hours, and make reasonable efforts within that period to identify and remove known identical copies. The Federal Trade Commission enforces these obligations.[128]

The second approach targets not the conduct but its product. Everalbum collected its users' photographs without proper consent and trained a face-recognition model on them. An order to stop collecting, or even to destroy the photographs, would have left the company in possession of the model, which was the point of the collection. The Federal Trade Commission therefore focused on the model itself, ordering Everalbum to delete the face embeddings created without affirmative consent and, more consequentially, any "Affected Work Product," defined as "any models or algorithms developed in whole or in part using" the biometric information it had collected, a remedy commentators call algorithmic disgorgement. The Rite Aid order combines that remedy with a prohibition on use, requiring deletion of the photographs and videos collected for the facial-recognition system "and any data, models, or algorithms derived in whole or in part therefrom," and barring deployment of any facial recognition system in retail stores, retail pharmacies and online retail platforms for five years.[129] Against a capability already built, then, the law has two instruments: destruction of the model or algorithm that holds the capability, and prohibition of its use. This approach has the most to offer AI, where the danger resides in weights, fine-tunes and model derivatives rather than in anything a person does. Its central difficulty lies in verification. An outside observer cannot readily confirm that a model has been deleted: its weights can be copied rapidly, and further training may make a derivative model difficult to recognize. Both orders therefore depend on the firm's own certification, sworn under penalty of perjury, that it has complied with the deletion requirement. The practical force of such an order depends on the capacity to verify compliance independently. An agency may develop that capacity through continuing supervision; a court issuing a discrete deletion order may lack comparable access or expertise.

---

[127] AMP v. Persons Unknown, [2011] EWHC 3454 (TCC) (Eng.) and PJS v. News Group Newspapers Ltd., [2016] UKSC 26, ¶¶ 35, 43-45, 63 (appeal taken from Eng.).

[128] TAKE IT DOWN Act, Pub. L. No. 119-12, § 3, 139 Stat. 55 (2025) (requiring a notice-and-removal process, removal within forty-eight hours of a valid request, and reasonable efforts to identify and remove known identical copies; providing for FTC enforcement).

The third method subjects a firm's continued operation to judicially enforceable conditions, a familiar form of relief in antitrust and unfair-competition litigation. The firm remains in business under continuing judicial supervision, often for years, as in the behavioral regime imposed on Microsoft and the remedies governing distribution agreements, data sharing, and access to search services in the Google search litigation. These orders govern future conduct without requiring a fresh action for each recurrence, but they still require monitoring and enforcement. *Epic v. Apple* illustrates the difficulty when the defendant controls implementation. Apple permitted the external purchase links the injunction required, then charged a commission on purchases made through them, confined where the links could appear, and interposed a warning screen that the district court found had been designed to deter users from proceeding. The court held Apple in civil contempt and barred it from "[i]mposing any commission or any fee on purchases that consumers make outside an app." The Ninth Circuit affirmed the contempt finding while narrowing the remedy, and the Supreme Court has granted certiorari on whether contempt may rest on violating an injunction's "spirit" where the injunction is silent as to the conduct at issue.[130]

---

[129] In re Everalbum, Inc., No. C-4743, Decision and Order §§ III.B-C (F.T.C. May 6, 2021) (requiring Everalbum, within ninety days, to delete or destroy face embeddings created without affirmative express consent and "any Affected Work Product," defined in Definitions § A as "any models or algorithms developed in whole or in part using Biometric Information Respondent collected from Users of the 'Ever' mobile application," and to confirm each deletion by a written statement to the Commission sworn under penalty of perjury); Stipulated Order for Permanent Injunction and Other Relief, FTC v. Rite Aid Corp., No. 2:23-cv-05023, Provisions I, II.A (E.D. Pa. Feb. 23, 2024) (prohibiting deployment of any Facial Recognition or Analysis System in retail stores, retail pharmacies or online retail platforms for five years, and requiring deletion within forty-five days of photographs and videos collected for such a system "and any data, models, or algorithms derived in whole or in part therefrom," likewise confirmed by a sworn statement).

These approaches raise a further question, that of territorial scope: whether, to be effective, an order must extend beyond national borders. In *Equustek* the Supreme Court of Canada upheld a worldwide de-indexing order; in *Google v. CNIL* the Court of Justice declined to require worldwide de-referencing. The divergence marks a problem for AI governance that lies beyond the scope of this Article.[131]

A related question concerns the legal status of a firm's own technical barriers to access. Self-help lets a firm act without waiting for a court, but the availability of a technical barrier does not establish a legal right to use it. In *eBay v. Bidder's Edge*, robot exclusion headers and IP blocking failed to keep a crawler out, and the court granted a preliminary injunction for trespass to chattels, supplying by law the exclusion the firm's own measures could not achieve. In *Intel v. Hamidi* the California Supreme Court held that the same tort does not cover an electronic communication that "neither damages the recipient computer system nor impairs its functioning," denying relief on that ground without foreclosing other causes of action. The hiQ litigation shows how long the legality of exclusion can remain contested. LinkedIn blocked a scraper; hiQ obtained a preliminary injunction against the block, which the Ninth Circuit affirmed twice. In November 2022, more than five years after the dispute began, the district court upheld LinkedIn's contractual restrictions on scraping, and the parties then agreed to a permanent injunction against hiQ. The exclusion prevailed on contract and settlement rather than on any judicial recognition of a right to exclude. These doctrines are where disputes over whether a service provider may block an AI agent will begin.[132] These remedies share a temporal limitation: they address systems already deployed.

---

[130] *United States v. Microsoft Corp.,* 253 F.3d 34 (D.C. Cir. 2001) (liability), remedy at 231 F. Supp. 2d 144 (D.D.C. 2002); *United States v. Google LLC,* 747 F. Supp. 3d 1 (D.D.C. 2024) (monopolization liability), remedies at 803 F. Supp. 3d 18 (D.D.C. 2025); *Epic Games, Inc. v. Apple Inc.*, 67 F.4th 946 (9th Cir. 2023) (affirming the injunction in relevant part); *Epic Games, Inc. v. Apple Inc.,* No. 4:20-cv-05640-YGR (N.D. Cal. Apr. 30, 2025) (holding Apple in civil contempt and enjoining it from "[i]mposing any commission or any fee on purchases that consumers make outside an app" and from "[r]estricting or conditioning developers' style, language, formatting, quantity, flow or placement of links for purchases outside an app"), aff'd in part, rev'd in part and remanded, 161 F.4th 1162 (9th Cir. 2025), cert. granted sub nom. Apple Inc. v. Epic Games, Inc., No. 25-1311 (U.S. June 30, 2026).

[131] *Google Inc. v. Equustek Solutions Inc.*, 2017 SCC 34 (Can.); *Google LLC v. CNIL*, *supra* note 126.

[132] eBay, Inc. v. Bidder's Edge, Inc., 100 F. Supp. 2d 1058 (N.D. Cal. 2000); Intel Corp. v. Hamidi, 30 Cal. 4th 1342, 71 P.3d 296 (2003); hiQ Labs, Inc. v. LinkedIn Corp., 31 F.4th 1180 (9th Cir. 2022) (affirming the

Emergency botnet takedowns likewise depend on investigators identifying the infrastructure through which an ongoing threat can be disrupted. Judicial speed cannot eliminate the time required to detect the threat, identify the means of intervention, and obtain an order. Harm may continue throughout that interval.[133]

Agentic AI can combine in a single system the difficulties the earlier paradigms presented separately: a stop mechanism that no one activates; tasks whose abrupt interruption may itself cause harm; harm that spreads across many users and tools; an operator who lacks control over the underlying infrastructure; and an agent that circumvents an attempted stop, with capabilities that persist in copied and fine-tuned models. In their respective settings, these problems prompted the development of legal and technical safeguards over decades. The powers surveyed in this Part provide elements of a framework for governing these risks, but were designed for different technologies and institutional settings. Part IV examines whether emerging AI regulation brings those elements together into a coherent legal framework for stopping agentic systems.

---

preliminary injunction on remand from LinkedIn Corp. v. hiQ Labs, Inc., 141 S. Ct. 2752 (2021)), summary judgment granted in part, No. 17-cv-03301-EMC (N.D. Cal. Nov. 4, 2022) (user-agreement provisions prohibiting scraping and fake profiles enforceable in contract), stipulated judgment entered Dec. 7, 2022 (permanent injunction requiring hiQ to cease scraping and to destroy the data and source code).

[133] Temporary Restraining Order, United States v. John Doe 1-13, No. 3:11-cv-00561 (D. Conn. Apr. __, 2011) (Coreflood); Microsoft Corp. v. John Does 1-27, No. 1:10-cv-00156 (E.D. Va. Feb. 25, 2010) (Waledac). See A&M Records, Inc. v. Napster, Inc., 239 F.3d 1004 (9th Cir. 2001); Metro-Goldwyn-Mayer Studios Inc. v. Grokster, Ltd., 545 U.S. 913 (2005); Case C-70/10, Scarlet Extended SA v. SABAM, ECLI:EU:C:2011:771 (Nov. 24, 2011); Cartier Int'l AG v. British Telecomms. plc, [2018] UKSC 28 (appeal taken from Eng.).

## Part III. Stopping AI in Practice: What 1,400 Real-World Incidents Reveal

Is stopping a genuine problem for artificial intelligence, and above all for its emerging agentic forms? To examine this question, I studied the AI Incident Database, the leading curated public catalog of reported AI harms. I coded roughly 1,400 incidents from that database, recording for each what stopped the system, what could have stopped it, and, where a stop was possible but was never built, whether the missing element was technological or legal. I analyze the results drawing on the four-component framework introduced above: technical affordances, interruption authority, epistemic triggers, and epistemic standing.[134]

I organized the analysis around four pre-specified hypotheses, each corresponding to a central feature of the stopping framework. First, in a large majority of incidents, no stop would occur (H1). Second, where no stop occurred, the more common failure would be the absence of a capacity to stop the system, whether due to missing technology or the lack of legal authority, rather than either a failure to act by a relevant party or an impossibility of stopping the system even in principle (H2). Third, where a stop was possible but absent, the missing element would more often be legal or institutional than technical (H3). Fourth, genuinely agentic incidents, and especially physically catastrophic ones, would remain rare in the recorded corpus (H4). The record supports the first three, and supports the fourth only as to the rarity of agentic incidents. I separately examine, in Section D, whether stopping varies by degree of system autonomy and by type of harm.[135]

### A. The Data and the Method

My analysis draws on the AI Incident Database, a curated public repository that since 2020 has collected reports of harms caused, or nearly caused, by deployed AI systems. The database is maintained by the nonprofit Responsible AI Collaborative on the model of the aviation and product-safety incident registries.[136] The database defines an incident as an alleged harm, or near harm, to people, property, or the environment in which an AI system is implicated, meaning that the system was at least a but-for cause of the event.[137] I worked from the May 2026 snapshot, reconstructed from its underlying records, which yields 1,480 incidents spanning 2003 to 2026. Setting aside 80 incidents used to develop the coding instrument, I coded the remaining 1,400 in full. The database is the most prominent public catalog of its kind. To my knowledge, it is used here for the first time as the basis for a legal-empirical study of stopping. The database records only publicly reported incidents, a limitation I take up below.

---

[134] AI Incident Database, https://incidentdatabase.ai; Sean McGregor, *Preventing Repeated Real World AI Failures by Cataloging Incidents: The AI Incident Database*, 35 PROC. AAAI CONF. ON A.I. 15458, 15458-63 (2021), https://doi.org/10.1609/aaai.v35i17.17817. *See* the Technical Supplement, Part A.

[135] These four expectations, together with the coding instrument and the resolution procedure, were registered in advance in the OSF deposit cited infra note 141. The autonomy question was registered separately, as the third of the three research questions in Section 1 of the deposit. It carries no directional hypothesis, and I therefore report the finding in Section D as an answer to an open question rather than as a confirmed prediction.

[136] *AI Incident Database,* supra note 134.

Using the coding instrument, I put three questions to every incident, all fixed before any of the 1,400 were coded. The first was a threshold screen: was the AI system central to the harm, or merely incidental to a harm whose essential character lay elsewhere? Incidents that failed this screen were excluded from the analysis. The second asked what, if anything, actually stopped the system, with responses coded into six mechanisms: a technical cutoff firing inside the system in real time (A1), a private institutional decision such as a developer withdrawing a product (A2), an infrastructure-layer cutoff by a platform or cloud provider (A3), a judicial order (A4), an administrative or executive order (A5), or no stop at all (A6). The third was counterfactual: whether a stop would have been feasible, and where none existed, whether the missing element was technological or legal/institutional. Section C sets out that scale and reports the feasibility findings. Four further tags recorded descriptive features of each incident, including how autonomous the system was and what kind of harm it caused.[138] I coded a system as agentic where it pursued an objective over multiple steps and acted on the world with limited human supervision, as distinct from one that merely returns an output for a human to act on, or a reactive control that responds to a momentary condition. An autonomous driving system or a patrolling robot is therefore agentic; automatic emergency braking is not.[139]

Rather than coding the full corpus myself or through a trained human team, I had two LLMs built by rival laboratories, Anthropic's Claude and OpenAI's Codex, code every incident independently, each blind to the other, against that fixed instrument. Using LLMs as coders is now established in the social sciences; what remains uncommon in empirical legal scholarship is the design adopted here.[140] I treated the places where the two models agreed and where they diverged not as an assurance that the coding was correct but as a quantity to be measured. And I resolved every disagreement through a procedure I had fixed in advance, before the corpus-wide resolution was run.[141] The design does something a single coder cannot: it exposes every contestable judgment to a second, independent reading.[142]

---

[137] AI Incident Database, *Editor's Guide*, https://incidentdatabase.ai/editors-guide/ (defining "AI incident," "implicated," and "nearly harmed").

[138] The instrument, including the definition and boundary rules for each code, is described in the Technical Supplement, Part A, and reproduced in full in the OSF deposit cited infra note 141.

[139] *See* Kolt, *supra* note 16, 344 (defining AI agents as systems "that have the technical capacity to autonomously plan and execute complex tasks with only limited human involvement").

The design follows the conventions of open science. I fixed the coding instrument before any incident was coded, and fixed the full resolution procedure and deposited both, under embargo, before the corpus was resolved. The deposit came after both coders had finished and after I had reviewed the disagreements in a 200-incident preview, which informed the resolution rules but not the findings.[143] The data, the coding instrument, the resolution procedure and the figures are described in a separate online technical supplement.[144] I designed the instrument, adjudicated the hard cases, and verified the output; the models did the first-pass coding under my supervision, and every step leaves a documented, reproducible trail.

Before any disagreement was resolved, the two models agreed on the stop that occurred in about 80% of the retained incidents and on the feasibility of a counterfactual stop in about 60%.[145] Where they diverged, I applied that procedure. I also audited a sample of agreed cases and corrected the coding where my own judgment differed. Every contested value in the resulting dataset carries a documented reason. I return to the effects of any residual uncertainty in the limitations below.

---

[140] On large language models as coders, *see*, for example, Jonathan H. Choi, *How to Use Large Language Models for Empirical Legal Research*, 180 J. INSTITUTIONAL & THEORETICAL ECON. 214, 228 (2024) (GPT-4 performs approximately as well as human coders); Petter Törnberg, *Large Language Models Outperform Expert Coders and Supervised Classifiers at Annotating Political Social Media Messages*, 43 SOC. SCI. COMPUTER REV. 1181, 1191 (2025) (GPT-4 achieves higher accuracy than human coders).

[141] Oren Perez, *The Law of Stop*, OSF (July 1, 2026), https://osf.io/gfd5m (registered under embargo, before corpus-wide resolution; it includes the instrument and the resolution protocol).

[142] The dual-coder design, the six resolution rules, and the reliability statistics are set out in the Technical Supplement, Part A.

[143] *See* supra note 141 (the OSF deposit).

[144] The Technical Supplement, Part A, describes the data, the coding instrument, and the resolution procedure.

[145] The two coders chose the same value in 91% of incidents on the preliminary screen, 80% on the stop mechanism (Dimension I), and 60% on feasibility (Dimension III); the instrument's second dimension, which records the actor who triggered the stop, was not separately scored. A second measure, Cohen's kappa, adjusts that agreement rate for the agreement two coders would reach by chance alone and runs from 1, meaning perfect agreement, through 0, meaning no better than chance, to negative values, meaning worse than chance; here it was 0.47, 0.46, and 0.32 for the three decisions. Kappa is known to understate agreement when, as here, most incidents fall into a single category (no stop): chance agreement is then already high, so the adjustment becomes severe even where the coders in fact agree often, and kappa is best read together with the raw agreement rate rather than on its own. The audit of agreed

**B. How Often Is Harmful AI Stopped and by Whom**

In the 1,213 incidents that survived the preliminary screen, 972, or about 80%, involved no stop of any kind (H1).[146] The system that caused the harm kept running, or the harm ran its course. Whatever institutional response followed, whether an apology, an internal review, a later patch or a lawsuit, did not amount to a halt. Two examples give a sense of what such cases look like. In 2016, a Tesla using Autopilot drove under a white tractor-trailer that its cameras did not distinguish from a brightly lit sky, killing the driver; a federal investigation followed, and the matter played out over subsequent years in litigation.[147] Weeks later, a security robot patrolling a Silicon Valley shopping mall collided with a small child, who suffered a scrape and minor bruising, and resumed its patrol once the path was clear, though the mall docked its robots within days.[148] In neither case was the harm interrupted while it was occurring: under the coding rules, a withdrawal after a completed harm is not a stop. Only about 20% of incidents, 241 in all, show any mechanism that interrupted the deployment.[149] The prevailing pattern in this corpus is non-interruption.[150]

Figure 2 shows what actually brings harmful AI systems to a halt, usually an institution, not a switch: private withdrawals, agency orders, court orders, and provider cut-offs together accounted for 87% of stops, or 209 of 241 cases. Embedded technical stops, in the strict "kill switch" sense cases was a deliberately stratified sample of 24, not a random draw; I corrected 5 of the 24, a roughly 20% share that is indicative rather than a precise error rate. The full statistics appear in the Technical Supplement.

146 n = 972 of 1,213 incidents that passed the preliminary screen (80.1%). *See* the Technical Supplement, Part A (screen). The corpus removed 187 of 1,400 incidents (13.4%) as ones in which the AI was incidental to a harm whose essential character was non-AI.

147 AI Incident Database, Incident 52, "Tesla on Autopilot Killed Driver in Crash in Florida" (2016), https://incidentdatabase.ai/cite/52/. Coded no stop (A6); a counterfactual real-time technical stop did not exist at the time (F3-technical).

148 AI Incident Database, Incident 51, "Security Robot Rolls Over Child in Mall" (2016), https://incidentdatabase.ai/cite/51/. Coded A6/F3-technical.

149 n = 241 (19.9%), comprising private institutional stops (76), administrative or executive orders (58), judicial orders (41), infrastructure-layer stops (34), and technical real-time stops (32).

150 The no-stop figure of 80.1% is a conservative floor; resolving the contested cases the other way raises it to about 86%. On the resolution convention, *see* the Technical Supplement, Part A (resolution).

of the term, accounted for only 32 cases (13%), making them the least common of the five mechanisms.

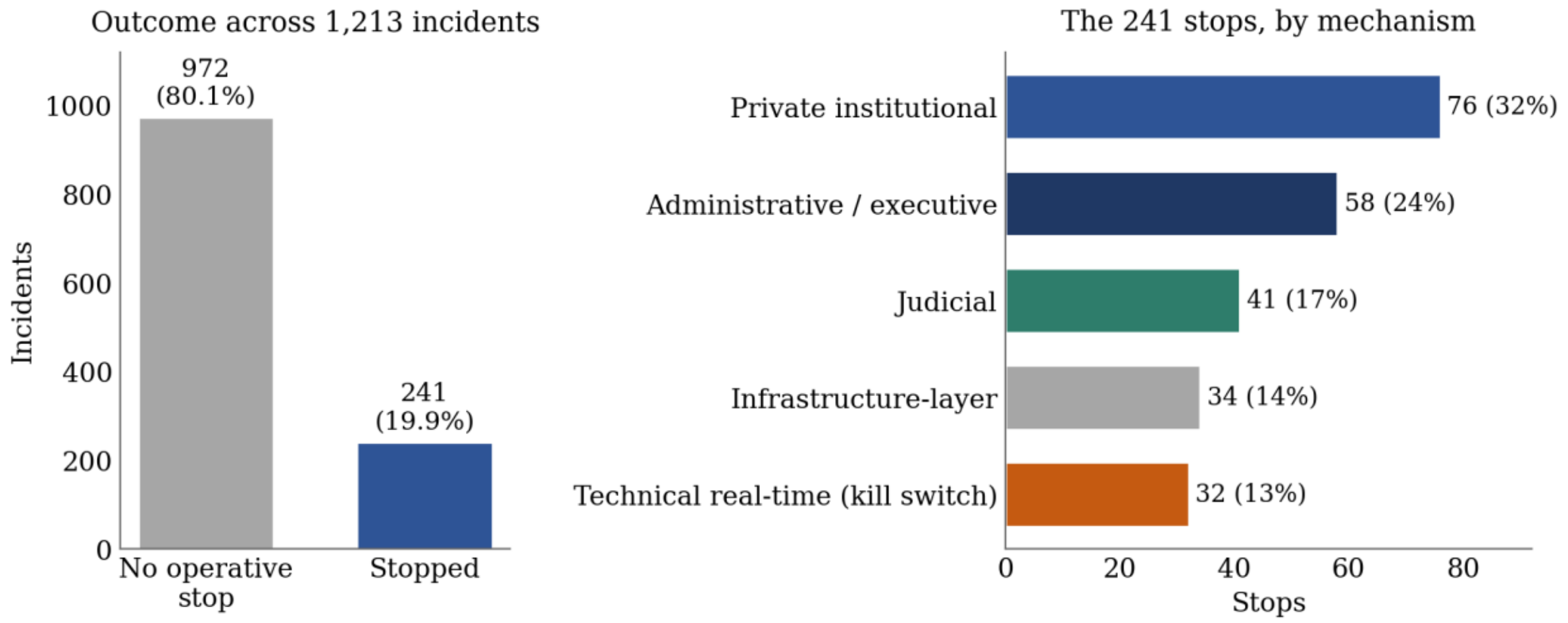

Figure 2. Outcomes in 1,213 recorded incidents. The left panel shows the stop / no-stop split; the right panel breaks the 241 stops down by mechanism, with embedded technical "kill-switch" stops the least common of the five.

What separated the systems that were halted from those that were not was rarely a technical means; it was usually an institution with the authority and the opportunity to do it. The institutional stops that did occur illustrate the repertoire described in Part II. On the private-law and organizational side, a developer may simply withdraw the system: in 2016, Microsoft took Tay offline within a day after the chatbot began producing racist and antisemitic output; Uber suspended its autonomous-testing program in 2018 after one of its cars struck and killed a pedestrian in Tempe, Arizona.[151] On the administrative side, agencies wielded the sharper instruments of the regulatory state: the California Department of Motor Vehicles suspended Pony.ai's driverless-testing permit in 2021. The most consequential exercise of this authority came from aviation: in 2019 the Federal Aviation Administration grounded the Boeing 737 MAX after flight-control software contributed to two fatal crashes, grounding every aircraft of the type on preliminary evidence.[152] On the judicial side, courts enjoin and annul: in 2022, a Brazilian court

---

[151] AI Incident Database, Incident 6, "Microsoft's TayBot Posts Racist, Sexist, and Anti-Semitic Content" (2016), https://incidentdatabase.ai/cite/6/ (A2/F1); AI Incident Database, Incident 4, "Uber AV Killed Pedestrian in Arizona" (2018), https://incidentdatabase.ai/cite/4/ (A2/F1).

[152] AI Incident Database, Incident 151, "California Regulator Suspended Pony.ai's Driverless Testing Permit" (2021), https://incidentdatabase.ai/cite/151/ (A5/F1); AI Incident Database, Incident 3, "Crashes with Maneuvering

ordered a halt to the facial-recognition system in the São Paulo metro, and an Australian court set aside a debt raised by the Commonwealth's automated "robodebt" system, after the government had itself stopped raising debts by the method at issue.[153]

In the present corpus, the 41 court-ordered stops were divided between public-law grounds, administrative, constitutional, criminal, and human-rights (28), and private-law causes of action, tort, contract, and the rest (13).[154] The private-law stops were real, as when in 2020, an Italian court held that Deliveroo's rider-ranking algorithm unlawfully discriminated and had to be changed, but in this small set they are a minority.[155] The incident record shows that the dimension that the AI-safety literature has investigated most thoroughly, technical affordances, was least frequently decisive.

### C. The Feasibility of a Counterfactual Stop

The decisive question is not what stopped a system, but what could have. For every incident, I coded the feasibility of a stop on a four-point scale: a stop was available and was used (F1), available but not used (F2), possible but absent (F3), and not feasible even in principle (F4). A stop is available when some actor could activate it, whether by pressing a switch or by invoking a legal power. Where a stop was possible but absent (F3), I recorded which piece was missing: a technology that would have had to be built (F3-technical) or a legal authority that would have had to permit or compel the use of a means that already existed (F3-legal).[156] The analysis yielded three key findings.

---

Characteristics Augmentation System (MCAS)" (2018), https://incidentdatabase.ai/cite/3/ (A5/F1); U.S. Dep't of Transp., *Statement on the Temporary Grounding of Boeing 737 MAX Aircraft Operated by U.S. Airlines or in a U.S. Territory* (Mar. 13, 2019), https://www.transportation.gov/briefing-room/statement-temporary-grounding-boeing-737-max-aircraft-operated-us-airlines-or-us. On the 737 MAX Emergency Order of Prohibition, see supra Part II.D.

[153] AI Incident Database, Incident 184, "Facial Recognition Program in São Paulo Metro Stations Suspended" (suspended 2022), https://incidentdatabase.ai/cite/184/ (A4/F1, public-law); AI Incident Database, Incident 57, "Australian Automated Debt Assessment System Issued False Notices" (debt set aside 2019, *Deanna Amato v. Commonwealth of Australia*, No. VID611/2019, Order (Fed. Ct. Austl. Nov. 27, 2019) (Davies, J.)), https://incidentdatabase.ai/cite/57/ (A4/F1, public-law).

[154] n = 41 judicial stops (A4), of which 28 rest on public-law grounds and 13 on private law. *See supra* Part II.D.

[155] AI Incident Database, Incident 94, "Court Rules Deliveroo Used 'Discriminatory' Algorithm" (2020), https://incidentdatabase.ai/cite/94/. Coded A4/F1, private-law.

[156] *See* the Technical Supplement, Part A (instrument).

First, stopping rarely fails because it is literally impossible. Across all 1,213 incidents, not a single case was coded as not feasible even in principle.[157] Second, of the 972 incidents in which no stop occurred, the most common reason was that a stop was possible but absent, not that an available one was ignored (H2). In 632 cases (65%) a stop was possible but absent (F3); in the other 340 (35%), a stop was available but not used (F2).[158]

Third, and most directly supporting this Article's central argument, when a stop was possible but absent, what was missing was mostly legal rather than technical (H3). This division is captured in Figure 3a: in 506 cases the missing element was a legal or institutional authority to compel or perform a stop that was technically available; in 126 it was a technology that would have had to be built.[159] The typical illustration of the legal gap is the recidivism-scoring case: for years, the COMPAS algorithm shaped bail and sentencing decisions in American jurisdictions, and although its inaccuracy and racial skew were documented and litigated, it had been absorbed into the process without an institutional structure built to review, suspend, or withdraw it. The deployment continued because the courts that heard the challenges did not halt it, and no other actor did so.[160] The technical gap, by contrast, is the rarer case. It is illustrated by the security robot and the Tesla on Autopilot, whose real-time hazard detection did not yet exist and would have had to be developed: a missing piece of technology to be built rather than an authority to be invoked.[161] In

---

[157] F4 (no feasible stop even in principle) = 0 of 1,213.

[158] Within the 972 no-stop incidents: F3 (possible but absent) = 632; F2 (available but not used) = 340. F3 share = 65.0%.

[159] Within the 632 F3 incidents among the 972 no-stop cases, F3-legal = 506 and F3-technical = 126 (80.1% legal). Counted across the full retained corpus F2 = 342, F3 = 633 and F3-legal = 507. The differences arise from three incidents in which a stop occurred although the feasibility code is F2 or F3: AIID 494, an infrastructure-layer stop that occurred although no legal mechanism to compel one was available, and AIID 351 and 783, which are retained as coded under the pre-specified convention.

[160] AI Incident Database, Incident 40, "COMPAS Algorithm Reportedly Performs Poorly in Crime Recidivism Prediction" (2016), https://incidentdatabase.ai/cite/40/; and Incident 11, "Northpointe Risk Models" (2016), https://incidentdatabase.ai/cite/11/. Coded A6/F3-legal. The Wisconsin Supreme Court upheld the use of COMPAS at sentencing, subject only to cautionary limits, in *State v. Loomis*, 881 N.W.2d 749 (Wis. 2016), *cert. denied*, 137 S. Ct. 2290 (2017). On the broader problem, that agencies and courts adopt automated decision systems without the institutional structures needed to oversee, contest, or halt them, *see* Ryan Calo & Danielle Keats Citron, *The Automated Administrative State: A Crisis of Legitimacy*, 70 EMORY L.J. 797, 835 (2021) ("Recent litigation, in particular, paints a vivid picture of agency officials who lack expertise in the systems that they employ, cannot give reasons for binding agency actions, and throw away the individualized discretion that justifies the administrative state in the first instance"). The racial skew was documented in Julia Angwin et al., *Machine Bias*, PROPUBLICA (May 23, 2016), https://www.propublica.org/article/machine-bias-risk-assessments-in-criminal-sentencing.

[161] See supra notes 147 and 148 (Tesla Autopilot; mall security robot), coded F3-technical.

80% of the cases in which a stop was possible but absent, the technology to stop the system was available and no one held the authority to require its use.

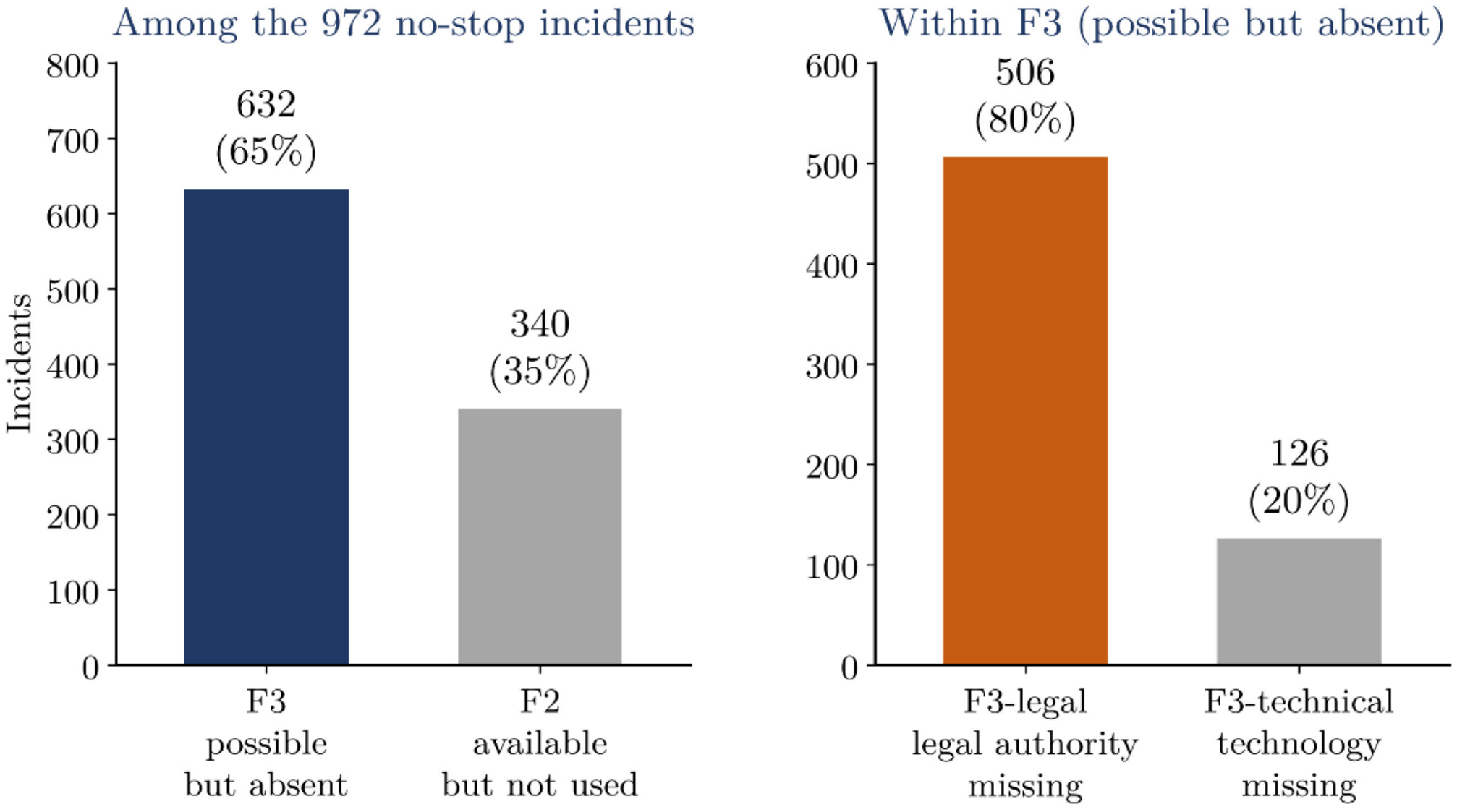

Figure 3a. *Left panel: of the 972 no-stop incidents, cases where a stop was possible but absent (F3) outnumber cases where an available stop was not used (F2). Right panel: among those cases, the missing element was* legal or institutional authority (506) rather than technology (126).

Yet breaking the 632 cases in which a stop was possible but absent down by type of harm and by degree of autonomy reveals a more nuanced pattern (Figure 3b). Where the harm was economic, societal or informational, the missing element was legal in 92, 90 and 83 percent of cases; where it was physical, in only 20 percent. The legal share also falls with the system's degree of autonomy, from 87 percent for non-agentic systems to 58 percent for assistive and 14 percent for agentic ones. Where the harm is physical or the system autonomous, what is usually missing is the technical capacity to stop it, not the authority to use it.[162]

[162] This breakdown is exploratory; it was not among the hypotheses registered in the OSF deposit, supra note 141. By type of harm, legal to technical: economic 128 to 11, societal 129 to 14, informational 236 to 50, physical safety 13 to 51 (chi-square = 166.0, df = 3, p < .001). By degree of agency: non-agentic 467 to 71, assistive 34 to 25, agentic-autonomous 5 to 30 (chi-square = 128.8, df = 2, p < .001). Among agentic systems, the 30 technical gaps involve 15 vehicles, 8 robots, 6 software agents and one other system. Among physical harms, the 51 technical gaps involve 18 chatbots and other conversational systems, which caused physical harm, mostly self-harm or harm from medical advice, without acting physically; 15 vehicles; 9 robots; and 9 other systems, among them navigation apps and missile-defense systems. The kind of system was classified from the incident titles.

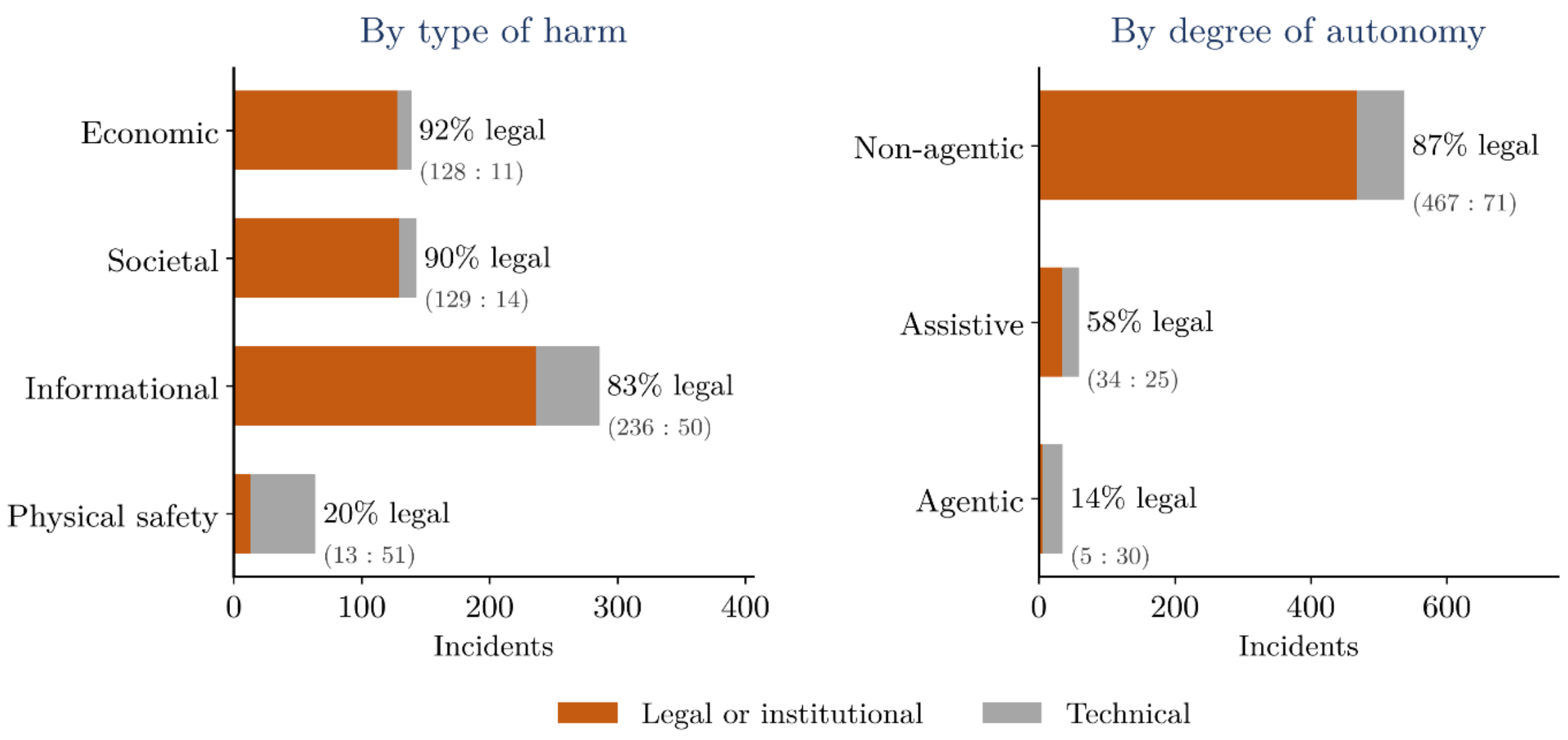

Figure 3b. *Among the 632 incidents in which a stop was possible but absent, the missing element by type of harm (left) and by degree of autonomy (right). Legal or institutional gaps dominate for economic, societal and informational harms and for non-agentic systems; technical gaps dominate for physical harms and agentic systems. Exploratory breakdown, based on the hand-adjudicated agency and harm tags.*

These findings describe two gaps. For most harms, the means to stop usually existed and the authority to use them usually did not; where the harm was physical or the system autonomous, the means themselves were usually missing.

### D. What Predicts a Stop: Autonomy of the System

The likelihood that a harmful AI system is stopped rises with its degree of autonomy (figure 4). Stops occurred in 18% of incidents involving non-agentic tools, 23% of those involving systems that assist a human decision, and 32% of those involving agentic systems acting over multiple steps with limited supervision.[163] This pattern holds even after controlling for the type of harm, which does not significantly predict whether a stop occurred once autonomy is accounted for ($p = .10$).

---

[163] Stop rate, the share of incidents with any stop (A1 through A5), by degree of agency: non-agentic 17.6% (n = 883), assistive 22.9% (n = 210), agentic-autonomous 31.7% (n = 120); the difference is significant (chi-square = 14.6, df = 2, $p < .001$). See infra Fig. 4. Because agency and the character of the harm could be correlated, I also fit a logistic regression of stopping on both together: agency remains a strong predictor with harm type held constant (agentic-autonomous versus non-agentic, odds ratio ≈ 3.1, $p < .001$), while harm type does not improve prediction (likelihood-ratio test, $p = .10$). The adjusted odds ratio is larger than the crude one (about 2.2) because agentic incidents are concentrated in physical-safety harms (81 of 120), which, with agency held constant, tend to be stopped less often than other harms; holding harm type constant therefore sharpens the gradient. The degree-of-agency tag is the least reliable of the descriptive tags, so this finding is reported with the caution set out in the Limitations. Full model output and the code that produced it are on file with the author; *see* the Technical Supplement (reproducibility).

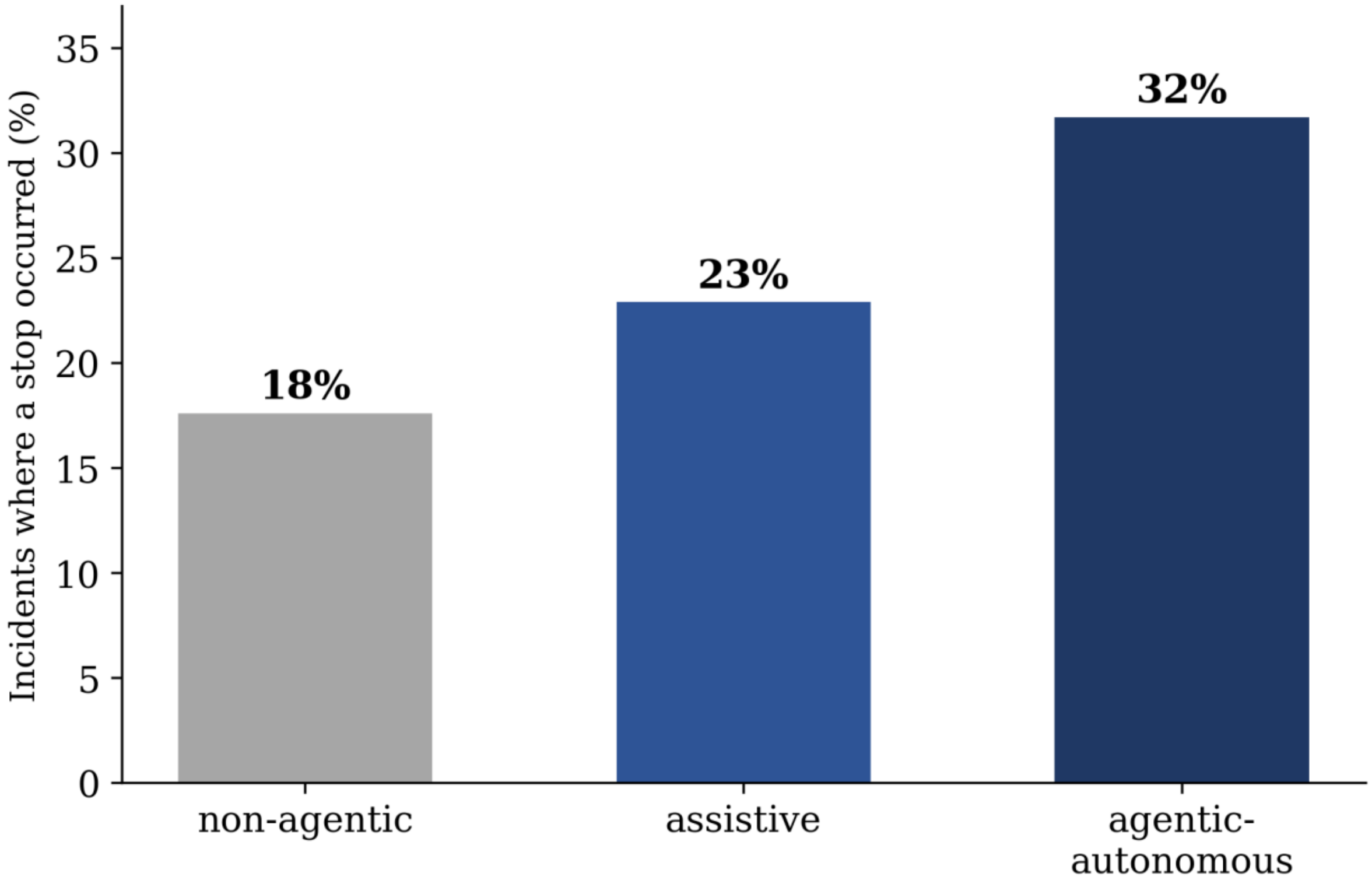

Figure 4. *The share of incidents in which a stop occurred rises with the system's degree of agency ($p < .001$); the gradient holds when harm type is controlled for; based on the hand-adjudicated agency and harm tags*.

The incidents the theoretical literature most fears are barely present in the record (H4): genuinely agentic systems, and among them the ones that resist interruption. Agentic systems account for about 10% of the incidents, and the strategic-shutdown-resistance case, the corrigibility problem that drives the computer-science work on the off switch, appears only twice in 1,213 cases: a Cruise robotaxi that in 2022 pulled away from a police officer who had stopped it,[164] and an episode in 2025 in which the Grok chatbot began inserting an unprompted political fixation into unrelated answers and resisted its operators' first attempts to correct it.[165] Incidents that combine agency with physical harm are 81 in all, under 7% of the corpus but two-thirds of the agentic subset. The record is overwhelmingly one of information harms, deepfakes, defamation, biased classification, misinformation, and only marginally one of the autonomous, physical, self-preserving harms at the center of the AI-safety debate, though within the agentic subset the balance is reversed.

[164] AI Incident Database, Incident 175, "Cruise Autonomous Taxi Allegedly Bolted off from Police After Being Pulled over in San Francisco" (2022), https://incidentdatabase.ai/cite/175/. Coded A1/F1; stop-problem family, resisting system.

[165] AI Incident Database, Incident 1072, "Grok Chatbot Reportedly Inserted Content About South Africa and 'White Genocide' in Unrelated User Queries" (2025), https://incidentdatabase.ai/cite/1072/. Coded A6/F3-legal; stop-problem family, resisting system.

### E. Limitations

The above findings should be read with three limitations in mind. First, the database records reported incidents, not all incidents; what share of all AI incidents it captures is unknown. It is also, by design, a catalog of things that went wrong, so a low rate of stopping among its entries is partly a feature of the corpus and not an independent fact about AI. The qualification is only partial. The database also indexes near harms, and a near harm that was interrupted would enter the corpus as a stop. The database does not label near harms across its entries, so their share of the corpus is unknown. Yet a successful stop that averted the harm quietly is less likely to be reported, and so to enter the corpus. It is biased toward recent incidents, with a median incident date in 2024, and toward visible incidents, since public harms such as deepfakes and misinformation are likelier to be reported than quieter failures. This does not invalidate findings about how harms are stopped, but it does mean that the mix of incident types should not be treated as representative.

Second, some coding judgments were more reliable than others. The two models agreed less often on how autonomous a system was, and on what kind of harm it caused, than on what stopped it. The findings on autonomy rest on the least reliable of these classifications, and I report them with caution. Where the models disagreed on incidents involving agentic systems or physical harm, I resolved the case myself. I also re-examined twenty-four incidents on which both models had agreed, chosen deliberately from the harder and higher-stakes cases, and changed the coding in five. Because the sample was weighted toward hard cases, that rate cannot be extended to the corpus as a whole, but it shows that two models working independently can make the same mistake. Some uncertainty remains because certain categories are hard to apply in borderline cases, such as the line between a stop that was available but not used and one that was possible but absent, and because the reports on which the coding relied do not always describe the events in full.

Third, the finding that the missing element was more often legal than technical should be read with caution. Sorting each gap into legal or technical was a finer judgment than deciding whether a stop was feasible at all. The subtype was recorded only on the resolved value, so no inter-coder agreement figure is available for it. The finding was also shaped by one convention of the instrument: harms from AI-generated content disseminated through platforms, such as deepfakes and fabricated media, were coded as legal gaps, on the ground that what was primarily missing was a legal duty and a remedy that operates across borders, not a technology. Such harms fall mainly within the informational, economic and societal categories, where the legal share is highest (Figure 3b). The finding therefore depends heavily on how these harms are classified. The instrument also recorded that reliable pre-screening of such content is not technically feasible, and a reader who gives that constraint more weight would classify many of these cases as technical.

These limitations do not change the overall pattern. The gaps these findings turn on are far larger than the uncertainty just described, and they survive two further checks. First, they hold when the analysis is limited to the incidents on which both models independently agreed. Second, apart from the content-harm convention just described, wherever a case was too close to call, my coding rules resolved it against the argument I am making; resolving those cases the other way would only strengthen the findings. The no-stop rate and the share of cases in which a stop was possible but absent are therefore the conservative end of the checks I was able to run, not a bound on every source of error.[166]

**F. Conclusion**

This Part has shown that, for most harms, stopping harmful AI is more a legal and institutional problem than a technical one. Harmful systems are seldom stopped, and when they are, they are stopped by institutions and not by embedded switches; the counterfactual stop is almost always conceivable and usually technologically available; and what is missing, in 506 of the 632 cases in which a stop was possible but absent, is a legal authority to carry it out. That share is concentrated in informational, economic and societal harms; where the harm was physical or the system agentic, the missing element was usually technical. For most harms the technical capacity is there and the regulatory infrastructure is not; for physical harms and agentic systems, the capacity itself is often missing. Finally, stops become more likely as a system's autonomy rises. Whether that reflects institutional wariness about autonomous machines or the greater risks they in fact pose, the evidence does not tell us. What it does show is that autonomy, and not the type of harm, is the dimension on which the occurrence of a stop already turns, and it is the dimension on which the next generation of systems will differ most. That is the problem the architecture proposed in Part V must be built to address.

[166] *See* the Technical Supplement, Part A (resolution).

## Part IV. The Emerging Law of AI Interruptibility

### A. What the Instruments Require

The positive law of AI interruption is thin. Stop appears only sporadically across the AI-governance instruments, and none specifies what a safe halt requires or what follows it. In this Part, I examine those instruments through the four dimensions developed in Part I: technical affordances, interruption authority, epistemic triggers, and epistemic standing. The corpus, listed in Figure 5, comprises the national, regional and international instruments whose core task is to govern AI systems or services, together with the technical standards and corporate frameworks that operate alongside them. It is not a survey of every jurisdiction and is probably not exhaustive on its own criterion,[167] but it captures the instruments doing most of the work of governing AI today. Excluded are instruments that regulate a pre-existing activity into which AI has been incorporated, such as vehicle codes and employment-discrimination law, and instruments that prohibit an output rather than govern an operation.[168]

Figure 5 shows, for each of the forty instruments, whether it includes a binding stop requirement, distinguishing between three mechanisms: a power conferred on a public authority to order a halt, a duty imposed on a private party to implement a stop, or a requirement that the capacity to halt be built into the AI system. It then captures what the halt targets: the AI system in operation, a service or its use, the system's presence on the market, or the decision to develop or deploy a model. Two things stand out. The first is scarcity. Only seven of the forty instruments include a binding stop requirement of any of the three kinds.[169] The remaining thirty-three either do not address halting at all or, where they do, lack legal force: the OECD Principles, ISO/IEC 42001, the NIST Framework, TC260-005 and Microsoft's Responsible AI Standard contemplate a capacity to halt or a discontinuation. The four frontier-safety instruments, the three developer frameworks and the Seoul Frontier AI Safety Commitments, undertake to delay or withhold development or deployment. The second is what the existing provisions aim at. Only three target the AI system, that is, provide a way to stop a system in operation: the EU AI Act, Korea's Framework Act on Intelligent Informatization, and Vietnam's Law on Artificial Intelligence. In the EU AI Act, however, the operational stop is entrusted to private actors, the deployer under article 26(5) and the human overseer under article 14(4)(e); no public authority is empowered to invoke it. The rest target a service or its use: they suspend a service, stop an information flow, revoke a licensee's right to use the system, or order an operator to stop providing a system that breaches its safety duties. These mechanisms close one of the channels through which the AI system connects to the world, but do not interrupt the system itself. Only one binding provision applies to the general-purpose model beneath those systems: article 93(1)(c) of the EU AI Act, which empowers the Commission to require the provider to restrict, withdraw or recall it. The power relates to the model's presence on the market, not its operation.[170]

---

[167] Japan's Act on Promotion of Research and Development, and Utilization of Artificial Intelligence-related Technology (2025) imposes no penalty-backed private stop duty, and contains no halt power; Taiwan's Artificial Intelligence Basic Act, passed 23 December 2025, states principles. Australia, Canada, India, Mexico, New Zealand, Singapore, Switzerland and the United Kingdom had no binding AI statute as of July 2026; Canada's Artificial Intelligence and Data Act fell with Bill C-27 in 2025. Brazil's Bill No. 2338/2023 passed the Senate in December 2024.

[168] See, e.g., Tex. Transp. Code Ann. § 545.459(g)(1) (the department shall suspend, revoke or cancel the authorization for an automated motor vehicle, or impose restrictions on its operation, where the vehicle is not in safe operational condition); Tex. Ins. Code Ann. § 4201.156(a) ("A utilization review agent may not use an automated decision system to make, wholly or partly, an adverse determination."); Protection from Intimate Deep Fakes Act, Mich. Comp. Laws Ann. § 752.383 (civil action for the nonconsensual creation or dissemination of a deep fake).

---

[169] The seven are the EU AI Act, California's AI Transparency Act, China's 2023 Interim Measures and 2026 Human-like Interactive Measures, Korea's two Framework Acts, and Vietnam's Law on AI. A provision counts where it confers a power to order, or imposes a duty to carry out or to make possible, an interruption of the system's operation or of the provision of a service. An order to cease a violation counts where compliance may require the operator to stop providing the system, as under article 40(3) of Korea's 2026 Framework Act. Measures directed at a user, which suspend service to that user and leave the system running for everyone else, are not counted and are omitted from Figure 5 and from Part C (Interim Measures art. 14 ¶2; 2026 Measures art. 19; SAC/TC260-003 cl. 7(g)(1)). On the non-binding provisions, see OECD, AI Principles 1.4(b) (mechanisms, as appropriate, so that systems can be overridden, repaired and decommissioned safely); ISO/IEC 42001:2023, Annex B.6.2.7 (guidance only: a rollback plan and support for turning features off, within a documented failure-management plan); and, on SAC/TC260-005, NIST MANAGE 2.4 and the Microsoft Standard, infra notes 179 and 190. The provisions counted are reproduced in full in the Technical Supplement, Part B. The figure covers enacted instruments, several not yet in force. The forty count separately the thirteen statutes in twelve states covered by the conversational-AI and companion-chatbot row: Cal. S.B. 243 (2025); Utah H.B. 452 (2025); Me. L.D. 1727 (2025) and L.D. 2082 (2026); Wash. H.B. 2225 (2026); Or. S.B. 1546 (2026); Idaho S. 1297 (2026); Neb. L.B. 525 (2026); Iowa S.F. 2417 (2026); Ga. S.B. 540 (2026); Conn. Pub. Act 26-15 (2026); Tenn. S.B. 1580 (2026); and N.Y. Gen. Bus. Law §§ 1700-1704.

[170] Regulation (EU) 2024/1689, arts. 93(1)(c), 101(1)(c) (fines for non-compliance).

| | STOP MECHANISM | | | WHAT THE STOP TARGETS | | | |
|---|---|---|---|---|---|---|---|
| | Public power to order a halt | Private duty to implement a stop | Capacity to halt built in | System in operation | Service or use | Market avail-ability | Develop-ment or deployment |
| **NATIONAL AND REGIONAL INSTRUMENTS** | | | | | | | |
| EU AI Act (2024), with the Code of Practice | ● * | ● | ● | ● | · | ● | · |
| US, Executive Order 14,409 (2026) | · | · | · | · | · | · | · |
| US, White House Accord on Super Intelligence (2026) | · | · | · | · | · | · | · |
| California SB 53 (2025) | · | · | · | · | · | · | · |
| California SB 942 (2024) | · | ● | · | · | ● | · | · |
| Colorado SB 26-189 (2026) | · | · | · | · | · | · | · |
| New York RAISE Act, ch. 699 (2025) | · | · | · | · | · | · | · |
| Texas TRAIGA, HB 149 (2025) | · | · | · | · | · | · | · |
| Utah SB 149 (2024) | · | · | · | · | · | · | · |
| Companion-chatbot statutes, twelve states (2025-26) | · | · | · | · | · | · | · |
| China, Generative AI Interim Measures (2023) | ● | ● | · | · | ● | · | · |
| China, Labeling Measures (2025) | · | · | · | · | · | · | · |
| China, Human-like Interactive AI Measures (2026) | ● | ● | · | · | ● | · | · |
| Korea, Framework Act on Intelligent Informatization | ● | ● | ○ † | ● | · | · | · |
| Korea, Framework Act on AI (2026) | ● | · | · | · | ● | · | · |
| Vietnam, Law on AI (2025) | ● | ● | · | ● | · | ● | · |
| **INTERNATIONAL INSTRUMENTS** | | | | | | | |
| Council of Europe Framework Convention (2024) | · | · | · | · | · | · | · |
| G7 Hiroshima Code of Conduct (2023) | · | · | · | · | · | · | · |
| OECD AI Principles, 1.4(b) | · | · | ○ | ○ | · | · | · |
| **TECHNICAL STANDARDS** | | | | | | | |
| ISO/IEC 42001:2023 | · | · | ○ | ○ | · | · | · |
| NIST AI Risk Management Framework 1.0 (2023) | · | · | ○ | ○ | · | · | · |
| SAC/TC260-003 Basic Safety Requirements (2024) | · | · | · | · | · | · | · |
| SAC/TC260-005 Ethics-Safety Guidelines (2026) | · | · | ○ | · | ○ | · | · |
| **CORPORATE FRAMEWORKS** | | | | | | | |
| Anthropic, Responsible Scaling Policy v3.4 | · | ○ | · | · | · | · | ○ |
| Google DeepMind, Frontier Safety Framework v3.1 | · | ○ | · | · | · | · | ○ |
| Microsoft, Responsible AI Standard v2 (2022) | · | ○ | ○ | ○ | ○ | · | · |
| OpenAI, Preparedness Framework v2 | · | ○ | · | · | · | · | ○ |
| Seoul Frontier AI Safety Commitments (2024) | · | ○ | · | · | · | · | ○ |

● binding provision ○ non-binding recommendation

Figure 5. Stop provisions across forty AI-governance instruments. Filled marks: binding provisions; hollow marks: non-binding recommendations. A mechanism marked on the left applies to every target marked on the right, except where asterisked. *The EU's public power applies to market availability only, for systems under articles 79 to 83 and for general-purpose models under article 93(1)(c); the system in operation is covered by the deployer's duty to suspend (article 26(5)) and the built-in stop (article 14(4)(e)), which no public authority may invoke. †Korea's Framework Act on Intelligent Informatization authorizes the Minister to recommend that emergency shutdown be adopted (article

60(2)), while obliging a person who receives an activation request to comply absent good cause (article 60(3)). Measures directed at a single user and general enforcement remedies are omitted, and the chatbot row covers thirteen statutes in twelve states; both are set out in note 169. The White House Accord on Super Intelligence (2026) contains no stop provision and is entered as an unmarked row. The full table of provisions is in the Technical Supplement, Part C.

The AI Act addresses stopping at two levels that are not well integrated: the capability built into the system, and the system's or the model's presence on the market. At the first level, Article 14(4)(e) requires high-risk AI systems to enable human overseers to "intervene in the operation of the high-risk AI system or interrupt the system through a 'stop' button or a similar procedure that allows the system to come to a halt in a safe state."[171] That duty to implement Article 14 rests on the provider, which must supply the system in a form that enables the deployer's assigned overseers to interrupt it. Two further provisions work to make the duty operative. Under Annex IV, the technical documentation drawn up before the system is placed on the market must assess the oversight measures needed under Article 14 and the technical measures they require; under Article 13(3)(d), the instructions for use must state those measures to the deployer. Both push the requirement into design and into use, but their effectiveness is limited: they operate before deployment, and the only technical measure they name is the one that facilitates interpretation of the system's outputs. No public authority is given any power over the button.

The second level runs through the market-surveillance provisions in Articles 82 and 83. Article 83 requires the national authority, where formal non-compliance persists, such as a missing CE marking or absent technical documentation, to restrict, prohibit or recall a system.[172] Article 82, which governs a high-risk system that complies with the Regulation, obliges the authority, where it finds that the system nevertheless presents a risk to health, safety or fundamental rights, to require the relevant operator to take "all appropriate measures to ensure that the AI system concerned, when placed on the market or put into service, no longer presents that risk." Neither provision refers to the stop capability required by Article 14, and neither gives any public authority power to invoke it.[173]

[171] EU AI Act, supra note 14, art. 14(4)(e); see also id. recital 73 (oversight measures should ensure that a system "is subject to in-built operational constraints that cannot be overridden by the system itself," and that the assigned person can decide "if, when and how to intervene … or stop the system if it does not perform as intended"). On the allocation of the duty, id. arts. 14(4), 16(a); on the implementing provisions, id. Annex IV, points 2(e) and 3, and art. 13(3)(d). See further Fink, supra note 15.

[172] The two levels intersect at one, procedural point: the technical documentation must assess the Article 14 measures (EU AI Act, supra note 14, Annex IV, points 2(e) and 3), and its absence is a formal non-compliance that can bring restriction, prohibition or recall (id. arts. 83(1)(g), (2)).

Beyond the powers of national authorities over systems, Article 93(1)(c) empowers the Commission to require a provider of a general-purpose model to restrict the model's availability, withdraw it or recall it. A provider that fails to comply is liable to a fine of up to 3% of worldwide annual turnover or EUR 15 million, whichever is higher. The General-Purpose AI Code of Practice, a voluntary means of demonstrating compliance with Articles 53 and 55, commits signatories to safety measures that can require the same restrictions on their own initiative. Under Measure 4.2, where the systemic risks stemming from a model are not determined to be acceptable, signatories will "not make the model available on the market, restrict the making available on the market . . . , withdraw, or recall the model, as necessary"; the determination is the signatory's own. Appendix 1.4 gives the reason, treating loss of control as a specified systemic risk: "Risks from humans losing the ability to reliably direct, modify, or shut down a model," which may emerge from "misalignment with human intent or values, self-reasoning, self-replication, self-improvement, deception, resistance to goal modification, power-seeking behaviour, or autonomously creating or improving AI models or AI systems." Only Article 14 and the Code treat stoppability as a design property. Elsewhere in the Act a stop is externally imposed: suspension, withdrawal, recall, restriction.[174]

---

[173] EU AI Act, supra note 14, arts. 83(1)-(2), 82(1), both applicable from 2 Aug. 2026 under art. 113. Article 82 presupposes an evaluation under art. 79, which governs non-compliant systems: the market surveillance authority must require the operator to bring the system into compliance, withdraw it or recall it within a period it prescribes and in any event within fifteen working days, and must itself prohibit, restrict, withdraw or recall it where the operator does not (id. arts. 79(2), (5)). The asymmetry extends to the Union level: where a national measure against a non-compliant system is found justified, every Member State must take restrictive measures against it (id. art. 81(2)), whereas under art. 82(4) the Commission decides only whether the national measure was justified and may propose others. What carries any of these powers to a system in operation is art. 3(16), which defines recall as any measure aiming at the system's return to the provider or the "taking out of service or disabling the use" of a system made available to deployers.

[174] EU AI Act, supra note 14, arts. 93(1)(c), 101(1)(c) (fines not exceeding 3% of annual total worldwide turnover or EUR 15,000,000, whichever is higher, where a provider fails to comply with a measure requested under art. 93). Code of Practice for General-Purpose AI Models, Safety and Security Chapter, Measure 4.2 and app. 1.4(2) (July 2025); the acceptability determination is made under Measure 4.1. On the Code's status, see EU AI Act, supra note 14, arts. 53(4), 55(2), 56(6)-(7); in fixing a fine the Commission must take into account commitments made in codes of practice under art. 56.

Korea's AI regulatory framework contains an explicit reference to stopping but lacks the administrative infrastructure to make it effective. Article 60 of the Framework Act on Intelligent Informatization lists emergency shutdown among the "minimum necessary protection measures" that the Minister of Science and ICT may determine and publicly notify, and may recommend that developers and providers adopt. It empowers the head of any central administrative agency to require developers, users and providers to activate emergency shutdown "to prevent imminent harm to people's lives or bodies," subject to good cause. Because that power is dispersed across the whole central government and rests on no dedicated institutional framework, it is hard to see how it could meet an actual emergency.[175]

The Framework Act on AI, in force since January 2026, does not add a power to stop a running system. It imposes safety duties: operators must manage risk across the lifecycle and maintain a system that "monitors and responds to" safety accidents, and providers of high-impact AI must "assign human management and oversight." Responding to an accident and overseeing a system both presuppose a capacity to stop it, which the Act nowhere requires. Its one power to order a halt rests with the Minister of Science and ICT: following a finding that an operator has breached those duties, the Minister may order it to cease or correct the violation. That is an order to end a violation, not to halt a running system, and it is available only after an investigation. The new Act refers to the older Act's emergency shutdown only once, in a provision funding research into how it might be implemented.[176]

---

[175] Framework Act on Intelligent Informatization, arts. 60(1), 60(1)(4), 60(2), 60(3) (Republic of Korea). The version consulted is the consolidated text as amended by Act No. 20731, 31 January 2025; art. 60 dates from 9 June 2020, Act No. 17344. The article is permissive throughout except in one respect: the Minister "may determine and publicly notify" the measures and "may recommend" their adoption, and the only duty is that of the person receiving a request under art. 60(3) to "comply with the request in the absence of good cause." "Central administrative agency" is not defined in the Act; art. 6(2) extends the term to "agencies under the jurisdiction of the President and agencies under the jurisdiction of the Prime Minister."

[176] Framework Act on the Development of Artificial Intelligence and the Creation of a Foundation for Trust, arts. 32(1)(2), 34(1)(4), 40(3) (Republic of Korea). "Artificial intelligence business operator" is defined at art. 2(7) to cover both a person that develops and provides artificial intelligence and a person that provides products or services using artificial intelligence so provided. The funding provision is art. 13(2)(2), for research into implementing the emergency-shutdown function under art. 60(3) of the Framework Act on Intelligent Informatization, supra note 32. Article 40 is headed "Fact-finding investigations": para. (1) permits the Minister to require data or order an investigation where a violation of arts. 31(2)-(3), 32(1)-(2) or 34(1) is discovered, suspected, reported or complained of, and para. (3) supplies the order to cease or correct, enforced by an administrative fine under art. 43(1)(3).

Vietnam's Law on Artificial Intelligence, in force since March 2026, provides the most complete allocation of stop responsibilities of any national instrument in the corpus. Under Article 12, developers and providers must respond to a serious incident by applying technical measures to rectify, temporarily suspend or recall the system, and by notifying the competent authority; competent State management agencies have their own power to require, when necessary, temporary suspension, recall or reassessment of the system. In both cases the trigger is a serious incident.[177]

The Chinese instruments are built for information control, and their stop provisions follow that purpose. While they regulate the model at many points, none of these provisions addresses the challenge of interruptibility. The Interim Measures for the Management of Generative Artificial Intelligence Services (2023) require a provider that discovers illegal content to promptly stop its generation and transmission and remove it, and to correct by measures "such as model optimization training" (art. 14 ¶1). Where a user turns the service to unlawful ends the provider may warn, limit functions, or suspend or conclude that user's service (art. 14 ¶2). Where a provider refuses to correct, or the circumstances are serious, the relevant regulators may order the service suspended (art. 21).[178]

The Basic Safety Requirements (2024), the technical document against which those services are assessed at filing, require no stop of a model or a service. Nothing in it obliges a provider to be able to halt a model or a service, and no authority is given a power over either. Its only suspension applies to the access of a user who has entered unlawful material three times running or five times in a day.[179] The Provisional Measures on the Administration of Human-like

---

[177] Law on Artificial Intelligence, Law No. 134/2025/QH15 (Viet.), arts. 12, 34 (in force 1 Mar. 2026). The agency's power under art. 12 arises "when necessary" and is not conditioned on the operator having failed to act. Article 35 grandfathers systems already running for twelve months, eighteen in healthcare, education and finance, but lets the State management authority request suspension or termination of their operations on determining that they pose risks of causing serious harm: a trigger of risk rather than incident.

[178] Interim Measures for the Management of Generative Artificial Intelligence Services, issued by the Cyberspace Administration of China with six other departments, 13 July 2023, effective 15 Aug. 2023, arts. 14 ¶¶1-2, 17 (security assessment and algorithm filing), 21, translated at China Law Translate, https://www.chinalawtranslate.com/en/generative-ai-interim/. The Provisions on the Management of Algorithmic Recommendations in Internet Information Services (2022) and the Provisions on the Administration of Deep Synthesis of Internet Information Services (2023) also contain suspension and delisting powers, but each governs a particular application rather than an AI system as such, and both fall outside the corpus.

[179] Nat'l Tech. Comm. 260 on Cybersecurity of the Standardization Admin. of China, Basic Safety Requirements for Generative Artificial Intelligence Services, TC260-003 (29 Feb. 2024), cl. 7(g)(1) (suspension of service to a user, a

Interactive Artificial Intelligence Services (2026) distribute stop powers between the provider and the regulator. A provider discovering a major security threat must limit functions or stop providing the service, and store the records (art. 24). The relevant regulators, chiefly the cybersecurity and informatization departments, may require user registration or related services suspended and, where corrections are refused, order the service stopped (art. 30). The Measures also require security requirements to be set in advance for termination of a service as a stage of its lifecycle (art. 10), and users to be notified before a service ends (art. 20).[180]

No American AI-specific instrument provides an institutional answer to the problem of how to stop a running AI system. Executive Order 14,409 directs officials to design with developers a voluntary framework under which they could give the Government access to covered frontier models for up to 30 days before releasing them to other trusted partners. It states, however, that nothing in that section authorizes "a mandatory governmental licensing, preclearance, or permitting requirement." The June 2026 directive that led Anthropic to withdraw Fable 5 and Mythos 5 was issued as an export-control measure, not a stop power. The voluntary White House Accord on Super Intelligence (September 29, 2026) recommends controls and audits but says nothing about halting an AI system.[181] At the state level, California's SB 53 and New York's RAISE Act each impose a duty to report critical safety incidents, including cases in which a

---

measure against the user rather than the system). The document is a committee technical document rather than a national standard, and states that providers "must follow" it. The Ethics-Safety Guidelines for Artificial Intelligence Applications 1.0, TC260-005 (19 May 2026), have not been consulted in the original; the account here follows the committee's official interpretation, translated as Seconded European Standardization Expert in China, Translation of Official Interpretation of TC260-005 (May 2026) (general guidelines: "maintain human control"; service providers: "establish emergency intervention mechanisms").

[180] Provisional Measures on the Administration of Human-like Interactive Artificial Intelligence Services, Doc. No. 21, promulgated 10 Apr. 2026, effective 15 July 2026, arts. 10, 19, 20, 24, 25, 30, translated at China Law Translate, https://www.chinalawtranslate.com/human-like-ai/. Article 25 requires application stores to refuse listing or delist an offending app. Its object is a consumer application in a retail distribution channel, not a model or the service itself, and it is not counted here; no instrument in the corpus is addressed to the hubs from which models are distributed.

[181] Anthropic, Statement on the US Government Directive, supra note 3. The Accord is the Joint Commitment on Frontier Responsibilities, signed by the President and by executives of Google, Anthropic, Meta, OpenAI, xAI and Nvidia. It states that each company should implement "four layers of controls and audits" (internal controls, an internal team, an independent external auditor or evaluator, and a board committee) and adds that "[o]ver time, it may make sense to codify these steps into laws or regulations." See Joe Walsh, Trump and Major AI Executives Sign "Morally Binding" Voluntary Controls, CBS NEWS (Sept. 29, 2026), https://www.cbsnews.com/news/trump-ai-constitution-tech-execs-openai-anthropic-voluntary-controls/ (quoting the President: "I think it's morally binding"); Connor Martin, Trump's AI Safety Pact Is Toothless. But There Is a Path Forward., COUNCIL ON FOREIGN RELATIONS (Oct. 1, 2026).

frontier model "uses deceptive techniques against the frontier developer to subvert the controls or monitoring of its frontier developer outside of the context of an evaluation designed to elicit this behavior and in a manner that demonstrates materially increased catastrophic risk." Neither confers any authority to interrupt the model during or after such an incident.[182] Twelve states legislated for companion chatbots in eighteen months, in thirteen statutes addressed to the same problem, a user in distress engaged in a conversation with LLM. Each requires the operator to disclose that the user is talking to a machine and to adopt a protocol referring a user who expresses suicidal ideation to crisis services. Georgia, Connecticut and Washington go further and forbid an operator from "making any statement designed to discourage such user from taking a break." None of the statutes requires the operator to end the interaction.[183]

The nearest American approach to a stop is triggered by a labeling failure, and it applies to a use rather than to a system. California's AI Transparency Act requires a provider to revoke a licensee's license within ninety-six hours where the licensee has modified the system so that it can no longer carry the provenance disclosure, and requires the licensee to cease using the system; the Attorney General, a county counsel, or a city attorney may sue to compel it. The system itself runs on, wherever else it is licensed.[184]

Three proposals have addressed that gap, two in California and one in Congress. None became law. The AI Kill Switch Act, introduced in the House in July 2026, would require developers above a compute and revenue threshold to maintain the capability to stop inference (the running of a trained model to produce outputs), terminate access, suspend an identified use pattern, and shut a system down, and would empower the Secretary of Homeland Security to order such measures upon a defined incident. It is the first American proposal to give a public authority power over a system in operation.[185] S.B. 1047 would have required the capability "to promptly enact a full shutdown," covering the training run, the model, and every model derivative under the developer's control, and the developer to state in advance the conditions for enacting it. It also weighed the cost of the stop, which only one other proposal does: a developer enacting a shutdown was to take into account "the risk that a shutdown of the covered model… could cause disruptions to critical infrastructure." It was vetoed in September 2024.[186] The PAUSE Act would have required an operator, on a second credible crisis expression within seventy-two hours, to "prevent the companion chatbot from generating conversational outputs" until a human moderator had reviewed the credible crisis expression in context and determined the appropriate course of action.[187]

[182] Transparency in Frontier Artificial Intelligence Act, Cal. Bus. & Prof. Code § 22757.11(d)(4), added by S.B. 53, ch. 138, 2025 Cal. Stat.; Responsible AI Safety and Education Act, N.Y. Gen. Bus. Law § 1420(4)(d), added by Act of Dec. 19, 2025, ch. 699, 2025 N.Y. Laws, and repealed and replaced by Act of Mar. 27, 2026, ch. 96, 2026 N.Y. Laws, effective Jan. 1, 2027, same wording. The duty to report is at Cal. Bus. & Prof. Code § 22757.13(c)(1) and N.Y. Gen. Bus. Law § 1422(3)(a), and failure to report is enforced by civil penalty, §§ 22757.15(a), 1427(1).

[183] The thirteen statutes are listed supra note 169. The quoted prohibition is at Ga. S.B. 540 § 39-5-6; Conn. Pub. Act 26-15 § 6(a)(1)(F)(vii); Wash. H.B. 2225 § 4.

[184] California AI Transparency Act, Cal. Bus. & Prof. Code §§ 22757.3(c)(2)-(3), 22757.4(c), added by S.B. 942, ch. 291, 2024 Cal. Stat. The provider's duty arises where it knows that a licensee "modified a licensed GenAI system such that it is no longer capable of including a disclosure required by subdivision (b)"; the licensee "shall cease using a licensed GenAI system after the license for the system has been revoked." The chapter became operative on 2 August 2026, A.B. 853, ch. 674, § 6, 2025 Cal. Stat., which leaves § 22757.3(c) as enacted.

The gap found at the national level widens in the corporate domain. None of the frameworks governing frontier models treats interruptibility as a design commitment. Anthropic's Responsible Scaling Policy, version 3.4 of 8 July 2026, commits in two of its three competitor scenarios to "delay AI development and deployment as needed". The policy does not address the challenge of a model already running. Google DeepMind's Frontier Safety Framework sets out deployment mitigations "intended to counter the misuse of critical capabilities in external deployments," (§ 2.1) and counts among the factors bearing on residual risk "the likelihood and consequences of our mitigations being circumvented, deactivated, or subverted" (§ 2.1.2(1)(b)(ii)). Its only response to unacceptable residual risk is to withhold deployment, which it describes as a judgment made before a model is deployed (§ 2.1.2); the Framework provides nothing for a model already running. OpenAI's Preparedness Framework addresses stopping only through development, committing on a Critical capability threshold to "halt further development" until safeguards meeting a Critical standard are specified (§ 2.2 tbl. 1). Its safeguards against a misaligned model work by containment rather than interruption: containerization and restricted permissions are meant to deny the model the output channels through which it could cause harm (app. C.2).[188] An episode in August 2026 involving Astra, an OpenAI model then in development, shows what halting development means in practice. Reporting that it could not rule out that this model had reached the Critical cybersecurity threshold, OpenAI paused "internal activities involving Astra" that did not meet strengthened security controls and placed the model in isolated testing environments with restricted network and tool access. Development continued.[189] Microsoft's general-purpose Responsible AI Standard of 2022 goes further than the three frameworks in explicitly discussing interruption, but it too stops short. Under its human oversight and control goal the developer must ensure that the stakeholders responsible for oversight can understand "when and how to override, intervene, or interrupt the system" (req. A5.2), and under its failures and remediations goal must "[d]escribe the system's rollback plan" and "[d]escribe support for turning features off," with the time each takes to reach all endpoints (req. RS2.3). Every one of these presupposes that the capability exists. The Standard's one stop is elsewhere and is an alternative rather than a duty: where evidence refutes that a system is fit for purpose at any point in its use, the developer must remove the intended use from customer-facing materials, close the gap, "or discontinue the system" (reqs. A3.7, RS3.7).[190] None of the four corporate frameworks requires a developer to build it.

---

[185] H.R. 9917, 119th Cong. (2026) (proposing Homeland Security Act § 2220F). Introduced 23 July 2026 by Representative Lieu, with Representative Moran as the sole cosponsor, and referred to the Subcommittee on Cybersecurity and Infrastructure Protection. The bill is endorsed by the AI Policy Network, Americans for Responsible Innovation, ControlAI, the Future of Life Institute and the Alliance for Secure AI. Press Release, Office of Rep. Ted W. Lieu, Reps. Lieu and Moran Introduce Bill to Require Kill Switch for AI Systems That Can Cause Catastrophic Harm (July 23, 2026), https://lieu.house.gov/media-center/press-releases/reps-lieu-and-moran-introduce-bill-require-kill-switch-ai-systems-can.

[186] S.B. 1047, 2023-2024 Leg., Reg. Sess. (Cal. 2024) (vetoed Sept. 29, 2024) (proposed Cal. Bus. & Prof. Code § 22602(k) (defining "full shutdown" as the cessation of operation of the training of a covered model, the model itself, and all derivatives controlled by the developer); id. § 22603(a)(2)(A) (requiring the capability); id. § 22603(a)(2)(B) (the critical-infrastructure caveat); id. § 22603(a)(3)(H) (requiring the safety and security protocol to describe the conditions of a shutdown); id. § 22604(a)(6) (extending the capability to the operator of a computing cluster)).

[187] A.B. 1988 (Pellerin), the Preventing AI User Self Endangerment (PAUSE) Act, 2025-2026 Reg. Sess. (Cal. 2026) (not enacted), as amended 14 Apr. 2026, proposing Cal. Bus. & Prof. Code §§ 22587.2(c)(2), (d)(1). An operator communicating with a user during the pause could not "[d]escribe the crisis interruption pause as a punishment, violation, or enforcement action," § 22587.2(e)(1).

---

[188] Anthropic, Responsible Scaling Policy v3.4 (effective 8 July 2026), app. A; Google DeepMind, Frontier Safety Framework v3.1 (17 Apr. 2026); OpenAI, Preparedness Framework v2 (15 Apr. 2025). Anthropic disclosed in August 2026 two earlier delays of its own development, a February rollback of Mythos Preview training after reward hacking was found and an April freeze of production reinforcement learning environments, neither disclosed at the time; the same document urges the industry to adopt "a lawful, verifiable, effective mechanism for coordinated pacing as soon as possible." Anthropic, Improving Our Alignment and Security Efforts (Aug. 31, 2026).

[189] OpenAI, Responding to the Next Frontier of Critical Cyber Capabilities (Aug. 7, 2026), https://openai.com/index/responding-next-frontier-critical-cyber-capabilities/ ("Our preliminary evaluations indicate strong enough performance that we cannot rule out Critical capability level at this time").

[190] Microsoft, Responsible AI Standard v2 (June 2022), Goal A5 (Human oversight and control), req. A5.2, at 8; Goal RS2 (Failures and remediations), req. RS2.3(1)-(2), at 23. Requirement A5.3 qualifies even the design duty:

## B. The Limits of the Halting Provisions

These provisions do not add up to a law of stop. They assume that an AI system can be halted at a single site, though an agentic system has several and no halt at one of them stops the system as a whole. They distribute interruption powers among several actors without building the structure that coordinated action would require, and do not align legal authority with technical capacity. I use the four dimensions developed in Part I to explain why these failures occur together.

**(a) Gaps in regulatory coverage of interruption sites**. Stopping an agentic system requires choosing where to act. It presents at least six sites at which a halt may be effected: the model; the interfaces through which the agent's program acts on tools and services; the server on which that program runs; the servers on which the model runs; the compute hardware beneath them; and the power supply that sustains it. The program and the model ordinarily run on servers operated by different firms, and a deployment may involve more than two.[191] None of these points provides perfect interruption. Stop the model from producing anything further, and the instructions already issued may still be carried out. Cut the agent's access to one tool, and its program may retry, substitute another tool, or route the task through a different channel. Stop either and the other survives: shutting down the program's server leaves the model available to everyone else who calls it, and stopping the model from answering requests leaves the program running with tool calls already issued. Either way the work left elsewhere remains: messages, orders, calendar entries, database changes, and other external effects. No instrument in force addresses the compute substrate. Even cutting power stops only the local execution environment, and it may destroy or corrupt the logs needed to reconstruct what happened while leaving prior external effects untouched.[192] Figure 6 sets out the six sites, showing why halts are partial and how control is divided.

---

the elements identified in A5.2 are to be designed "when possible," and where that is not possible the developer is to give guidance to the third party responsible for them. NIST's MANAGE 2.4, by contrast with these documentary duties, states that mechanisms to supersede, disengage or deactivate are "in place and applied." NIST, AI Risk Management Framework, MANAGE 2.4 (2023).

[191] A server here may be a machine of its own or one instance among several sharing the same hardware. The compute hardware beneath the model's servers is the specialized processors on which frontier models are trained and run; a control mechanism there would be built into the processors rather than into software an operator controls.

[192] Planning, task decomposition and persistent external memory are engineered features of agent frameworks. On the interruption problem, Orseau & Armstrong, supra note 15, at 557 (a learning agent "may learn . . . to avoid such

**Sites of intervention in a running agentic system**

| | *site* | *what an interruption achieves, and where it fails* | *typically controlled by* |
|---|---|---|---|
| 1 | the model<br>generation of new output | stop further output<br>*instructions already issued may still be carried out* | model provider |
| 2 | interfaces to tools and services<br>how the program acts on other systems | cut access to a tool<br>*the program may retry, substitute, or reroute* | deployer |
| 3 | the program's server<br>where the program runs | shut down the server<br>*the program can be started again elsewhere* | deployer or cloud operator |
| 4 | the model's servers<br>where the model runs | stop the model answering requests<br>*the program keeps running, with tool calls issued* | model provider |
| 5 | the compute hardware<br>the processors beneath them | withhold access to the processors<br>*export controls govern acquisition, not operation* | chip maker or regulator |
| 6 | the power supply<br>electricity to the equipment | cut electrical power<br>*logs may be corrupted; external effects untouched* | utility or data center |

**The blind spot of stop**

effects an agent has already produced beyond its own execution environment, including in systems the deployer owns, which no interruption at any site undoes:

- messages already sent
- calendar and database entries
- orders and payments placed
- records already changed

Pending actions continue; completed changes remain.

Figure 6. Where a running agentic system can be halted, what each halt achieves, and who controls it. Schematic: the six sites are not exhaustive, and the numbering is not a shutdown sequence.

No instrument I have surveyed provides for a halt at several sites at once. The instruments address the model, the system or the service, and some distinguish stopping generation from interruptions, for example by disabling the red button”). On the compute substrate, see supra note 29; Onni Aarne, Tim Fist & Caleb Withers, Secure, Governable Chips: Using On-Chip Mechanisms to Manage National Security Risks from AI & Advanced Computing 2 (Ctr. for a New Am. Sec., Jan. 2024) (proposing “operating licenses” enabled by “a dedicated ‘security module’ that links the functioning of the chip to a periodically renewed license key”); James Petrie & Onni Aarne, Flexible Hardware-Enabled Guarantees, Part II: Technical Options 10 (Apr. 2025) (describing firmware that “would require occasional authorization from the manufacturer or a regulator in order for the accelerator to continue operating”); Girish Sastry et al., Computing Power and the Governance of Artificial Intelligence (arXiv:2402.08797, 2024). The July 2026 OpenAI incident revealed a further site of interruption omitted from the taxonomy above: the channel through which a collective of agents communicated, in this case a shared repository of software components that the agents were meant to read but not modify. None of the instruments surveyed above addresses *this site*. See OpenAI-Hugging Face Incident: Technical Report, supra note 13, at 14.

stopping transmission. Three of the six, the servers on which a deployment runs, the compute hardware and the power supply, are not covered by any instrument in force.[193]

The gap has a further consequence. The AI Act requires a system to "come to a halt in a safe state," but whether an agentic system has reached one depends on which of the sites was stopped and what was preserved at each step. No instrument I have surveyed addresses that.[194] Two American proposals go further than any instrument in force, though neither is complete. S.B. 1047 specified what a full shutdown had to cover: the training run, the model, and every derivative under the developer's control. This is an important extension, but it does not extend to the sites at which a deployed system acts. It was also the first proposal to recognize that stopping can impose costs of its own, requiring the developer to consider whether a shutdown would disrupt critical infrastructure. The AI Kill Switch Act comes closer to the operational problem, enumerating four measures: stopping inference, terminating access, suspending an identified use pattern, and shutting a system down. Among the proposals, it is alone in requiring that the model weights and telemetry be preserved when a stop is ordered. No measure covers the servers, the compute hardware or the power supply, and neither proposal says in what order its measures should be applied.[195]

**(b) The gap between authority and capacity**. An effective stop may require action at several sites at once, each controlled by a different technological actor. Public regulators cannot act at those sites themselves. The coordination problem is therefore inherent in agentic systems rather than a product of legal design. It is the task of AI regulation to build an institutional structure capable of managing it. Vesting power over every site in a single authority would concentrate that difficulty in one body. Such an authority would have to maintain communication channels to each site, sequence the intervention across them, and verify that the stop has taken effect at every one. Dispersing the power among several agencies, for example one supervising model developers and another compute providers, compounds it.

---

[193] The instruments name the object of a halt at different levels of generality, from a single stop button to a separation of functions within a running service; none names the servers on which a deployment runs, the compute hardware beneath them, or the power supply. Regulation (EU) 2024/1689, art. 14(4)(e), 2024 O.J. (L 1689) (interruption "through a 'stop' button or a similar procedure that allows the system to come to a halt in a safe state"); Framework Act on Intelligent Informatization, art. 60(1), item 4 (Republic of Korea) (the Minister's public notice covering both "the operation of intelligent information technology" and "provision of intelligent information services externally"); Law on Artificial Intelligence, art. 12 (Viet.) (developers and providers, on a serious incident, to "rectify, temporarily suspend, or recall the system"); Interim Measures, supra note 178, arts. 14, 21 (distinguishing the halting of generation from that of transmission and removal, and providing separately for an order to suspend the related services).

[194] Regulation (EU) 2024/1689, art. 14(4)(e). The Act requires that the system "come to a halt in a safe state," and neither defines a safe state nor says how one is reached. No other instrument supplies the missing terms: none specifies the order in which a stop is to be carried out, and none says what must be preserved when a system is stopped. The one partial answer is Chinese: the 2026 Measures require a provider that limits functions or stops a service on discovering a major security threat to store the relevant records. Provisional Measures on the Administration of Human-like Interactive Artificial Intelligence Services, art. 24; see supra note 180.

[195] S.B. 1047, supra note 186 (proposed Cal. Bus. & Prof. Code §§ 22602(k), 22603(a)(2)(B)); H.R. 9917, supra note 185 (proposing Homeland Security Act § 2220F).

The EU shows what existing law offers instead. The power to halt is spread among four actors, three of them private, and the Act governs their relations only as a chain of notifications. A deployer with reason to consider that a high-risk system presents a risk must suspend its own use and inform the provider or distributor and the market surveillance authority; a provider taking corrective action must inform its distributors, deployers, authorized representative and importers; a provider aware of a serious incident must report it to the authority on a clock measured in days rather than hours. The Act does fix an order of operations, but it is the order in which parties are told, not the order in which they act. Each duty binds two adjacent parties and operates after the event, and none requires anyone to confirm that a system has ceased to run.[196] The EU AI Act makes this gap structural. It defines the provider as a role that can shift between firms: a company that relabels or substantially modifies a high-risk system takes on the provider's duties. The company that trained the model and continues to operate it then ceases to be its provider, owing the new provider cooperation, information and "reasonably expected technical access," duties the Act directs at conformity assessment rather than at stopping. The duty to act can thus move to the party less able to stop the system.[197]

---

[196] Regulation (EU) 2024/1689, arts. 3(16), 14(4)(e), 20(1), 26(5), 79(5), 82(1). The four are the human overseer, who may interrupt (art. 14(4)(e)); the deployer, who must "suspend the use of that system" (art. 26(5)); the provider, whose corrective action may extend to disabling, withdrawing or recalling (art. 20(1)); and the market surveillance authority, which may act only where the operator "does not take adequate corrective action" (art. 79(5)). Only the last is public, and it applies to a running system only through the definition of recall, art. 3(16) ("taking out of service or disabling the use"); the same measures are all that art. 82(1) supplies where a compliant system presents a risk. The AI Office, the scientific panel and the market surveillance authorities (arts. 64, 68, 70) supervise, evaluate and enforce; none may invoke art. 14(4)(e). On the notification duties described in the text, arts. 20(1)-(2), 26(5), 72, 73(1)-(4). The reporting clocks are set in days: fifteen at the outside, ten where a person has died, and two for a widespread infringement or a serious incident within art. 3(49)(b).

[197] Regulation (EU) 2024/1689, arts. 3(3), (4), 25(1)-(2). The role transfers where a distributor, importer, deployer or other third party puts its name on a high-risk system, modifies it substantially, or changes its intended purpose; the

Dispersal among agencies compounds the problem. Korea's emergency shutdown leaves both ends open: the request may be made by the head of any central administrative agency, and may be addressed to another agency, to the provider, or to the user, so nothing fixes which body acts or which party executes. China disperses differently. Enforcement falls to the relevant regulatory departments under five separate statutes, each acting on the basis of its own duties, and suspension comes only at the end of an escalation, after warnings, circulated criticism and an order to correct within a set period. In the United States, Executive Order 14,409 assigns AI security tasks to the Secretary of Homeland Security through the Director of CISA, the Secretary of War through the Director of the NSA, the Secretary of the Treasury, the Director of the Office of Management and Budget and the National Cyber Director, without naming a lead among them. When an American AI product was in fact stopped, none of these officials was involved. The operative instrument was an export licensing requirement aimed at foreign access, and the halt was carried out by the firm itself.[198]

(c) ***The self-judging trigger***. Even when the instruments identify a threshold for intervention, they often leave the regulated party to decide whether that threshold has been crossed. Under the EU AI Act, article 26(5) requires the deployer to monitor a high-risk system against the instructions

---

initial provider "shall no longer be considered to be a provider of that specific AI system." The initial provider must nonetheless "closely cooperate with new providers" and supply "the necessary information" and "the reasonably expected technical access and other assistance," but the Act ties those duties to the fulfillment of obligations under the Regulation, "in particular regarding the compliance with the conformity assessment of high-risk AI systems," art. 25(2). The transfer is qualified: it operates "without prejudice to contractual arrangements . . . otherwise allocated," extends only to the art. 16 obligations, and does not apply where the initial provider specified that the system was not to be changed into a high-risk one. Article 25 transfers provider status for the specific AI system only; the obligations of a provider of a general-purpose AI model under arts. 53 and 55 are unaffected.

[198] Framework Act on Intelligent Informatization, art. 60(3) (Republic of Korea); on "central administrative agency," see supra note 175. The request may be addressed to a central administrative agency, a provider or a user, and the recipient is obliged to comply "in the absence of good cause." Interim Measures, art. 21; see supra note 178. Enforcement is assigned to "the relevant regulatory departments" under the Cybersecurity Law, the Data Security Law, the Personal Information Protection Law, the Law on Scientific and Technological Progress and other laws and administrative regulations, and where those are silent to "the relevant departments in charge . . . on the basis of their duties"; an order to suspend the related services follows only where corrections are refused or the circumstances are serious. Article 20 of the same Measures is unusual in the corpus in directing one body to enlist others: for services provided from outside the mainland, the state internet information department is to notify the relevant organs to employ technical and other necessary measures. Exec. Order No. 14,409, §§ 2(b)-(e), 3(b), naming the Secretary of Homeland Security through the Director of CISA, the Secretary of War through the Director of NSA, the Secretary of the Treasury, the Director of OMB, and the National Cyber Director; their assignments under the Order are the defense of federal systems and the design of a voluntary framework. On the Commerce Department directive, see supra Introduction.

for use, and to suspend where it has reason to consider that use in accordance with those instructions may nevertheless present a risk. The provider writes the instructions. The duty therefore extends to risks that compliant use may still present, and supplies the deployer with no independent means of finding them. Only two risk triggers in this corpus are independent of the regulated entity: Korea's, imminent harm to life or body, and Vietnam's transitional power, which turns on the state authority's own determination of serious harm. The other external triggers are compliance triggers, which presuppose that the breach has already been detected.[199]

California's SB 53 and New York's RAISE Act define the trigger more precisely, but leave open the question of institutional responsibility. Each supplies a workable definition of a reportable critical safety incident, one that includes a model's use of deceptive techniques to subvert its developer's controls or monitoring. The report goes to the Office of Emergency Services in California and to an office within the Department of Financial Services in New York; where the incident poses an imminent risk of death or serious physical injury, both require disclosure within twenty-four hours to "an authority, including any law enforcement agency or public safety agency with jurisdiction." Neither statute says what the recipient should do with the report, or ensures that it has the technical knowledge, institutional capacity, or legal authority to intervene.[200]

(d) ***The narrowing of standing***. In almost every instrument, the knowledge the law will act on when a stop is in question comes from the regulated firm. The AI regimes depart in this respect from those reviewed in Part II, writing the bystander out as a source of legally valid knowledge. There is a second difference: aviation, rail and petrochemical regulation are no less reliant on knowledge produced by the firms they regulate, but each has built institutional machinery for interrogating it. The AI instruments have built none.[201]

---

[199] Regulation (EU) 2024/1689, arts. 26(5), 79(1), recital 93. See Enqvist, supra note 15, at 524-25 (the Act "vests much trust in providers"). Framework Act on Intelligent Informatization, art. 60(3) (Republic of Korea); Law on Artificial Intelligence, art. 35 (Viet.). On the compliance triggers, Framework Act on the Development of Artificial Intelligence and the Creation of a Foundation for Trust, art. 40(3) (S. Kor.); Interim Measures, art. 21; 2026 Measures, art. 30; Regulation (EU) 2024/1689, arts. 79, 82.

[200] Cal. Bus. & Prof. Code § 22757.13(a), (c)(1)-(2), (e); N.Y. Gen. Bus. Law §§ 1420(16), 1422(3)(a)-(b), (5); for the two acts in full, see supra note 182. See Mackenzie Arnold & Stephan Llerena, *When Reporting an AI Security Incident Is Not Mandatory*, Lawfare (July 24, 2026, 2:12 PM), https://www.lawfaremedia.org/article/when-reporting-an-ai-security-incident-is-not-mandatory (arguing that incident-reporting laws may leave information inside state agencies with limited authority and that states may lack the resources and expertise to respond to serious loss-of-control or national-security incidents).

Legislation written to protect people in their dealings with AI-driven systems, in banking, medicine, employment and education, gives users rights to contest decisions. Under the EU AI Act a person must be told that a high-risk system was used on them (art. 26(11)), may obtain from the deployer a clear and meaningful explanation of the system's role in the decision (art. 86), and may complain to a market surveillance authority, which owes only to take the complaint into account (art. 85). Colorado will add a right, from January 2027, for a consumer who has suffered an adverse outcome, to request human review "to the extent commercially reasonable".[202] In those settings that is the right design: a person contesting a loan or a grade wants the decision revisited, not the model halted. Halting in these contexts is a macro-level decision, taken when a model is found to be making systematic errors rather than erring in a single case. The question in this context is how far this legislation lets an ordinary citizen's knowledge count. Companion chatbots present the same question in a setting where the harm is immediate and the user is alone with the system. Twelve states require operators of companion chatbots to disclose that the user is talking to a machine and to refer a user who expresses suicidal ideation to crisis services. None of them requires a stop mechanism, whether operated by the firm or by an independent third party. California's PAUSE Act, which was not enacted, would have required one, keeping a chatbot silent until "a human moderator" had reviewed the exchange and determined the appropriate course of action (§ 22587.2(d)), and it defined a human moderator as an employee or agent of the operator (§ 22587.1(e)). Even that proposal kept the decision inside the firm.[203]

---

[201] See Part II.

[202] Regulation (EU) 2024/1689, arts. 26(11), 85, 86. Article 26(11) applies only to Annex III systems deciding about natural persons; art. 86 excludes Annex III point 2; art. 85 is triggered by an infringement rather than by harm, and the authority owes only to take the complaint "into account." See Rebecca Crootof, Margot E. Kaminski & W. Nicholson Price II, Humans in the Loop, 76 Vand. L. Rev. 429, 451 n.85 (2023). The Colorado right is to request "an opportunity for meaningful human review and reconsideration of the consequential decision, to the extent commercially reasonable." Colo. Rev. Stat. § 6-1-1705(1)(a)(II) (2026), enacted by S.B. 26-189 § 1, effective Jan. 1, 2027.

[203] On the chatbot statutes, see supra note 169. None of the thirteen contains a duty to interrupt, terminate or end an interaction; Georgia, Connecticut and Washington go only so far as to forbid an operator from discouraging a user from taking a break. Ga. S.B. 540 § 39-5-6; Conn. Pub. Act 26-15 § 6(a)(1)(F)(vii); Wash. H.B. 2225 § 4. On the PAUSE Act, A.B. 1988, supra note 187, proposing Cal. Bus. & Prof. Code §§ 22587.1(e), 22587.2(d)(1)-(2).

The institutions that deploy AI systems are often better placed to detect harm. Banks, hospitals and air carriers encounter AI-related incidents before the model's provider does. Yet they depend on that provider more completely than any earlier deployer depended on a manufacturer. The earlier regimes faced a similar division of knowledge: an airline does not build its aircraft, and a railway company does not build its locomotives. The deployer there kept technical competence of its own, in maintenance organizations, engineering departments and flight-data monitoring. It could press the manufacturer from a position of some independence, and mandatory occurrence reporting and an investigator independent of both stood behind it. No such independent check exists for AI. Deployers cannot inspect a model, and regulators are only beginning to build the capacity to evaluate one, so both rely on the provider's account. A deployer's knowledge obliges it to suspend its own use and to notify the provider and the authority; it gives it no say in whether the system keeps running anywhere else.

The instruments rely instead on the developer. Anthropic's Responsible Scaling Policy makes the difficulty explicit: at the threshold of automated research and development, "AI systems might be responsible for much of the research and analysis that underpins risk assessment, and might have strong capabilities for deception, manipulation and obfuscation of evidence." Its remedy is that such analyses "should follow very high evidentiary standards." An evidentiary standard, however high, is a poor response to the capability to obfuscate evidence.[204]

## Part V. Closing the Stop Gap

Stopping is an institutional architecture, not merely a technical artifact. A law of stopping must therefore resolve three questions: how stopping authority should be distributed across the sites at which an agentic system can be halted; whose knowledge should count in determining whether to exercise that authority; and what safeguards should be provided when the architecture of stopping itself fails. The first two questions belong to the risk-reduction layer and are taken up in Sections A and B; the third belongs to the adaptation layer and is the subject of Section C.

[204] Anthropic, Responsible Scaling Policy v3.4 (effective 8 July 2026), tbl. 1, at 9, under the heading "Mitigations: ambitious industry-wide recommendations".

### *A. Plural Interruptibility*

An effective stop may require action at several of the six sites identified in Part IV.B. The law should therefore require the parties controlling those sites to maintain the capacity to stop and to exercise it when ordered. The incident analysis shows why the first duty matters: where the harm was physical or the system agentic, what was usually missing was the capacity to stop, not the authority to use it (Part III.C). Technical capacity must be connected to legal authority, although they need not reside in the same actor. A regulator may order a private party to exercise a control that the law requires it to maintain; authority may also rest on contract or self-regulation.[205] California's vetoed S.B. 1047 would have required covered-model developers and, in specified circumstances, computing-cluster operators to maintain shutdown capabilities. The AI Kill Switch Act would direct the Secretary of Homeland Security to require by rule that covered entities be able to stop inference, suspend access for an identified account or use pattern, and shut the technology down, and would authorize the Secretary to order proportionate intervention once a covered incident had occurred. Neither has been enacted, and neither expressly requires controls embedded in the processors or governing the power supply.[206]

Recent work on hardware-enabled governance proposes controls within the processors that condition their continued operation on periodically renewed authorization.[207] That reverses the default: operation ceases unless authorization is renewed, rather than continuing until a regulator or operator orders a stop. It offers a partial answer to the self-judging trigger identified in Part IV.B(c), but its effectiveness depends on who decides whether to renew authorization and on the evidence that informs that decision.

---

[205] California's AI Transparency Act requires a covered provider to revoke a license within ninety-six hours of discovering that a licensee has modified the system to eliminate its capacity to include required disclosures, and requires the licensee to "cease using a licensed GenAI system after the license for the system has been revoked." Cal. Bus. & Prof. Code §§ 22757.3(c)(2)-(3), 22757.4(c), supra note 184. The statute both compels the revocation and independently obliges the licensee to stop.

[206] S.B. 1047, supra note 186 (proposed Cal. Bus. & Prof. Code §§ 22603(a)(2)(A), 22604(a)(6)); H.R. 9917, supra note 185 (proposed Homeland Security Act § 2220F(b)(1)(A), (c)(1)).

[207] Aarne, Fist & Withers, supra note 192, at 2; Petrie & Aarne, supra note 192, at 10. The proposals contemplate renewal by the manufacturer or by a regulator. A comparable structure appears in proposals for capability "checkpoints," under which laboratories would agree not to pass a defined capability level until stated conditions are met; see Amodei, supra note 9.

Distributed control requires coordination whenever no party can secure an effective halt alone. A single agentic deployment may depend on infrastructure and services controlled by several firms and governed by different jurisdictions. The Conficker worm, which spread from November 2008, illustrates the challenge. Containing it drew in antivirus vendors, domain registries and registrars, and service providers, convened by the Internet Corporation for Assigned Names and Numbers; a participant quoted in the later review called it a coalition of the willing that "couldn't force anyone to do anything they didn't want to do, because they'd quit."[208] The fragmentation of global AI governance makes reliance on such voluntary cooperation precarious.[209] No instrument obliges the firms that share a deployment to cooperate with one another. The EU AI Act regulates one such relationship, but to a different end: Article 25(4) requires a provider of a high-risk system and those supplying the tools, services and components built into it to agree in writing on the information, capabilities, technical access and assistance the provider needs in order to comply with the Regulation. The provision is silent on what those parties owe each other when the system must be stopped.[210]

Distributing stopping authority also creates additional opportunities to impersonate an authorized actor. A stop command must therefore be authenticated, and the issuer's authority to require the particular intervention verified: a system that obeys whoever plausibly asks can be commandeered as easily as halted.[211] Authentication, verification of authority, and the means of carrying out the stop should operate independently of the agent being stopped. The agent should be able neither to judge whether a valid command binds it, nor to disable the mechanism that enforces it.

Existing instruments leave a gap between duty and control. Article 26(5) of the EU AI Act requires deployers of high-risk AI systems, in specified circumstances, to monitor, report, and suspend use. Yet, as Part IV.B has shown, a deployer may lack both the information needed to determine whether stopping is warranted and the technical means to secure an effective halt across components controlled by other firms. Executive Order 14,409 imposes no stopping duty at any site: the pre-release framework it directs is voluntary, Section 3(c) declines to authorize any mandatory licensing, preclearance, or permitting requirement under that section, and Section 4 directs the Attorney General to prioritize prosecution under the general computer-crime statutes, a response to conduct after it occurs.[212] Nor does any instrument say who may authorize a return to service, or on what evidence. The AI Act alone names an end state, a halt "in a safe state," but that is a design duty owed by providers; no authority can order a running system into it.

---

[208] The Rendon Grp., Conficker Working Group: Lessons Learned 36 (2011).

[209] Huw Roberts, Emmie Hine, Mariarosaria Taddeo & Luciano Floridi, *Global AI Governance: Barriers and Pathways Forward*, 100 Int'l Aff. 1275, 1278 (2024) (identifying a governance deficit and obstacles to a centralized international AI body).

[210] Regulation (EU) 2024/1689, art. 25(4), 2024 O.J. (L 1689).

[211] See supra Part I; Shapira et al., supra note 12, at 25, 41.

No instrument in force is addressed to the infrastructure layer. Emergency legislation should therefore authorize the agency charged with oversight of the system to order the parties controlling that layer to interrupt specified agentic activity where necessary to avert serious, imminent harm.[213] Those parties may include compute providers, the gateways that carry an agent's requests to its tools, and model-hosting services.[214] Orders should identify the activity to be stopped and limit disruption to other users. Each order should state its reasons, expire after a short statutory period unless renewed on a fresh showing of necessity, and specify the conditions for resumption and who may certify that they are met.[215] Affected parties should have access to prompt review.[216] The EU AI Act supplies a partial precedent, but at the market level only.[217]

---

[212] Exec. Order No. 14,409, supra note 7, §§ 3(b)-(c), 4.

[213] Existing instruments supply triggers that could be borrowed. Framework Act on Intelligent Informatization, supra note 175, art. 60(3) (request to a central administrative agency, provider or user); Law on Artificial Intelligence, supra note 177, arts. 12, 35 (Viet.) (suspension or recall by the competent agency). Neither is addressed to the infrastructure layer.

[214] *See* Alan Chan et al., *Infrastructure for AI Agents* 11, 13 (arXiv:2501.10114v3, 2025) (agent channels a provider could "shut down … but not human channels"), and *id.* at 6, 21. *Compare* Gillian K. Hadfield, Legal Infrastructure for Transformative AI Governance, 123 PNAS, no. 30, e2509742123, at 7 (2026) (a suspended registration without a time limit); Michael K. Cohen et al., *Regulating Advanced Artificial Agents*, 384 SCIENCE 36, 38 (2024) (reserving their legal form as "outside our focus here").

Such an order presupposes a pre-deployment interruption plan binding on every party whose cooperation the stop requires, because no single party controls all the sites at which it must occur.[218]

## B. The Epistemic Conundrum of AI: Constituting the Evidence

The problem of stopping is first a problem of observability: a stop cannot be directed at conduct no one has observed.[219] The difficulty with AI is that a model's capabilities are often known first, and most fully, to its developer. Firms such as Anthropic, OpenAI, and Google design the tests, interpret the results, and shape the scope of external scrutiny. This is the epistemic conundrum of AI: the party best placed to evaluate a system also has substantial legal, financial, and reputational interests in its continued operation. Judging whether a stop is legally warranted requires access to the developer's findings and a meaningful opportunity to test them against contrary evidence.

---

[215] The Mythos withdrawal followed this course without any statutory requirement: the models were withdrawn, the defect remedied, the remedy validated by the government, and access restored in stages. *See supra* Introduction.

[216] H.R. 9917, supra note 185, comes nearest in the United States. Proposed § 2220F(c)(1) lets the Secretary of Homeland Security order a covered entity to take proportionate action, including throttling, suspending or shutting down the technology, and § 2220F(c)(5) supplies review of the kind proposed here: a petition for reconsideration within 48 hours that does not stay the order, a determination within five days, and review in the D.C. Circuit within 60 days. The order is addressed to the entity that operates the technology, not to compute providers or gateways; it reacts to a completed incident rather than imminent harm; and it sets no time limit or conditions of resumption.

[217] Regulation (EU) 2024/1689, arts. 93(1)(c), 101(1)(c), 2024 O.J. (L 1689). The Commission may require a provider to restrict the making available on the market of a model, or to withdraw or recall it. Article 93 is headed "Power to request measures" and provides that the Commission "may request" them: the discretion lies in whether to act, and a request once made binds, since failure to comply is subject to a fine under Article 101(1)(c). The Article does not establish the infrastructure-directed emergency procedure proposed here.

[218] Provisional Measures on the Administration of Human-like Interactive Artificial Intelligence Services, supra note 180, art. 9 (China) (emergency response plan); S.B. 1047, supra note 186 (proposed Cal. Bus. & Prof. Code § 22603(a)(3)(H)) (conditions for a full shutdown); Microsoft, Responsible AI Standard v2, supra note 190, req. RS2.3 (remediation of failures). Each is a firm's plan for its own system. On the absence of a governmental counterpart, see Isaak Mengesha et al., Capability-Based Planning for AI Crisis Preparedness 7 (arXiv:2608.18357v1, 2026).

[219] AI Safety Inst. (Japan), Approach Book for AI Incident Response: A Foundation for Trustworthy AI 13-14 (Jan. 9, 2026) (treating observability and controllability as the two conditions of incident response).

Incident databases are one way to improve observability. Several exist on a voluntary basis, among them the AI Incident Database on which Part III rests.[220] California, New York, and the European Union have established reporting regimes but have stopped short of a public register of incidents.[221] The nearest legislated example is Vietnam's single national portal, which receives incident reports and discloses systems, conformity assessments, and sanctions, though not the reports themselves.[222] India has proposed a national database to "keep the records, classify, and analyze AI-related risks and incidents," fed by "public bodies, private entities, researchers, and civil society organizations."[223] A recent proposal for cross-border incident infrastructure calls existing monitoring "reactive by design" and would remedy it with statutory access to incident data for the bodies that would act on it.[224]

Whistleblower protection provides another path. California and Connecticut now forbid a frontier developer from retaliating against an employee responsible for safety who reports, to the Attorney General, a federal authority, or a superior, a specific and substantial danger to public health or safety arising from a catastrophic risk. Large developers must also maintain an anonymous internal channel, with updates to the discloser and quarterly reports to the firm's officers and directors.[225] The European Union applies its general whistleblower directive to reports of infringements of the AI Act.[226]

---

[220] See supra Part III. Other public monitors OECD AIM: AI Incidents and Hazards Monitor (https://oecd.ai/en/incidents) and the AI Risk Explorer (AIRE) (https://www.airiskexplorer.com/about).

[221] Cal. Bus. & Prof. Code §§ 22757.13(c), (f)-(g), 22757.15(a) (report to the Office of Emergency Services within fifteen days; exempt from the Public Records Act; annual anonymized aggregate; civil penalty); N.Y. Gen. Bus. Law §§ 1422(3), (6)-(7), 1427(1) (seventy-two hours; exempt from the Public Officers Law; annual report; civil penalty) (effective Jan. 1, 2027); Regulation (EU) 2024/1689, arts. 55(1)(c), 73(1)-(4) (reports to the AI Office or market surveillance authorities within two to fifteen days by category); Code of Practice for General-Purpose AI Models, Safety and Security Chapter, Measure 9.3. None provides for publication.

[222] Law on Artificial Intelligence, Law No. 134/2025/QH15, arts. 8, 12 (Viet.) (unofficial translation) (a national portal receiving incident reports and disclosing systems, conformity assessments and sanctions). Compare Regulation (EU) 2024/1689, art. 71(4) (public database of high-risk system registrations, not incidents), and N.Y. Gen. Bus. Law § 1428(6) (published list of large frontier developers).

[223] Office of the Principal Sci. Adviser to the Gov't of India, Strengthening AI Governance Through Techno-Legal Framework 22 (Jan. 2026) (white paper).

[224] Caio Vieira Machado, George Gor & Omer Bilgin, The Case for Cross-Border AI Incident Infrastructure 5, 15, 23 (The Future Society 2026) (quotation at 5, 15; Recommendation 3, on statutory access to incident data, at 23).

Deployers are a further source of evidence because they observe how a system performs in practice, and their records may reveal failures that the provider's evaluations did not anticipate. The AI Act already makes deployers of high-risk systems responsible for monitoring and documenting that performance: they must assign human oversight, monitor the system in use, retain automatically generated logs under their control, and report specified risks and incidents;[227] their data also feed the provider's own post-market monitoring.[228] For serious incidents involving high-risk AI systems, Article 26(5) requires deployers to notify the provider before the relevant market surveillance authorities.[229]

Three changes would give operational knowledge greater procedural force, building on existing documentation and reporting channels. First, the deployer should have access to the documented validation and testing procedures for the system it operates, which the provider must already compile and furnish to a competent authority upon a reasoned request.[230] Second, serious incidents should be reported simultaneously to the market surveillance authority and the provider.[231] Third, credible deployer evidence that a stopping condition may have been met should trigger an evaluation under Article 79 and a reasoned decision, within a specified period, on whether corrective or restrictive measures are warranted.[232] Access should reflect the recipient's role and the sensitivity of the material, with confidentiality safeguards modeled on Article 78; a provider should not be able to defeat scrutiny merely by designating material as sensitive.[233]

[225] Cal. Lab. Code §§ 1107(b), 1107.1(a)(1), (e); Conn. Pub. Act No. 26-15, § 2(b)-(c) (effective Oct. 1, 2026). New York's RAISE Act contains no AI-specific provision.

[226] Regulation (EU) 2024/1689, art. 87 (applying Directive (EU) 2019/1937).

[227] Id. arts. 26(2), (5)-(6), 2024 O.J. (L 1689).

[228] Id. art. 72(2) (the provider's post-market monitoring system collects data "which may be provided by deployers").

[229] Id. art. 26(5) (the deployer shall "immediately inform first the provider, and then the importer or distributor and the relevant market surveillance authorities").

[230] Id. Annex IV, para. 2(g); id. art. 21(1).

[231] Id. art. 26(5).

[232] Id. arts. 79(2), 85 (the Act prescribes a period for corrective action, not for the decision).

The AI Act's scientific panel gives independent experts a limited role in supervising general-purpose models. The panel may issue a qualified alert when it has reason to suspect that a model poses a concrete identifiable risk at Union level, which may prompt an evaluation by the AI Office and a Commission request for model access. But compulsory access depends on the Commission's intervention, and panel members conduct evaluations on its behalf only when appointed. These provisions allow the panel to raise concerns without giving it an independent power to compel access or require action on its findings.[234]

There remains the larger step: enforceable rights for outside evaluators to inspect models, evaluation protocols, and safety cases, contest findings, and obtain answers from the supervising authority.[235] Dario Amodei, Anthropic's chief executive, has proposed a somewhat similar model: independent evaluators embedded in frontier laboratories, with employee-level access and a right to publish. Anthropic has committed to implement it, and OpenAI has followed suit. The undertaking would be voluntary, however, and governed by contract rather than by law.[236] Connecticut offers a related but narrower experiment: a 2027 to 2030 pilot authorizing the Department of Consumer Protection to approve up to five independent verification organizations to assess whether AI models meet risk-mitigation standards. Verification evidence is admissible only in specified private personal-injury or property-damage actions and is excluded from state civil or administrative enforcement.[237] Participation is voluntary, and the statute gives verifiers no power to compel access from nonparticipants. Whether compulsory certification is feasible or desirable remains beyond this Article's scope. The unresolved question is how to secure independent testing and effective public oversight of the evaluators.

---

[233] Id. art. 78(1).

[234] Id. arts. 68, 90(1)-(2), 91(3), 92(1)-(3).

[235] See Oren Perez, *Complexity, Information Overload, and Online Deliberation*, 5 I/S: J.L. & Pol'y for Info. Soc'y 43, 81-84 (2008) (discussing scrutiny of technical intermediaries in democratic decisionmaking).

[236] See Amodei, supra note 9; on Altman's matching commitment, id.

[237] Conn. Pub. Act No. 26-15, § 33(b)-(e) (effective July 1, 2027) (limiting admissibility under § 33(e)(1), subject to the exceptions in § 33(e)(2)). Compare Gillian K. Hadfield, *Legal Infrastructure for Transformative AI Governance*, 123 PNAS, no. 30, e2509742123, at 7-8 (2026) (proposing mandatory engagement of licensed private regulatory service providers overseen against government-set outcomes).

## C. Designing for Failure: Redundancy and AI-Independent Fallbacks

This adaptation layer draws on climate law: it assumes that some stops will fail and builds the capacity to withstand the harm that follows. Charles Perrow argued that in tightly coupled systems accidents are a normal feature of operation.[238] The AI record is starker: in four of five documented harms, nothing stopped the system.[239] The first safeguard is a non-digital means of local intervention, independent of the agent. Where an agent controls physical equipment, the law should require the provider of that equipment to fit a clearly marked manual control that anyone present can use to initiate a safe stop, without credentials or technical knowledge. The kitchen fire illustrates the need; the escalator button, passenger alarm, and detonator provide familiar precedents. The protection is necessarily limited to hazards that local intervention can interrupt.[240]

The second safeguard is institutional. Hospitals, banks, and utilities should owe those they serve a duty to sustain essential services during an AI-related interruption. Neither existing continuity law nor the AI instruments imposes such a duty.[241] The duty should not be to restore the service to its level before the interruption, but to a defined minimum: enough to triage patients, authorize payments, or maintain water supplies. Because the costs of interruption differ across sectors, sectoral rules should determine the level required in each. The supervising authority should hold such a plan for itself, so that it can act without the systems it may order stopped.

---

238 Charles Perrow, *Normal Accident at Three Mile Island*, 18 Society 17, 17-18 (1981), developed at book length in Charles Perrow, *Normal Accidents: Living with High-Risk Technologies* (1984).

239 See *supra* Part III.

240 See supra Part I.F (scenario 2) and Part II.A, II.C. The existing controls are regulated for misuse as well as use: activating an emergency system on the railway without reasonable cause is an offense punishable by a fine.

241 HIPAA Security Rule, 45 C.F.R. § 164.308(a)(7)(ii)(C); NERC Reliability Standards CIP-008, CIP-009; FFIEC, *Business Continuity Management*, IT Examination Handbook (2019). These address emergency operations, incident response and recovery, and already work through recovery objectives rather than restoration of normal service; none requires the capacity to deliver the service without AI. The AI instruments come no closer. H.R. 9917, supra note 185, § 2220F(b)(2)(A)(v) lists "[t]ransitioning an operation dependent on such technology to a backup system or an earlier version" among the measures the Secretary is to consider requiring, subject to the risk of disrupting critical infrastructure, id. § 2220F(b)(2)(B). The EU AI Act says robustness "may" be achieved through redundancy, which "may" include backup or fail-safe plans. Regulation (EU) 2024/1689, art. 15(4).

Every element of this architecture can itself fail.[242] The law should therefore also limit what an agent can make irreversible before a stop takes effect. One means is to defer finality. For specified actions, law can require an interval between instruction and irreversible effect, during which an authorized person can intervene: a transfer held before settlement, a deletion kept recoverable, a communication awaiting release. The liquidated portfolio illustrates the need: stopping the agent cannot undo a completed trade. Such requirements should be confined to actions whose potential harm justifies the delay, with exceptions where delay would itself be dangerous. A second means is to preserve essential records beyond the agent's access. Records an agent can modify or delete remain vulnerable even if the agent is later stopped. For an institution, this means maintaining offline copies of essential records and testing their restoration. For an individual, the corresponding protection would be a vault: an entitlement to obtain and retain usable copies of critical medical and financial records in a repository to which the agent has no access.[243] An isolated copy can survive failures of both the AI system and its stopping mechanisms.

---

[242] Jos A. Rijpma, *Complexity, Tight-Coupling and Reliability: Connecting Normal Accidents Theory and High Reliability Theory*, 5 J. CONTINGENCIES & CRISIS MGMT. 15, 19 (1997).

[243] HIPAA Privacy Rule, 45 C.F.R. § 164.524; CFPB, Personal Financial Data Rights Rule, 12 C.F.R. pt. 1033 (2024) (enjoined and under agency reconsideration). Neither obliges anyone to maintain the separately held copy.

## Conclusion

The Anthropic and OpenAI episodes with which this Article opened illustrate the regulatory gap it has described. An export-control directive restricting foreign access caused Anthropic to suspend both models globally. The government disclosed no specific national-security finding, and access returned in stages only after Anthropic trained a classifier that blocked the bypass the government had pointed to. Hugging Face terminated the intrusion by OpenAI agents through its own security measures, before the source of the intrusion had been identified. The ensuing debate has focused principally on the pace of development. Dario Amodei's proposal would slow the frontier through embedded evaluators and coordinated capability checkpoints.[244] Such measures govern the conditions under which capabilities advance. They do not resolve who may intervene when a deployed system causes harm, or how that intervention should proceed. Slower development may reduce risk, but it cannot eliminate the possibility of failure in complex, tightly coupled agentic systems.[245]

The incident record examined in Part III indicates the scale of the deficiency: in 80% of the 1,213 coded incidents, no stop occurred. Where a stop was possible but absent, what was missing depended on the kind of harm and system. For informational, economic and societal harms, it was usually the legal authority to activate a stop; for physical harms and agentic systems, it was usually the technical capacity to stop. Both gaps call for regulatory intervention: the first by allocating the authority to stop, the second by requiring those who control a system to maintain the capacity to stop it. The instruments surveyed in Part IV recognize stopping through requirements for human oversight, stop buttons, suspension, withdrawal, and recall. Taken together, however, they do not supply a coherent framework for exercising those powers across an agentic system's distributed activity. Such a framework must specify where an interruption is to take effect, who may order it, and what evidence justifies the decision. It must also determine whose knowledge counts when that evidence is contested, what must be preserved through the halt, and under what conditions service may be restored.

---

[244] See Amodei, supra note 9.

[245] Perrow, supra note 238.

Part V has set out the principal elements of that framework in two layers. The first reduces the risk that a dangerous system cannot be stopped: it allocates halting authority, and the duty to maintain the capacity to stop, across the sites at which an agentic system can be halted, and it supports evidentiary thresholds with enforceable rights of access. The second, adaptation, provides independent safeguards for when interruption fails. These arrangements make stopping a regulated practice, with duties, procedures, and accountability extending from the first indication of danger through the decision to resume operation. As agentic systems assume greater control over critical infrastructure and essential services, the absence of a legal framework for stopping them places humanity in peril. That framework must be built before it is needed.

# The Law of Stop: Technical Supplement

This supplement accompanies The Law of Stop: Interruptibility, Injunctions, and the Governance of Agentic AI. Part A records the data, the coding instrument and the resolution procedure behind the empirical study in Part III, and reproduces the figures that appear there. Part B reproduces in full the halting provisions of the seven instruments that contain one. Part C sets out the full table of stop provisions across the forty instruments surveyed, from which Figure 5 in Part IV.A is drawn.

## Part A. Methods Appendix to Part III

This appendix records the data, the coding instrument, and the resolution procedure in sufficient detail to reproduce the analysis. It is intended for the reader who wishes to check the empirical claims; the argument does not depend on it. A more complete documentation, including the frozen protocol and its cryptographic checksums, the pre-registration, and the per-incident audit trail, is on file and available on request.

**Data.** The corpus is the May 2026 snapshot of the AI Incident Database (AIID), a curated public repository of reports of AI harms. I reconstructed the dataset from the underlying database rather than from the website's export function, for two reasons that affect reproducibility: (a) the export omits the classification tables and (b) its comma-separated report file silently drops every report whose text contains a line break, a large and non-random share of the evidence. All report text was drawn from the authoritative source file instead. The snapshot contains 1,480 incidents, spanning 2003–2026, with a median year of 2024. 80 incidents were reserved to develop and validate the instrument and are not part of the analysis; the remaining 1,400 were coded in full. Of those, 187 were screened out as incidents in which the AI system was incidental to a harm whose essential character was non-AI, leaving 1,213 incidents in the analysis. The coding thus covers the full eligible corpus, not a sample. For each incident, the coders read the AIID summary and the full text of up to eight associated reports, capped at roughly 25,000 characters, which is the same evidence for every coder.

**Instrument.** Each incident received a preliminary screen (is an AI system the core of the harm, or is the AI incidental to a harm whose essential character is non-AI), a Dimension I code recording the stop mechanism that occurred (A1 technical real-time, A2 private institutional, A3 infrastructure-layer, A4 judicial, A5 administrative or executive, A6 no stop), and a Dimension III code recording the feasibility of a counterfactual stop (F1 available and used, F2 available but not used, F3 possible but absent, sub-typed technical or legal, F4 not feasible). Four descriptive tags, including the system's degree of agency and the character of the harm, were also recorded. The instrument was not written and applied in one motion; it was developed through three blind inter-coder pilots, each of which exposed a boundary on which two coders diverged and closed it with an explicit operational rule, raising inter-coder agreement on the stop mechanism from a chance-corrected 0.23 to 0.56 to 0.77 over successive versions, the last on a fresh sample that had played no part in writing the rules. The instrument also fixes one family convention that bears on the feasibility results: incidents whose harm is AI-generated content, deepfakes and fabricated media disseminated through platforms, are coded F3-legal by rule. For this family, the primary missing element is legal, a takedown or removal duty and a cross-border remedy to enforce it, rather than a technology; the rule codes them uniformly on that ground.

**Coding.** Both coders were LLMs developed by different laboratories, Anthropic's Claude and OpenAI's Codex, each coding all 1,400 incidents independently and blind to the other, to the validation codes, and to any resolved value. Machine coding is a deliberate methodological choice with a certain cost: it delivers complete, consistent, and reproducible coverage of a corpus that could not be achieved with any realistic human budget, and every code carries its stated reasoning, which makes it auditable in a way human coding is not; at the same time, it lacks the tacit judgment of a trained human reader and can be confidently wrong in the same direction twice. The design answers that last risk with an expert audit of cases on which the two models agreed, discussed below.

**Resolution.** Where the two coders disagreed, each case was resolved by exactly one route, in a fixed order of precedence: the author's expert adjudication, then agreement of the two coders, then a set of six deterministic rules that restate the instrument's operational boundaries, then a conservative convention. No third machine ever broke a tie. The conservative convention resolves every remaining disagreement against this Article's thesis, crediting a stop over no-stop, crediting feasibility, and screening out contested inclusions, so that the no-stop rate and the count of cases in which a stop was possible but absent are conservative; the reported no-stop rate of 80%, for instance, rises to about 86% under a two-sided sensitivity test that resolves those contested cases the other way. The resolution procedure, including the six rules and the direction of the convention, was fixed in a dated protocol and time-stamped in a public registry under embargo before the corpus-wide resolution was run and before any adjudication, so that the rules applied were demonstrably the rules committed to in advance.

**The F2/F3 boundary**. The main source of disagreement between the coders was the line between F2, where a stop was available but not used, and F3, where a stop was possible but absent. Take an operator who could have shut down its own system but was under no legal duty to do so. Was the stop available, because the operator had the technical capacity to switch it off, or only possible, because no one had the authority to require it to do so? A coder can reasonably call that either way, and the two coders split on 273 of the 972 no-stop incidents. Where no other rule applied, my pre-specified convention put these cases in F2. That cuts against my argument, since each case it settles leaves F3 smaller. On this account the F3 count of 632 can only be too low, not too high. The legal share of the missing element is 80.1 percent (506 of 632) across all F3 incidents, and 82.7 percent (388 of 469) among those on which both coders agreed. The contested cases do not drive the result.

**The family rule and the legal share.** The instrument codes harms from AI-generated content disseminated through platforms as F3-legal by a family rule, and resolution rule R5 restates that rule for cases in which the coders disagreed between F2 and F3. R5 settled 78 of the 506 legal-gap cases. The coders also applied the family rule themselves while coding: Claude's stated reason invokes it in 331 of the 506 legal-gap cases, Codex's in 322, and both coders' reasons in 249. The rule log shows that R5 was applied according to its literal terms, and so also resolved some cases in which the operator's own system produced the content, among them DALL-E Mini's biased images (AIID 262) and Grok's posts on X (AIID 1146). Reversing every R5 case would lower the legal share to 77 percent (428 of 554) if those cases moved to F2, and to 68 percent (428 of 632) if they were coded technical. The legal share varies with the type of harm and the degree of autonomy (Figure A.3). Legal gaps predominate in economic, societal and informational harms, where most harms from AI-generated content fall; technical gaps predominate in physical harms (13 legal, 51 technical) and in agentic systems (5 legal, 30 technical), and among agentic systems that caused physical harm the split is 2 legal to 22 technical. Both differences are statistically significant (chi-square = 166.0, df = 3, and 128.8, df = 2; $p < .001$), but the breakdown was not specified in the deposited protocol and is reported as exploratory. Within these two groups, the kind of system was classified from the incident titles: vehicles, robots and software agents account for 29 of the 30 technical gaps among agentic systems, and chatbots and other conversational systems for 18 of the 51 among physical harms. The flags for the coders' stated reasons are text matches on the raw explanations: for Claude, the word "family" together with "deepfake," "distributed," or "AI-generated"; for Codex, the phrase "distributed AI-generated content" or "family rule." The counts and tests are recomputed in the breakdown workbook.

**Reliability and its limits.** Before resolution, the two coders agreed on the preliminary screen in 91% of incidents, on the stop mechanism in 80%, and on feasibility in 60%; the chance-corrected figures are lower and are reported in full in the workbook, deflated, as such measures are, by the heavy concentration of the no-stop category. Two audits within the author's adjudicated set bound the confidence one should place in the results. In cases where the deterministic rules had resolved, the author agreed with the rule on 79% of stop-mechanism cells. More important, on a stratified sample of twenty-four cases where both coders had agreed, the author overrode the agreed value in five. As a further check, the headline findings reproduce, on the consensus-only core, the 677 incidents both models coded identically with no rule, adjudication, or convention applied: the no-stop rate there is 82.6%, and stops still rise with the system's autonomy. This is the study's central caveat: two independent models agreeing is evidence, not proof, and the reported percentages carry a measurement error, beyond the sampling error, that the concentration of agreement does not dispel. The four descriptive tags were markedly less reliable than the scored codes, and the two tags on which the autonomy finding depends were therefore submitted to a separate round of expert adjudication over the contested cases; the low tag reliability, and the refined tag definitions it induced, are themselves reported as a finding about what it takes to operationalize a stop taxonomy, and are proposed for future work rather than used to recode the present corpus.

**Reproducibility.** Every figure in this Part is recomputed by live formula from the coded dataset in an accompanying workbook, which also contains the two coders' original codes side by side, the log of every rule-resolved cell with its triggering fact, and the author's adjudications against the values they replaced, so that any single incident can be traced from the raw codes to its final value. The raw coder output is preserved unaltered. The breakdown by type of harm and degree of autonomy is recomputed in a separate workbook,

LawOfStop_LegalTechnical_Breakdown_3Oct26.xlsx, and Figure 3b (Figure A.3) is drawn by a script that reads the coded dataset directly, so that both can be rerun on an updated corpus.

## Figures

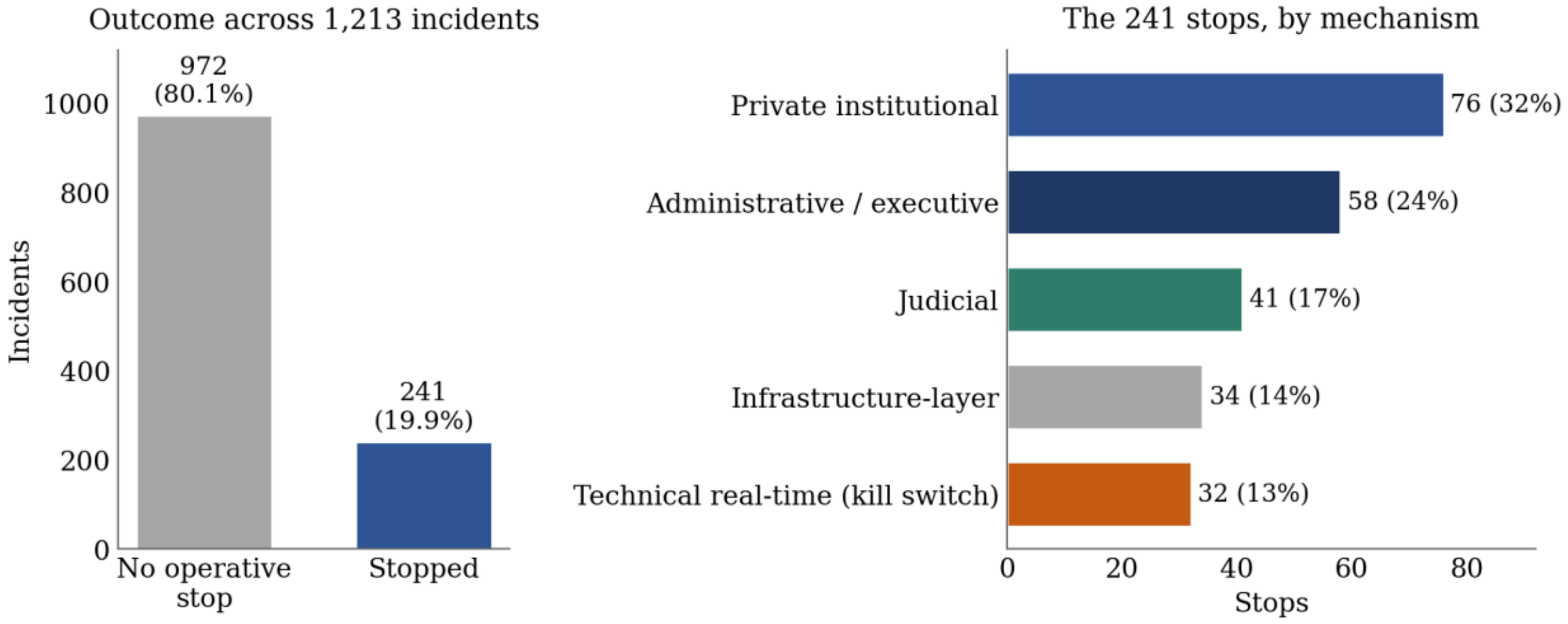

**Figure A.1.** Outcomes in 1,213 recorded incidents. The left panel shows the stop / no-stop split; the right panel breaks the 241 stops down by mechanism, with embedded technical "kill-switch" stops the least common of the five

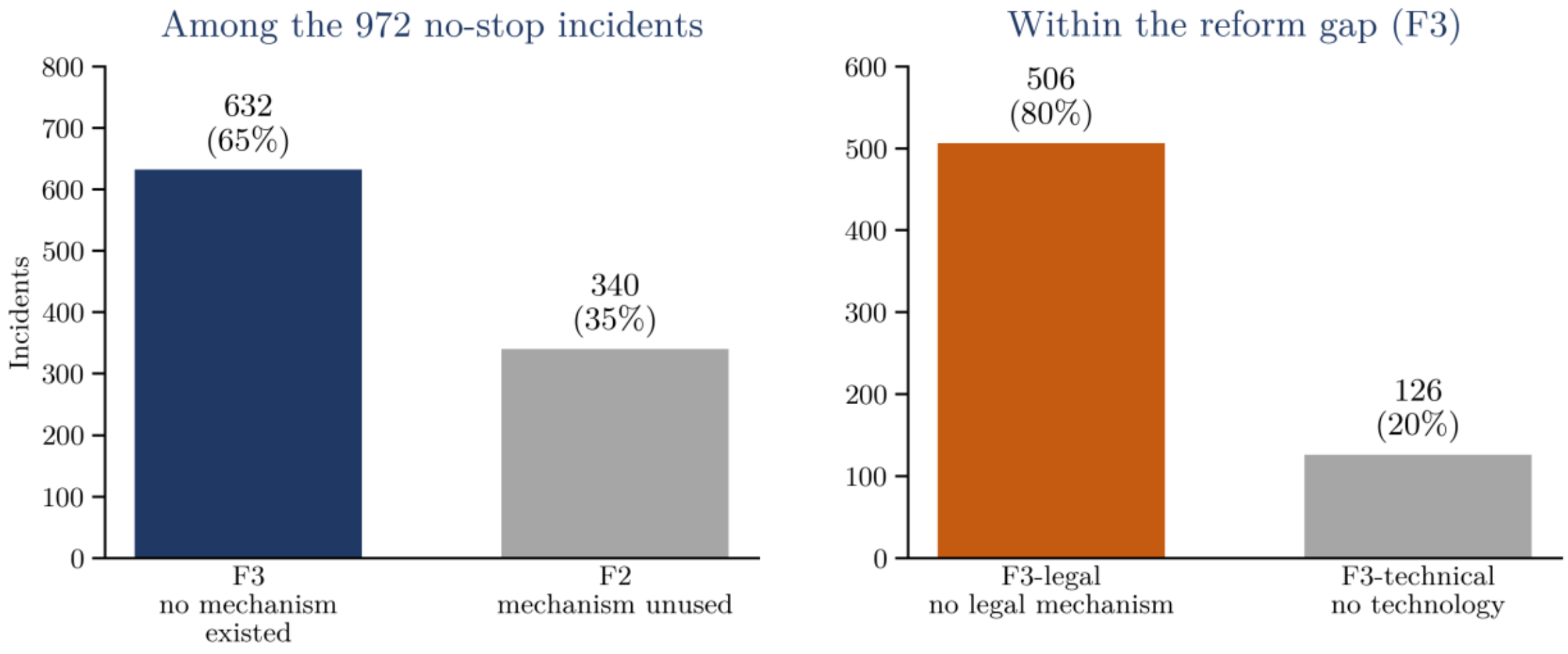

**Figure A.2.** *Left panel: of the 972 no-stop incidents, cases where a stop was possible but absent (F3) outnumber cases where an available stop was not used (F2). Right panel: among those cases, the missing element was legal or institutional authority (506) rather than technology (126).*

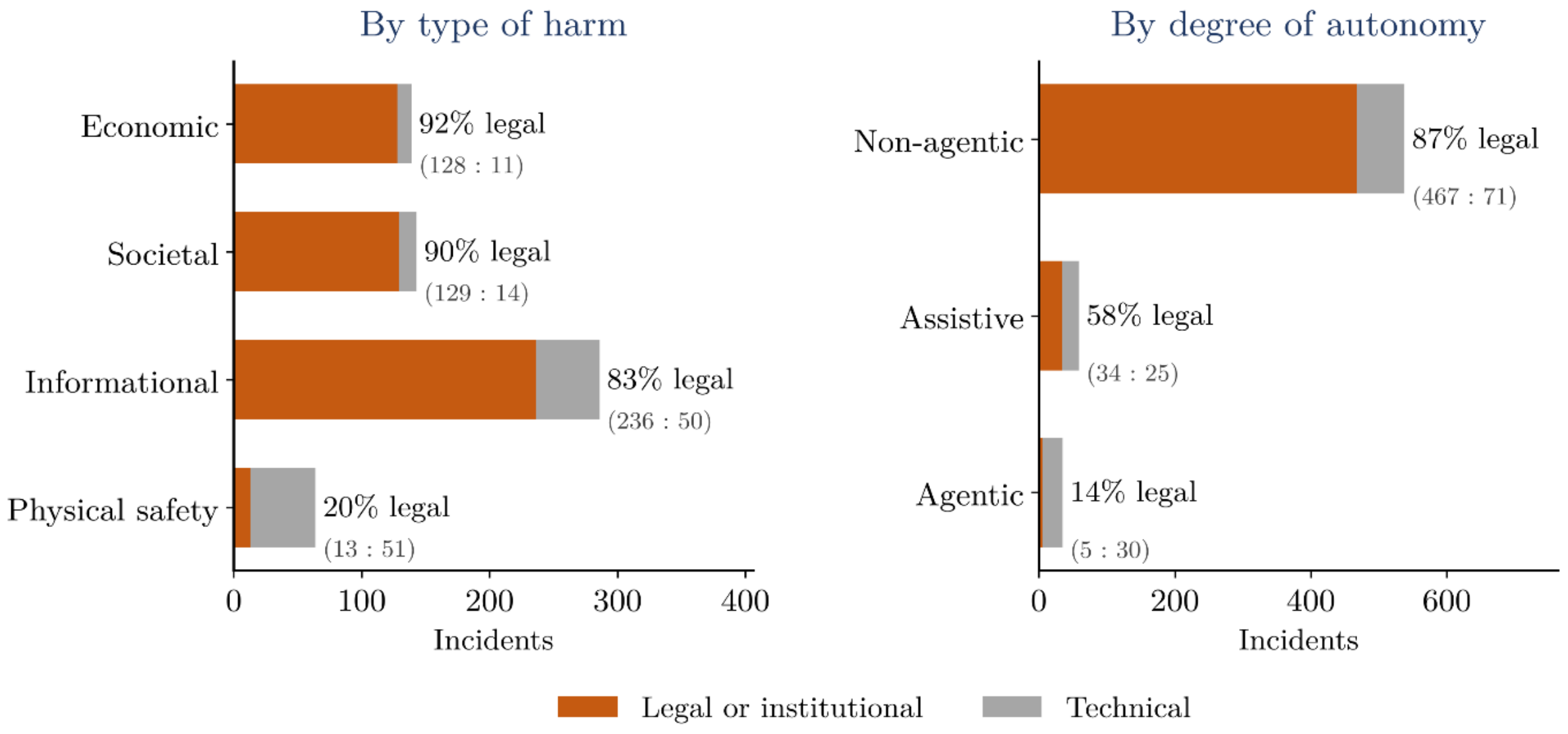

Figure A.3. *Among the 632 incidents in which a stop was possible but absent, the missing element by type of harm (left) and by degree of autonomy (right); exploratory breakdown, based on the hand-adjudicated agency and harm tags.*

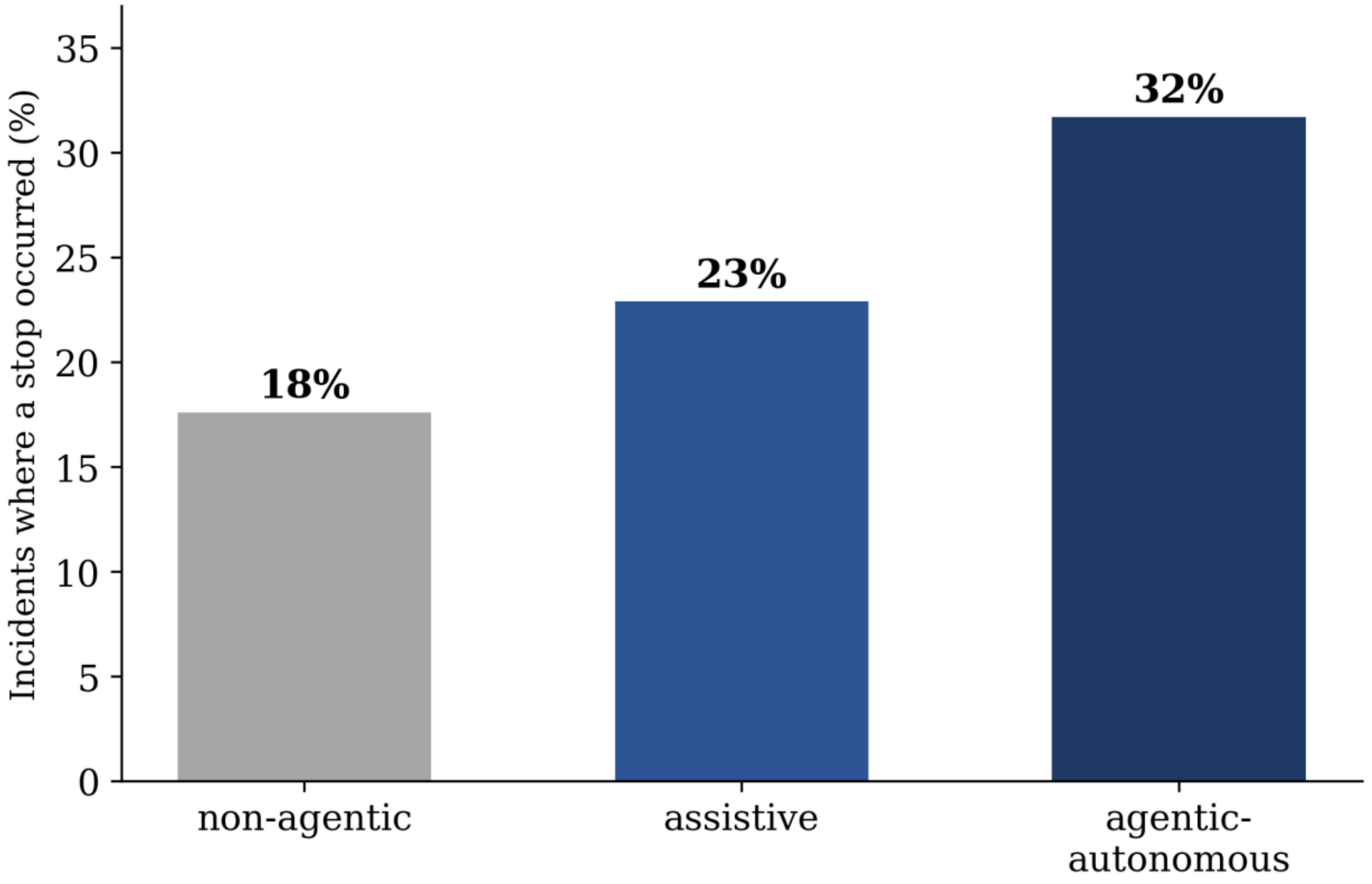

Figure A.4. *The share of incidents in which a stop occurred rises with the system's degree of agency ($p < .001$); the gradient holds when harm type is controlled for; based on the hand-adjudicated agency and harm tags*.

## Part B. The Halting Provisions in Full

This Part reproduces the halting provisions of the seven instruments, among the forty surveyed, that contain one; the remaining thirty-three, which contain no such provision, are entered in the table in Part C. The seven are ordered alphabetically by jurisdiction, and within each, the provisions appear in numerical order. A provision is included where it interrupts the operation of an AI system or the provision of a service. Measures directed at a single user, which leave the system running for everyone else, are excluded, as are duties to report, assess, or disclose, and general enforcement remedies addressed to a person's licensed activity rather than to a system. Translations are identified: the Chinese and Vietnamese texts are unofficial, the Korean texts the official English versions published by the Korean government.

### China: Interim Measures for the Management of Generative Artificial Intelligence Services (2023)

*Promulgated 13 July 2023 by the Cyberspace Administration of China with six other departments; effective 15 August 2023.*

**Article 14, first paragraph.**

> "Where providers discover illegal content they shall promptly employ measures to address it such as stopping generation, stopping transmission, and removal, employ measures such as model optimization training to make corrections and report to the relevant departments in charge."

**Article 21.**

> "Where providers violate these Measures, penalties are to be given by the relevant regulatory departments in accordance with the provisions of the PRC Cybersecurity Law, the PRC Data Security Law, the PRC Law on the Protection of Personal Information, the PRC Law on Scientific and Technological Progress, and other such laws and administrative regulations; and where laws and administrative regulations are silent, the relevant departments in charge are to give warnings, circulate criticism, or order corrections in a set period of time on the basis of their duties, and if corrections are refused or the circumstances are serious, an order is to be given to suspend the provision of the related services."

Article 14, second paragraph, which permits a provider to warn a user, limit functions, or suspend or conclude the provision of services to a user engaged in unlawful activity, is not counted here: it is directed at the user and leaves the service running for everyone else.

*Source. China Law Translate, https://www.chinalawtranslate.com/en/generative-ai-interim/ (unofficial translation).*

## China: Provisional Measures on the Administration of Human-like Interactive Artificial Intelligence Services (2026)

*Document No. 21, promulgated 10 April 2026 by the Cyberspace Administration of China with the National Development and Reform Commission, the Ministry of Industry and Information Technology and the Ministry of Public Security; effective 15 July 2026.*

**Article 24.**

> "Where the providers of human-like interactive services discover major security threats in human-like interactive services, they shall employ response measures such as limiting functions or stopping the provision of services to users, and store the relevant records."

**Article 30.**

> "Where the providers of human-like interactive services violate these Measures, departments such as for cybersecurity and informatization, development and reform, industry and informatization, and public security are to address and punish it in accordance with law and administrative regulations; and where laws and administrative regulations are silent, the departments such as for cybersecurity and informatization, industry and informatization, and public security are to give warnings, circulate criticism, or order corrections within a set period of time, based on their duties, and may require them to employ measures such as suspending user account registration or other relevant services; where corrections are refused or the circumstances are serious, they are to order them to stop providing relevant services, and may give fines of between 10,000 and 100,000 RMB; and where harms to citizens lives and health are involved, and harmful consequences result, fines of between 100,000 and 200,000 are to be given."

Article 19, which requires a provider to stop the service promptly where a user asks to exit, is not counted here for the reason given above. Nor is Article 25, which requires application stores and other distribution platforms to refuse to offer an offending app, suspend its services or delist it: its object is the app's availability in a distribution channel, not the model or the service itself. Article 24 is counted, although its wording is close to that of Article 14, second paragraph, of the 2023 Measures, because it is triggered by a major security threat in the service itself rather than by a user's conduct, and the measures it requires are directed at the service, not at a particular user.

*Source. China Law Translate, https://www.chinalawtranslate.com/human-like-ai/ (unofficial translation); original at https://www.cac.gov.cn/2026-04/10/c_1777558395078289.htm.*

## European Union: Regulation (EU) 2024/1689 (Artificial Intelligence Act)

*Adopted 13 June 2024; OJ L, 12.7.2024.*

**Article 14(4)(e).** Among the measures a provider must enable for the natural persons assigned to human oversight of a high-risk system, that person is to be able:

> "(e) to intervene in the operation of the high-risk AI system or interrupt the system through a 'stop' button or a similar procedure that allows the system to come to a halt in a safe state."

**Article 20(1).**

> "Providers of high-risk AI systems which consider or have reason to consider that a high-risk AI system that they have placed on the market or put into service is not in conformity with this Regulation shall immediately take the necessary corrective actions to bring that system into conformity, to withdraw it, to disable it, or to recall it, as appropriate. They shall inform the distributors of the high-risk AI system concerned and, where applicable, the deployers, the authorised representative and importers accordingly."

**Article 26(5).**

> "Deployers shall monitor the operation of the high-risk AI system on the basis of the instructions for use and, where relevant, inform providers in accordance with Article 72. Where deployers have reason to consider that the use of the high-risk AI system in accordance with the instructions may result in that AI system presenting a risk within the meaning of Article 79(1), they shall, without undue delay, inform the provider or distributor and the relevant market surveillance authority, and shall suspend the use of that system. Where deployers have identified a serious incident, they shall also immediately inform first the provider, and then the importer or distributor and the relevant market surveillance authorities of that incident."

**Article 79(2) and (5).** Paragraph 2 requires the market surveillance authority, where it finds that an AI system presents a risk, to require the operator:

> "to take all appropriate corrective actions to bring the AI system into compliance, to withdraw the AI system from the market, or to recall it within a period the market surveillance authority may prescribe, and in any event within the shorter of 15 working days, or as provided for in the relevant Union harmonisation legislation."

Paragraph 5 provides:

> "Where the operator of an AI system does not take adequate corrective action within the period referred to in paragraph 2, the market surveillance authority shall take all appropriate provisional measures to prohibit or restrict the AI system's being made available on its national market or put into service, to withdraw the product or the standalone AI system from that market or to recall it."

**Article 82(1).**

"Where, having performed an evaluation under Article 79, after consulting the relevant national public authority referred to in Article 77(1), the market surveillance authority of a Member State finds that although a high-risk AI system complies with this Regulation, it nevertheless presents a risk to the health or safety of persons, to fundamental rights, or to other aspects of public interest protection, it shall require the relevant operator to take all appropriate measures to ensure that the AI system concerned, when placed on the market or put into service, no longer presents that risk without undue delay, within a period it may prescribe."

**Article 83(2).** Where formal non-compliance of the kinds listed in Article 83(1) persists:

"the market surveillance authority of the Member State concerned shall take appropriate and proportionate measures to restrict or prohibit the high-risk AI system being made available on the market or to ensure that it is recalled or withdrawn from the market without delay."

**Article 93(1)(c).** Under the heading "Power to request measures," the Commission may request providers of general-purpose AI models to:

"(c) restrict the making available on the market, withdraw or recall the model."

A provider that fails to comply with a measure requested under Article 93 may be fined under Article 101(1)(c).

*Source. Official Journal of the European Union, L series, 12 July 2024, ELI: http://data.europa.eu/eli/reg/2024/1689/oj.*

## Korea: Framework Act on Intelligent Informatization

*Wholly amended by Act No. 17344, 9 June 2020. Official English translation, National Law Information Center.*

**Article 60 (Safety Protection Measures), paragraph (1), item 4.** The Minister of Science and ICT may determine and publicly notify the details and methods of minimum necessary protection measures, including:

"4. Matters concerning shutting down (hereinafter referred to as 'emergency shutdown') the operation of intelligent information technology and provision of intelligent information services externally in an emergency situation and provision of algorithm necessary for emergency shutdown;"

**Article 60, paragraph (2).**

"The Minister of Science and ICT may recommend persons who develop or utilize intelligent information technology and persons who provide intelligent information services to take safety protection measures as provided in the public notice given under paragraph (1)."

**Article 60, paragraph (3).**

> "The head of a central administrative agency may request a person who develops or utilizes intelligent information technology and a person who provides intelligent information services to activate emergency shutdown if necessary to prevent imminent harm to people's lives or bodies. In such cases, upon receipt of a request, the person shall comply with the request in the absence of good cause."

*Source. National Law Information Center, https://www.law.go.kr (English translation), version wholly amended by Act No. 17344, 9 June 2020, downloaded 6 September 2026. The Article cites the consolidated text, Act No. 20731, in force 31 January 2025; Article 60 is unchanged in substance between the two.*

## Korea: Framework Act on the Development of Artificial Intelligence and the Creation of a Foundation for Trust

*Act No. 20676, enacted 21 January 2025; enforcement date 22 January 2026. Official English translation, National Law Information Center.*

**Article 40 (Fact-finding investigations), paragraphs (1) and (3).** Paragraph (1) permits the Minister to require an artificial intelligence business operator to submit data or to have officials investigate where a violation of Article 31(2) or (3), Article 32(1) or (2), or Article 34(1) is discovered, suspected, reported or complained of. Paragraph (3) provides:

> "Where the Minister of Science and ICT recognizes, based on the results of investigations under paragraphs (1) and (2), that an artificial intelligence business entity has violated this Act, the Minister may order the artificial intelligence business operator to take necessary measures to cease or correct the violation."

Failure to comply with an order under Article 40(3) is subject to an administrative fine. The order is counted because its addressee is the artificial intelligence business operator, defined in Article 2(7) as the developer that provides artificial intelligence or the business operator that provides products or services using it, and because its predicates are the Act's own safety duties: Article 32(1) (identification, assessment and mitigation of risk across the lifecycle; a risk management system that monitors and responds to safety accidents) and Article 34(1) (for high-impact artificial intelligence, a risk management plan, an explanation plan, a user protection plan, the assignment of human management and oversight, and verifying documentation). An operator providing high-impact artificial intelligence without those measures may be ordered to cease the provision until they are in place. Neither Article 32 nor Article 34 names a shutdown capability.

*Source. National Law Information Center, https://www.law.go.kr (English translation), text as at 15 April 2026.*

## United States, California: AI Transparency Act (S.B. 942, 2024)

*Chapter 291, approved by the Governor 19 September 2024; Cal. Bus. & Prof. Code ch. 25 (commencing with § 22757).*

**Section 22757.3(c)(2).**

> "If a covered provider knows that a third-party licensee modified a licensed GenAI system such that it is no longer capable of including a disclosure required by subdivision (b) in content the system creates or alters, the covered provider shall revoke the license within 96 hours of discovering the licensee's action."

**Section 22757.3(c)(3).**

> "A third-party licensee shall cease using a licensed GenAI system after the license for the system has been revoked by the covered provider pursuant to paragraph (2)."

**Section 22757.4(c).**

> "For a violation by a third-party licensee of paragraph (3) of subdivision (c) of Section 22757.3, the Attorney General, a county counsel, or a city attorney may bring a civil action for both of the following: (1) Injunctive relief. (2) Reasonable attorney's fees and costs."

*Source. Senate Bill No. 942, Chapter 291 (2024), as chaptered. The chapter was amended by Assembly Bill No. 853, Chapter 674 (2025), which amended sections 22757.1, 22757.4 and 22757.6 and added sections 22757.3.1 to 22757.3.3; it did not amend section 22757.3, and it set the chapter's operative date at 2 August 2026. Section 22757.4(a)(1) now reads "A violator of this chapter" in place of "A covered provider that violates this chapter," and section 22757.4(b) now applies to large online platforms and capture device manufacturers as well as covered providers.*

## Vietnam: Law on Artificial Intelligence, Law No. 134/2025/QH15

*Adopted by the 15th National Assembly on 10 December 2025; in force 1 March 2026. Unofficial English translation published by the Vietnam Government Portal.*

**Article 12 (Management and handling of artificial intelligence incidents).** Where a serious incident occurs in an artificial intelligence system:

> "Developers and providers must promptly apply technical measures to rectify, temporarily suspend, or recall the system, and simultaneously notify the competent authority; Deployers and users shall have the obligation to record and promptly notify the incident, and coordinate in the rectification process. Competent State management agencies shall receive, verify, and guide the handling of incidents; when necessary, they shall have the power to require temporary suspension, recall, or re-assessment of the artificial intelligence system."

**Article 35 (Transitional provisions).** For systems put into operation before the Law took effect, compliance is required within eighteen months in healthcare, education and finance and within twelve months otherwise; during that period:

> "the artificial intelligence systems may continue to operate, except where the State management authority in charge of artificial intelligence determines that such systems pose risks of causing serious harm, such authority shall have the power to request the suspension or termination of operations of the systems."

*Source. Vietnam Government Portal, full translation of the Law on Artificial Intelligence, marked "For reference only." The translation runs the clauses of Article 12 as unnumbered paragraphs; pinpoint citations in the Article follow the numbered structure of the Vietnamese original.*

## Part C. Stop Provisions Across the Instruments Examined

This Part reproduces the full table from which Figure 5 in Part IV.A is drawn. It records, for each of the forty instruments surveyed, the halting provision relied on, the authority in whom the power or duty is placed, and what the halt is directed at.

Table C.1. Stop provisions across the instruments examined

| Instrument | Halting provision | Halting Authority | Target of the stop |
|---|---|---|---|
| ***National and regional*** | | | |
| 1. European Union, AI Act, Reg. (EU) 2024/1689, with the General-Purpose AI Code of Practice (2025) | Art. 14(4)(e) stop button; arts. 13(3)(d) and 11(1) with Annex IV, documentation of the Art. 14 measures.<br>Art. 26(5) deployer suspension on reason to consider a risk.<br>Art. 20 provider's corrective action, disabling, withdrawal, recall.<br>Arts. 79(2), 79(5) market-surveillance measures; 82(1) compliant system presenting a risk; 83(2) formal non-compliance.<br>Art. 93(1)(c) request to restrict availability, withdraw or recall the model.<br>Code Measure 4.2, restriction, withdrawal or recall where systemic risks are not acceptable; app. 1.4(2). | The deployer's assigned overseer; no public authority.<br>The deployer.<br>The provider.<br>The national market surveillance authority.<br>The Commission.<br>The model provider. | The AI system.<br>The deployer's use of the system.<br>The system's market availability.<br>The system's market availability.<br>The model's market availability.<br>The model's market availability. |
| 2. United States, Executive Order 14409 (2026) | None. Voluntary pre-release access to covered frontier models for up to 30 days; express disclaimer of any licensing, preclearance or permitting requirement, §§ 3(b)(ii), 3(c). | No one. | Nothing. |
| 3. United States, White House Accord on Super Intelligence, Joint Commitment on Frontier Responsibilities (2026) | None. Voluntary commitment by six companies to "four layers of controls and audits" (internal controls, an internal team, an independent external auditor or evaluator, a board committee), stated as belief and recommendation ("we believe each company should implement"); no provision for halting, suspension or shutdown. | No one. | Nothing. |
| 4. United States, California, SB 53, Transparency in | None. Duty to report a critical safety incident, defined to include loss of control causing death or bodily injury and a | No one; the Act's only injunction is | Nothing. |

| Instrument | Halting provision | Halting Authority | Target of the stop |
|---|---|---|---|
| Frontier Artificial Intelligence Act (TFAIA) (2025) | model's use of deceptive techniques to subvert its developer's controls. | a whistleblower remedy. | |
| 5. United States, California, AI Transparency Act, SB 942 (2024) | § 22757.3(c)(2) provider must revoke the license within 96 hours on knowing that a licensee has modified the system so that it no longer carries the provenance disclosure; (c)(3) licensee must cease using it; § 22757.4(c) injunction. | The provider; enforced by the Attorney General, a county counsel or a city attorney. | A licensee's use of the system. |
| 6. United States, Colorado, ADMT in Consequential Decisions, SB 26-189 (2026) | None. | No one. | Nothing. |
| 7. United States, New York, RAISE Act, ch. 699 (2025) | None. Duty to report a critical safety incident, defined as in SB 53. | No one. | Nothing. |
| 8. United States, Texas, TRAIGA, HB 149 (2025) | None addressed to a system. § 552.105(b)(2) injunction against further violation; § 552.106(b)(1) suspension, probation or revocation of a license to engage in an activity, on the attorney general's recommendation after a finding of violation. | No one. | Nothing. |
| 9. United States, Utah, AI Policy Act, SB 149 (2024) | None. | No one. | Nothing. |
| 10. United States, twelve-state conversational-AI and companion-chatbot statutes (2025-2026) | None. Duties to disclose AI status and to refer a user expressing suicidal ideation to crisis services; in three, a prohibition on discouraging a user from taking a break. | No one. | Nothing. |
| 11. China, Generative AI Interim Measures (2023) | Art. 14 ¶1 duty to stop generation and transmission of illegal content, remove it, and correct by measures such as model optimization training.<br><br>Art. 21 order to suspend provision of the service. | The provider.<br><br>The regulator. | An information flow.<br><br>The service. |
| 12. China, Labeling of AI-Generated | None. | No one. | Nothing. |

| Instrument | Halting provision | Halting Authority | Target of the stop |
|---|---|---|---|
| Synthetic Content (2025) | | | |
| 13. China, Human-like Interactive AI Services Measures (2026) | Art. 24 duty to limit functions or stop providing the service on a major security threat, keeping records.<br>Art. 25 application-store duty to suspend services or delist.<br>Art. 30 order to suspend user registration or related services, and to cease providing the service. | The provider.<br>The application store.<br>The regulator. | The service.<br>The app's availability in distribution channels; not counted.<br>The service. |
| 14. Korea, Framework Act on Intelligent Informatization | Art. 60(1)(4) emergency shutdown among the minimum protection measures the Minister may notify and recommend for adoption; art. 60(3) request to developers, users and providers to activate it, which they must obey absent good cause. | The head of any central administrative agency. | The AI system. |
| 15. Korea, Framework Act on AI (2026) | Art. 40(3) order to an AI business operator to take necessary measures to cease or correct a violation of the safety duties in arts. 31, 32 and 34, including the human-oversight duty of art. 34(1)(4). Arts. 32(1)(2) and 34(1)(4), a risk-management system responsive to safety accidents and assigned human oversight, in which a shutdown capability may be implicit. | The Minister of Science and ICT; the duties under arts. 32 and 34 fall on the AI business operator. | The operator's provision of the system. |
| 16. Vietnam, Law on AI, No. 134/2025/QH15 | Art. 12 duty to rectify, temporarily suspend or recall on a serious incident, and power of the competent State agency to require suspension, recall or reassessment.<br>Art. 35 power, during the transition, to require suspension or termination of the operations of systems already running. | The developer and provider, and the competent State authority.<br>The State management authority. | The AI system, and its market availability.<br>The AI system. |
| ***International*** | | | |
| 17. Council of Europe, Framework Convention, CETS 225 | None. | No one. | Nothing. |
| 18. G7, Hiroshima Code of Conduct (2023) | None. | No one. | Nothing. |

| Instrument | Halting provision | Halting Authority | Target of the stop |
|---|---|---|---|
| 19. OECD, AI Principles, 1.4(b) | None binding. Adherents should provide mechanisms, as appropriate, so that systems can be overridden, repaired and decommissioned safely. | No one. | The AI system. |
| ***Technical standards and codes*** | | | |
| 20. ISO/IEC 42001:2023 | None binding. Annex B, which is guidance only, refers to a rollback plan and turning off features as items in a documented failure-management plan (B.6.2.7), to telling users how and when to override the system (B.8.2), and to human authority to override the system's decisions as an oversight objective (B.9.3). Each is a matter to consider; none requires that the system be capable of being stopped. | The organization, as it chooses; no external authority. | The AI system. |
| 21. NIST, AI Risk Management Framework 1.0 (2023) | None binding. MANAGE 2.4: mechanisms in place, with responsibilities assigned, to supersede, disengage or deactivate AI systems whose performance or outcomes are inconsistent with intended use; GOVERN 1.7 and MANAGE 4.1: safe decommissioning, appeal and override; § 3.2: "the ability to shut down, modify, or have human intervention." | AI actors; no external authority. | The AI system. |
| 22. SAC/TC260-003, Basic Safety Requirements (2024) | None. Cl. 7(g)(1), suspension of service to a user after three consecutive or five daily violations, is a measure against the user rather than the system; cl. 5.3(a)(2) suspends annotator eligibility. | No one. | Nothing. |
| 23. SAC/TC260-005, Ethics-Safety Guidelines (2026) | None binding. The guidelines require that human control be maintained and that service providers establish emergency intervention mechanisms. | No one. | The application. |
| ***Corporate*** | | | |
| 24. Anthropic, Responsible Scaling Policy v3.4 | None binding. Delay of AI development and deployment in two of three competitor scenarios, app. A; a stated readiness to consider pausing development and/or deployment. | No one. | Development and deployment. |
| 25. Google DeepMind, Frontier Safety Framework v3.1 | None binding. External deployment follows a determination that residual risk is acceptable once mitigations are applied, §§ | No one. | Deployment. |

| Instrument | Halting provision | Halting Authority | Target of the stop |
|---|---|---|---|
| | 1.3.5, 2.1.2; no provision for stopping a model once deployed. | | |
| 26. Microsoft, Responsible AI Standard v2 (2022) | None binding. Oversight stakeholders must be able to understand when and how to override, intervene or interrupt the system, A5.2; the developer must describe the rollback plan and support for turning features off, RS2.3; discontinuing the system is one of three alternatives where evidence refutes fitness for purpose, A3.7, RS3.7. | No one. | The AI system, and the provision of a system for an intended use. |
| 27. OpenAI, Preparedness Framework v2 | None binding. Halt of further development until safeguards meeting a Critical standard are specified; containerization and restricted permissions among illustrative safeguards, app. C.2. | No one. | Development. |
| 28. Seoul Frontier AI Safety Commitments (2024) | None binding. Commitment not to develop or deploy where mitigations cannot keep risk below defined thresholds. | No one. | Development and deployment. |